\documentclass[journal]{IEEEtran}

\ifCLASSINFOpdf
\else
\fi
\usepackage[utf8x]{inputenc}
\usepackage{amssymb}
\usepackage{graphicx}
\usepackage{url}
\usepackage{amsmath}
\usepackage{comment}
\usepackage{subfigure}
\usepackage{balance}
\usepackage{amsfonts}
\usepackage{multirow}

\usepackage{balance}

\usepackage{color, soul}
\usepackage[dvipsnames]{xcolor}

\usepackage{amsthm}

\begin{document}




\onecolumn
This is an author-submitted to IEEE Communications Surveys and Tutorials article.

\vspace{2mm}
© 2021 IEEE.  Personal use of this material is permitted.  Permission from IEEE must be obtained for all other uses, in any current or future media, including reprinting/republishing this material for advertising or promotional purposes, creating new collective works, for resale or redistribution to servers or lists, or reuse of any copyrighted component of this work in other works.
\twocolumn
\newpage

\title{A Tutorial on Mathematical Modeling of Millimeter Wave and Terahertz Cellular Systems}

\author{{Dmitri~Moltchanov, Eduard~Sopin, Vyacheslav Begishev,\\ Andrey Samuylov, Yevgeni Koucheryavy, and Konstantin~Samouylov}
\thanks{D. Moltchanov, and Y. Koucheryavy are with Tampere University, Finland. Email:~{firstname.lastname}@tuni.fi}
\thanks{E. Sopin, V. Begishev,  A. Samuylov, and K. Samouylov are with Peoples' RUDN University, Moscow, Russia. Email:~{begishev-vo}@rudn.ru, \{sopin-es, begishev-vo, samuylov-ak, samuylov-ke\}@rudn.ru}
\thanks{E. Sopin and K. Samouylov are also with Institute of Informatics Problems, Federal Research Center Computer Science and Control of Russian Academy of Sciences, Moscow, Russia.}
\vspace{-0mm}}

\maketitle	

\begin{abstract}
Millimeter wave (mmWave) and terahertz (THz) radio access technologies (RAT) are expected to become a critical part of the future cellular ecosystem providing an abundant amount of bandwidth in areas with high traffic demands. However, extremely directional antenna radiation patterns that need to be utilized at both transmit and receive sides of a link to overcome severe path losses, dynamic blockage of propagation paths by large static and small dynamic objects, macro- and micromobility of user equipment (UE) makes provisioning of reliable service over THz/mmWave RATs an extremely complex task. This challenge is further complicated by the type of applications envisioned for these systems inherently requiring guaranteed bitrates at the air interface. This tutorial aims to introduce a versatile mathematical methodology for assessing performance reliability improvement algorithms for mmWave and THz systems. Our methodology accounts for both radio interface specifics as well as service process of sessions at mmWave/THz base stations (BS) and is capable of evaluating the performance of systems with multiconnectivity operation, resource reservation mechanisms, priorities between multiple traffic types having different service requirements. The framework is logically separated into two parts: (i) parameterization part that abstracts the specifics of deployment and radio mechanisms, and (ii) queuing part, accounting for details of the service process at mmWave/THz BSs. The modular decoupled structure of the framework allows for further extensions to advanced service mechanisms in prospective mmWave/THz cellular deployments while keeping the complexity manageable and thus making it attractive for system analysts.
\end{abstract}

\section{Introduction}\label{sect:intro}


As the standardization of fifth-generation (5G) New Radio (NR) technology operating in microwave and millimeter wave (mmWave) bands is over, the focus of the research community is shifting towards performance optimization of these systems for future cellular deployments \cite{volk20215g,antoniou2021quality,hutajulu2020two}. At the same time, seeking for even more capacity at the air interface, the researchers already start to entertain the challenge of utilizing terahertz (THz) frequency band for cellular communications having tens of even hundreds of GHz of consecutive bandwidth available \cite{petrov2020ieee,giordani2020toward,polese2020toward,akyildiz2014terahertz}.


The utilization of mmWave and THz frequency bands promises to not only bring the extreme capacity to the air interface enabling novel rate-greedy applications such as virtual reality (VR), augmented reality (AR), and holographic telepresence but to enable truly multi-service access networks delivering guarantees to applications sensitive to various parameters such as delay and throughput \cite{parkvall2017nr,antoniou2021quality}. However, the inherent properties of these bands including the need for extremely high radiation patterns at both sides of communications link to compensate for extreme path losses \cite{andrews2014will,sadhu2018128,akyildiz2016realizing}, blockage of propagation paths by large objects such as buildings \cite{ruiz2020analysis,mohebi2021sectors,tafintsev2020handling} or small dynamic obstacles such as human bodies and vehicles \cite{dynamicBlockage,maccartney2016millimeter,petrov2019exploiting}, as well as macro- and micromobility of user equipment (UE) \cite{orsino2016direct,petrov2020capacity,petrov2018effect,moorthy2020beam} may drastically affect the performance of these technologies frequently leading to either drastic rate degradation or even outage situations. In fact, the performance of these radio access networks (RAT), tailored specifically towards service provisioning in crowded environments with extreme traffic demands, will suffer most in these conditions. To overcome these problems, novel mechanisms maintaining the advertised performance are needed. To evaluate the performance of these mechanisms system-level performance evaluation methodologies are required.


The system-level performance evaluation of cellular communications systems is conventionally performed utilizing either a purely mathematical approach or via computer simulations. As discussed in \cite{gkonis2020comprehensive,begishev2018connectivity,gapeyenko2017temporal} the multi-path propagation and as well as dynamic blockage phenomena drastically affect the efficiency of simulation techniques. At the physical layer, the extreme mmWave and THz propagation sensitivity to different surfaces naturally call for ray tracing techniques for precise modeling \cite{yaman2021ray,solomitckii2019evaluation} inducing accuracy-complexity trade-off \cite{lecci2020accuracy,lecci2020simplified}. Further, accounting for blockage requires capturing not only static scenario geometry and tracking UEs having active sessions but accounting for all of the dynamic objects in the channel, e.g., humans, vehicles. In these conditions, the utilization of mathematical frameworks may provide a viable way for the first-order assessment of the novel mechanisms improving the performance of mmWave and THz systems. In doing this, the frameworks proposed in the past for LTE systems \cite{elsawy2016modeling,elsawy2013stochastic} have to be properly extended to account for specifics of mmWave and THz communications including highly directional antenna radiation patterns, static and dynamic blockage, atmospheric attenuation, etc. Notably, those frameworks have been developed assuming adaptive (full buffer) traffic patterns inherently adaptive to network state and thus mainly utilized the elements of stochastic geometry \cite{haenggi2012stochastic,haenggi2009stochastic,blaszczyszyn2018stochastic}.
Contrarily, mmWave and THz RATs, having extreme capacity at their disposal, are expected to primarily target applications generating non-elastic/adaptive traffic and requiring high and guaranteed bitrates at the air interface and having no or limited application layer adaptation capabilities \cite{vannithamby2017towards,bariah2020prospective,ghosh2020nr}. Thus, the performance evaluation frameworks tailored towards mmWave and THz RATs have to take into account not only the specifics of the radio part and stochastic factors related to the randomness of UE locations, but the traffic service dynamics at BSs by joining the tools of stochastic geometry and queuing theory.



The goal of this manuscript is to provide a comprehensive tutorial on mathematical performance analysis of prospective mmWave and THz RAT deployments supporting non-elastic/adaptive traffic and implementing advanced capabilities for improving service reliability at the air interface. To this aim, we first provide an exhaustive survey of analytically tractable models of various components utilized for building the modeling scenarios including deployment, propagation, antenna, blockage, micromobility, beamsearching, traffic, and service models for a different system and environmental conditions. We discuss the abstraction levels and parameterization techniques as well as comment on the potential pitfalls and applicability of these models to considered RATs. We also define and evaluate both user-centric and system-centric key performance indicators (KPI). For non-elastic/adaptive traffic the former are mainly related to the service reliability and include the new session drop probability and the ongoing session drop probability while the latter is defined in terms of the ratio of utilized to available resources at BSs and characterizes the efficiency of resource utilization. More detailed discussion on these KPIs is provided in Sections \ref{sect:metrics} and \ref{sect:approaches}.


Then, we introduce the structure of the generic performance evaluation framework capable of simultaneously capturing radio part details of mmWave/THz systems as well as traffic service dynamics at BSs. For flexibility reasons, the contributed framework is divided into two parts: (i) service and (ii) radio abstraction. Specifically, the latter characterizes the type of deployment and abstracts the stochastic effects of radio channel via a separate parameterization part, i.e., UE locations, blockage, and micromobility, and represents them via a predefined set of parameters in the form suitable for the queuing part. The queuing part accepts the probability mass function (pmf) of the amount of requested resources by arriving sessions and the temporal intensity of the UE stage changes induced by blockage and micromobility processes and utilizes them to produce performance metrics. 

The modular structure of the framework allows for the reuse of its core models for studying various mmWave and THz deployments characterized by different scenario geometry, blocker types and their mobility, antenna arrays, micromobility patterns of applications, network associations, reliability and rate improvement mechanisms, etc. In addition to the baseline model having a single traffic type, no priorities, and no reliability improvement mechanisms, we consider in detail more sophisticated traffic service processes at BSs in incremental order of complexity, including systems with resource reservation, multiconnectivity, and models with multiple UE types and priorities. Note that additional models can be built on top of these, e.g., by uniting multiconnectivity and priorities and defining additional rules to utilize the former, one could produce a model of a network segment with multiconnectivity capabilities servicing more than a single traffic type. In overall, in this tutorial we provide our readers with building blocks that can be selected and then combined to form a comprehensive performance evaluation framework suitable for a given deployment type and use-case of mmWave/THz cellular systems.



Our main contributions are:
\begin{itemize}
  \item{a comprehensive review of analytically tractable models utilized in system-level mathematical performance evaluation frameworks targeting mmWave and THz communications systems including deployment, propagation, antenna, blockage, micromobility, beamsearching, traffic and service models;}
  \item{compound performance evaluation methodology tailored at evaluating user- and system-centric KPIs of mmWave and THz communications systems capturing their critical specifics and consisting of two independent parts interfaced with each via a predefined set of parameters;}
  \item{detailed treatment of several cases of specific service processes at BS side (baseline, multiconnectivity, resource reservation, explicit priorities) as well as examples of parameterization of the framework.}
\end{itemize}


The rest of the manuscript is organized as follows. First, in Section \ref{sect:specifics} we provide an outlook of state-of-the-art in 5G/6G cellular systems utilizing mmWave and THz band as well as overview the applications, use-cases, and traffic specifics. Further, in Section \ref{sect:approaches} we review the conventional stochastic geometry approach to performance evaluation of cellular systems, discuss its limitations for traffic pattens that are expected to be inherent for mmWave/THz5G/6G systems and formulate the basics of joint approach accounting for radio and service parts simultaneously. In Section \ref{sect:system} we introduce the mathematically tractable models of individual components of mmWave and THz communications systems. These models are further combined together in Section \ref{sect:param} providing abstraction of the mmWave/THz radio channel specifics. Queuing-theoretic service models for such systems are introduced in Section \ref{sect:perf}. Finally, conclusions and future challenges are drawn in the last section.

\section{Millimeter Wave and Terahertz Systems}\label{sect:specifics}

In this section, we start by briefly reminding the state-of-the-art and specifics of mmWave and THz systems that differentiate them from 4G LTE systems operating in microwave band. Then, we discuss two challenges requiring development of new mechanisms: (i) reliable communications over mmWave/THz systems and (ii) support of multiple traffic types in these RATs. Finally, we provide overview of applicarions, use-cases and traffic specifics for mmWave/THz cellular systems.

\subsection{5G NR mmWave Systems}

Directional wireless communications in mmWave bands is one of the most significant novelties introduced in the fifth-generation (5G) wireless networks. To this aim, 3GPP and ITU-R presumes the use of the two frequency ranges, up to 6 GHz (FR1) and higher than 6 GHz (FR2), where the latter mainly utilizes frequencies in the mmWave band. MmWave communications at frequencies between 24.25 GHz and 52.6 GHz allow for data exchange at the rates of several gigabits per second (Gbps) \cite{ericsson5G}. To overcome the limitation of LTE networks, the 3GPP has recently defined NR (New Radio) technology. NR exploits the mmWave band and supports new techniques such as massive Multiple Input Multiple Output (MIMO), flexibility in terms of the frame structure, target different use cases, and multiple deployment options for the Radio Access Network (RAN) \cite{polese20183gpp}. Another  NR novelty with respect to LTE is the support for ultra-low latency communications at the air interface to target the sub-1 ms latency requirement.

The initial phase of the standardization process for mmWave NR was finished in 2018 and resulted in the first NR release, Release 15, standardized by 3GPP, where the support of mmWave band was limited to non-stand-alone (NSA) operation. Since that, the evolution of 5G New Radio (NR) has progressed swiftly. Release 16 was completed in early July 2020. The most notable enhancements to the existing features in Release 16 lie in the areas of MIMO design and beamforming enhancements, dynamic spectrum sharing (DSS), dual connectivity (DC), carrier aggregation (CA), and user equipment (UE) power saving functions \cite{ericssonReport1718}. Also, in July 2020, 3GPP 5G Release 15 and 16 were formally endorsed as ITU IMT-2020 5G standard \cite{ieee5GmmWave}. In addition to the utilization of sub-6GHz spectrum, Release 16 also enables stand-alone (SA) mmWave band operation.

Release 17 will introduce new features for the three main use-case families: (i) enhanced mobile broadband (eMBB), (ii) ultra-reliable low latency communications (URLLC), and (iii) massive machine-type communications (mMTC). The purpose is to support the expected growth in mobile-data traffic and customize NR for automotive, logistics, public safety, media, and manufacturing use cases. Some of the new functionalities added in Release 17 are, for instance, extended NR frequency range to allow exploitation of more spectrum, utilization of the 60 GHz unlicensed band via CA functionality, the definition of new OFDM numerology and channel access mechanism, anything reality (XR) evaluations to support various forms of AR/VR, support of reduced-capability NR UEs for mMTC.

Nowadays, 3GPP has also started to plan Release 18 and beyond. To enable large-scale mmWave network deployment, the research community still faces significant challenges, such as massive MIMO beamforming for outdoor and indoor coverage, cost-effective deployment solutions, seamless mobility with effective beam management, network and device energy efficiency, mobile device complexity, enabling bands higher than 52.6 GHz, etc. In addition to the existing challenges, there are numerous new emerging use cases for 5G mmWave NR, such as uplink-centric traffic, real-time immersive communications, positioning for IIOT, sensing. Some other notable challenges to solve are outlined in the pre-Release 18 presentation \cite{3gpp5gplans}. They include NR sidelink enhancements related to the use of unlicensed bands, power saving, relay enchantments and co-existence with LTE V2X and NR V2X technologies. Non-Terrestrial Networks evolution, including both NR and IoT aspects, together with the evolution for broadcast and multicast services, including both LTE based 5G broadcast and NR MBS (Multicast Broadcast Services), are also targeted.


For both microwave and millimeter wave bands NR provides flexible resource allocations by utilizing different sets of the so-called numerologies defining the flexible time and bandwidth allocation unit for efficient operation over large bandwidth \cite{nrmcs}. On top of this, NR received significant enhancements as compared to LTE with respect to the initial access phase, synchronization mechanism, channel modulation and coding, etc. For an in-depth description of NR radio interface, we refer to specialized two recent monographs \cite{dahlman20205g, bhattacharyya2013handbook}. For physical channel models, NR design considerations, antenna constructions, and link-budget calculation we also refer our readers to an excellent survey paper in \cite{hemadeh2017millimeter}.

\begin{figure*}[!t]
\hspace{-0mm}
\subfigure[{Different BS antenna arrays, 28 GHz}]{
	\includegraphics[width=0.3\textwidth]{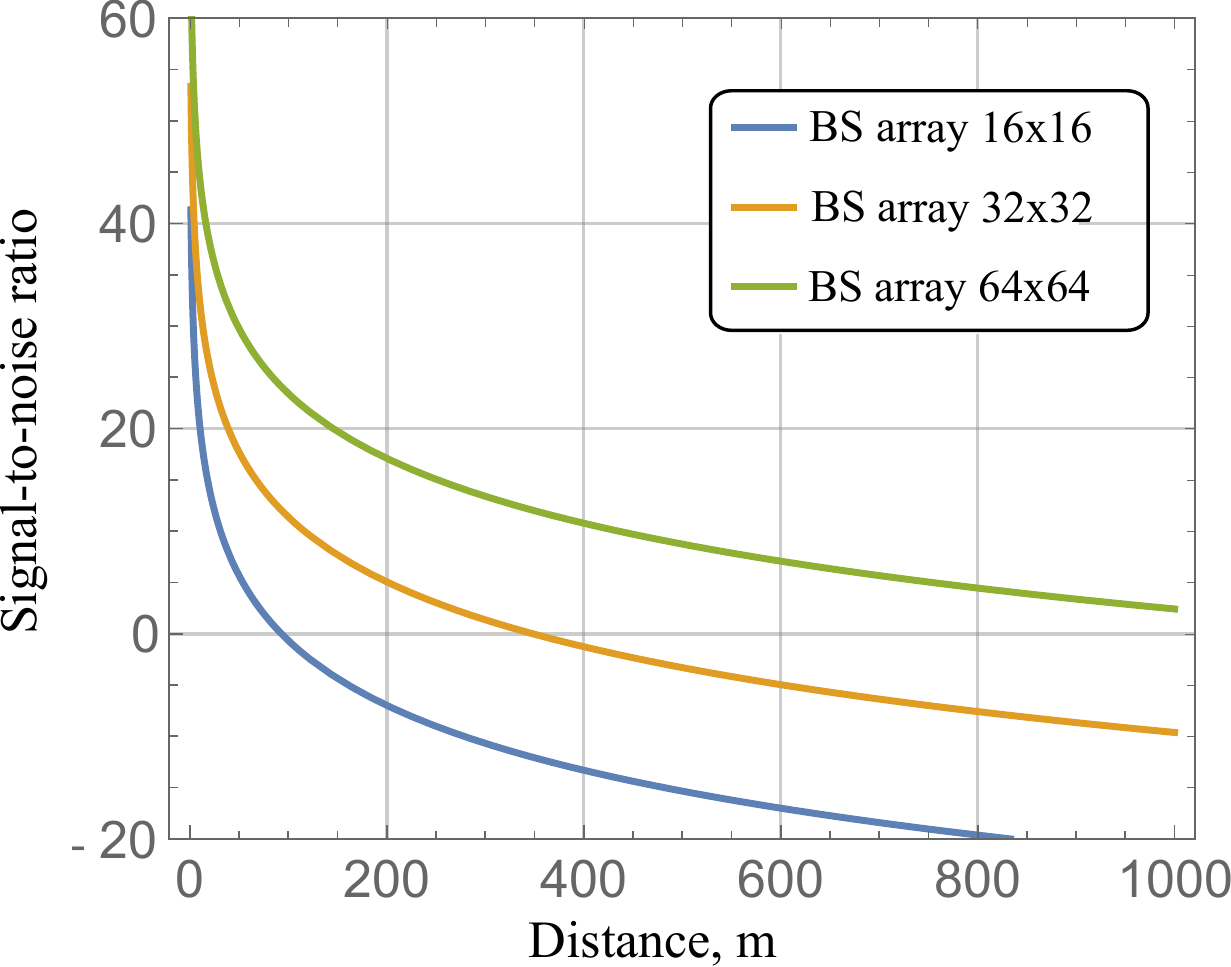}
	\label{fig:effectsChannel_antenna}
}
\subfigure[{Different absorption compounds}]{
	\includegraphics[width=0.3\textwidth]{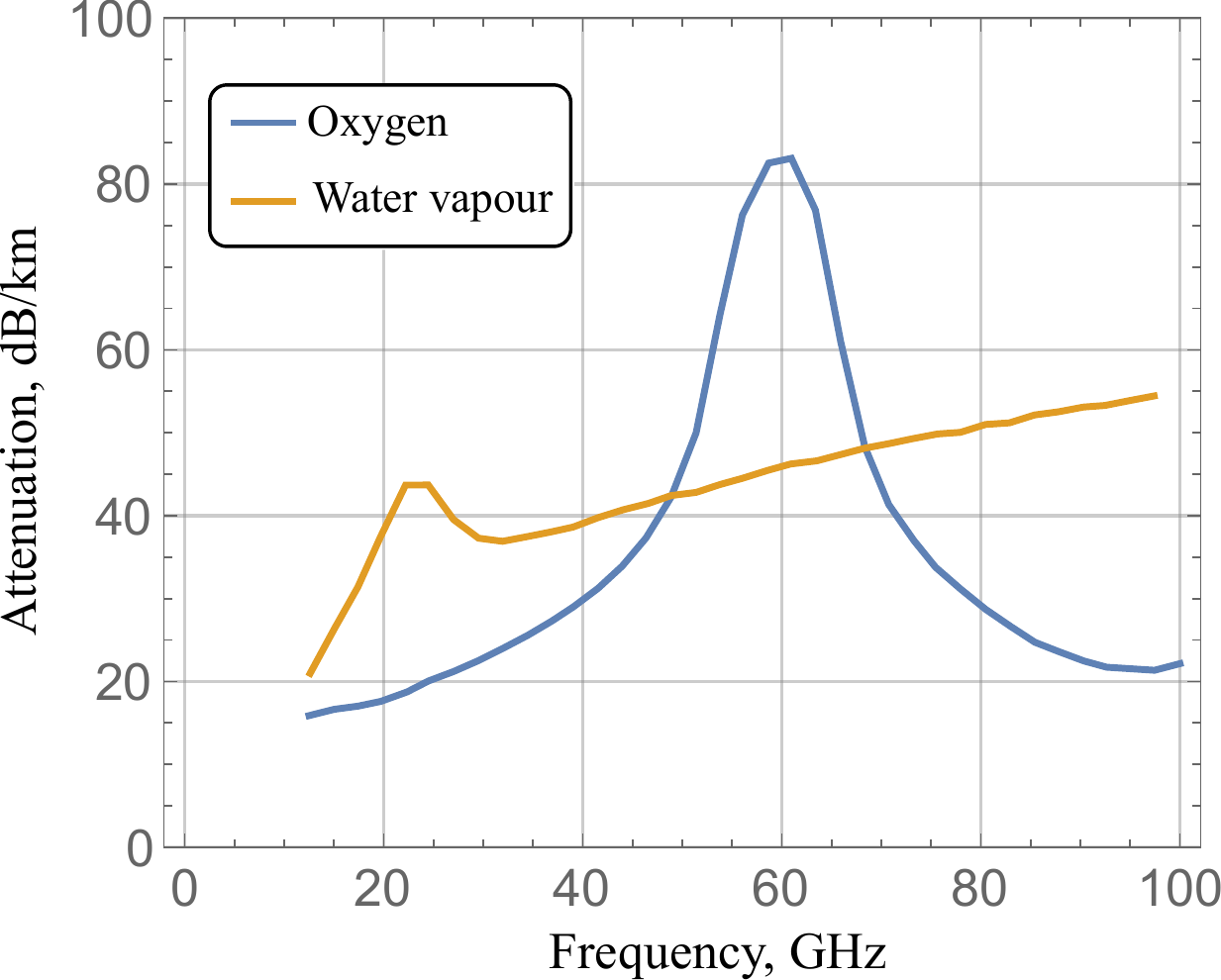}
	\label{fig:effectsChannel_absorption}
}
\subfigure[{Different whether conditions, 28 GHz}]{
	\includegraphics[width=0.3\textwidth]{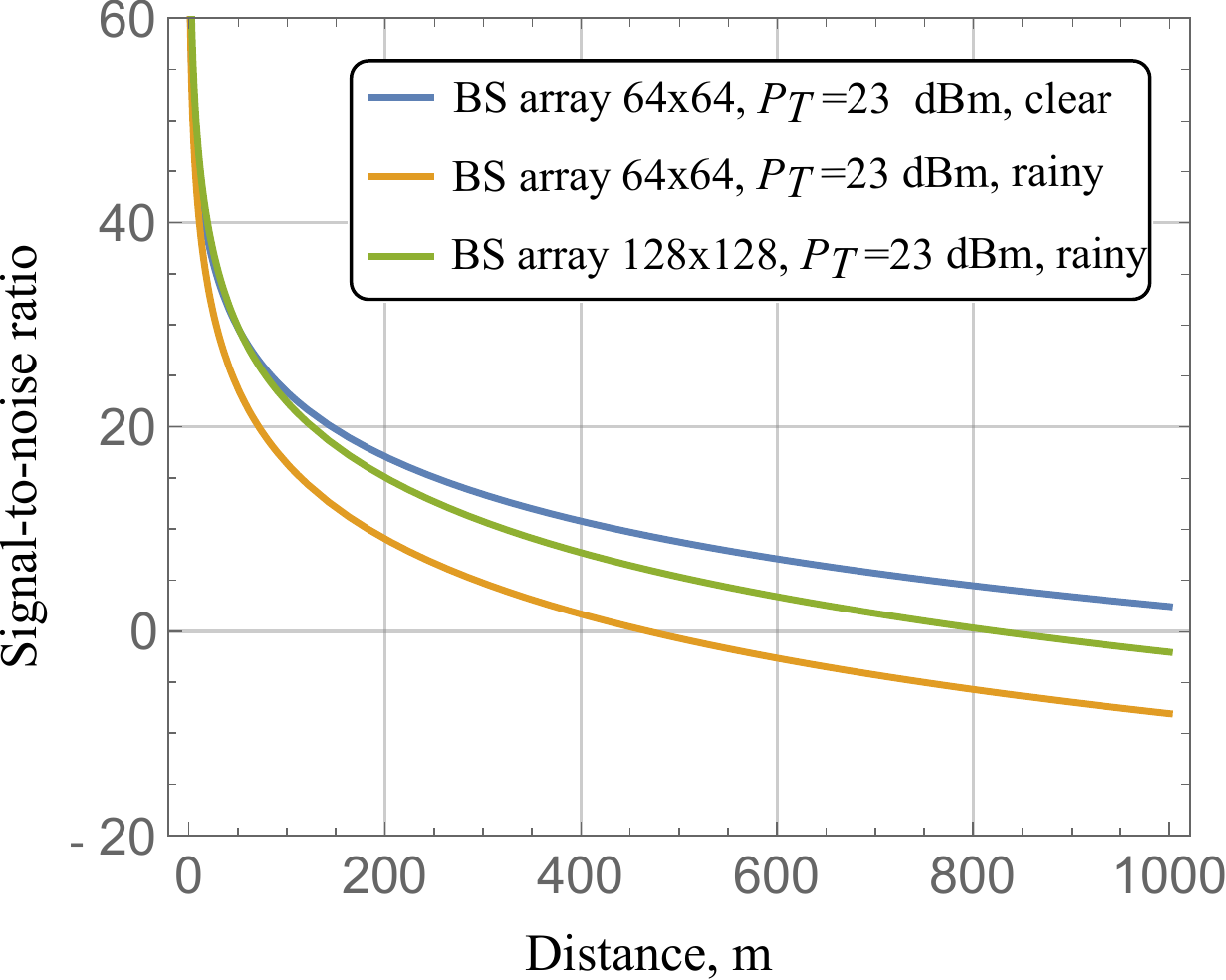}
	\label{fig:effectsChannel_whether}
}
\caption{Effect of environmental and system parameters in mmWave/THz channel performance.}
\label{fig:effectsChannel}
\vspace{-0mm}
\end{figure*}

\subsection{6G THz Systems}

Terahertz (THz) communications systems utilizing the range of frequencies  from 100 GHz up until 3 THz are seen as a key technology for 6G wireless systems to meet the demands for extremely high data rates and ultra-low end-to-end latency in the next decade \cite{han2019terahertz}. However, there are still significant limitations in the ability to efficiently and flexibly handle the enormous amount of QoS/QoE-compliant data. This data will be exchanged in a future Big-data-driven society, along with the requirements of ultra-high data rates and near-zero latency. Thus, terabit-per-second wireless communication and the support of backhaul network infrastructure are expected to be the leading technological trends over the next ten years and beyond \cite{terranova}.


The standardization process of THz cellular air interface is still in its infancy. Nevertheless, the standardization process of future wireless communication systems in THz bands had already been initiated by the Interest Group on THz communication under the IEEE 802.15 umbrella in early 2008. By 2013, the IEEE 802.15 WPAN Task Group3-D 100 Gbit/s Wireless (TG 3d 100 G) was established to create a 100 Gbit/s wireless communications standard in the 275-325 GHz band. As a result of this group's work, the IEEE 802.15.3d-2017 wireless standard \cite{ieee802153d}, operating in the 300 GHz band, was formally approved and published in fall 2017 \cite{petrov2020ieee}.

The standard includes a new physical layer based on IEEE 802.15.3-2016, MAC based on IEEE 802.15.3e-2017, 8 different channels with bandwidth multiples of 2.16 GHz to 69.12 GHz. It also allows for multiple modulation and coding schemes. The standard covers applications of THz communication systems such as kiosk loading, wireless backhauling and fronthauling, in-device communications, and wireless channels in data centers. The Application Requirements document within IEEE 802.15.3d also defines use-cases and system performance and functionality requirements for the application-based approach. The standard also supports single carrier mode and OOK mode at the physical layer. In addition, the Channel Modeling Document summarizes channel propagation characteristics for target scenarios and proposes application-based channel models highlighting that propagation losses are much heavily affected by atmospheric absorption as compared to mmWave band \cite{tekbiyik2019terahertz}. Future steps for the standard should investigate interference in the bands defined by the ITU-R for use by applications such as radio astronomy, Earth exploration satellites, and space research services. 

THz bands allow high-speed data transmission of up to 100 Gbps for short-range communications. However, system designers still face numerous challenges such as channel modeling, transceiver design, antenna design, signal processing, upper layer protocols, macro- and micro-mobility implications, and security. These challenges are expected to be explored within 3GPP over the next decade to specify a new air interface technology for 6G cellular systems.

\subsection{mmWave/THz Design Challenges}

\subsubsection{mmWave/THz Propagation Challenges}

In addition to the advantages, the use of mmWave/THz brings a number of problems stemming from the frequency band in use. The reason lies in the specific characteristics of radio wave propagation including high propagation losses, atmospheric and rain absorption, low diffraction, higher scattering due to the roughness of materials, high penetration losses through objects. However, many of these disadvantages can be effectively solved, which will allow the use of a new frequency spectrum for 5G communication networks.

First, according to the standard Friis propagation model, an increase in carrier frequency leads to a significant increase in propagation losses \cite{rappaport1996wireless, steinberg1976principles}. Increasing the carrier frequency by an order of magnitude increases propagation losses by 20 dB, which can be effectively offset by the means of beamforming. This effect is illustrated in Fig. \ref{fig:effectsChannel_antenna} as a function of distance for a given set of antenna array elements. Note that the use of antenna arrays also makes it possible to significantly increase the potential coverage area of one BS.

The additional component in signal attenuation is caused by  absorption in the atmosphere \cite{smith1982centimeter, liebe1987millimeter}. The main components responsible for absorption in the considered frequency range are oxygen and water vapor. The absorption data for mmWave and THz bands are shown in Fig. \ref{fig:effectsChannel_absorption}. The absorption by oxygen is exceptionally high reaching 15 dB/km at 60 GHz frequency \cite{reber1968oxygen}. However, in general, absorption is insignificant both for indoor communications and for urban cellular deployments, where the distance between BSs is limited to a few hundred of meters. Furthermore, to some extent, this kind of absorption is beneficial since it allows you to reduce the interference from BSs utilizing the same frequency channel.

\begin{table}[b!]
\vspace{-2mm}
\renewcommand{\arraystretch}{1.0}
\centering
\caption{Impact of environment on mmWave propagation \cite{ometov2019packet}}
\begin{tabular}{lll}
\hline
Type&Value&Measurements\\
\hline\hline
Rain &$50$ mm/hr&$<$10 GHz: 1-6 dB/km, $>$10 GHz: 10 dB/km\\
\hline
Fog&0.5g/m$^3$&50 GHz: 0.16 dB/km, 81 GHz: 0.35 dB/km\\
\hline
Snow&$700$g/m&35-135 GHz: 0.2-1 dB/km\\
\hline
Foliage&0.5$m^2$/m$^3$& 28.8, 57 GHz: 1.3-2.0 dB/m, 73 GHz: 0.4 dB/m\\
\hline
\end{tabular}   
\label{tab:effects}
\end{table}

Additionally, effects of weather conditions on the propagation of millimeter waves are fairly well investigated so far \cite{brooker2007seeing}, see Table \ref{tab:effects}. The most significant influence is exerted by foliage, in the presence of which in the channel, the magnitude of the signal drop reaches 2 dB/m. Losses caused by heavy snow, fog and clouds are quite insignificant (less than 1 dB/km). Rain usually has an additional attenuation of about 10 dB/km, which can seriously affect the wireless channel performance, which is illustrated in Fig. \ref{fig:effectsChannel_whether}. High  directionality partially addresses this problem.

Finally, mmWave/THz bands are characterized by much smaller diffraction as compared to microwave frequencies, leading to the frequent blockage situations induced by large static objects in the channel such as buildings \cite{umi}, and mobile entities such as vehicles and humans \cite{gapeyenko2019spatially,gapeyenko2016analysis,gapeyenko2019spatially}.
Particularly, dynamic blockage introduces additional losses of about 15-40 dB \cite{slezak2018empirical, maccartney2017rapid}. The received signal power degradation caused by a human body blockage of the line-of-sight (LoS) path is demonstrated further in Fig. \ref{fig:pathGain}, in Section \ref{sect:system}, where we discuss blockage modeling principles. Notably, the signal power drop is quite sharp, taking about 20-100 milliseconds. Blockage duration depends on the density of dynamic blockers and their speed and can take longer than 100 ms \cite{gapeyenko2017temporal}. It should be noted that when there is no LoS path, the use of reflected signal paths may not provide sufficient signal power. For example, reflection from rough surfaces, such as concrete or bricks, may attenuate mmWave/THz signal by 40-80 dB \cite{etsitr138}.




The extreme propagation losses in mmWave/THz bands increasing as a power function of the frequency naturally call for massive antenna arrays featuring tens and even hundreds of elements \cite{akyildiz2016realizing,bjornson2019massive}. These arrays will operate in a beamforming regime \cite{ali2017beamforming} and be capable of creating very directional radiation patterns with the half-power beamwidth (HPBW) values reaching few degrees for mmWave systems down to a fraction of degree for THz links \cite{singh2020design,han2021hybrid}. The later feature is vital for mmWave/THz communications systems not only allowing to overcome severe path losses but ensuring an almost interference free environment even for extreme deployment density of base stations (BS) \cite{petrov2017interference}. However, as a side effect, these high directivities bring additional challenges. First of all, they lead to large beamforming codebooks, especially for THz systems, drastically increasing the beamsearching time. Secondly, in THz communications systems, in addition to macromobility, micromobility manifesting itself in fast UE displacements and rotations have to be considered \cite{hur_dynamic_mmWave,petrov2018effect,kokkoniemi2020impact}. This phenomenon happens during the communications process may cause frequent misalignments of the highly-directional THz beams, resulting in fluctuations of the channel capacity and outages \cite{peng_dynamic_thz}.

\subsubsection{Reliable Communications and Session Continuity}


As opposed to previous generations of cellular systems, mmWave and THz RAT in 5G/6G systems will mostly target bandwidth-greedy new applications such as VR/AR, uncompressed video, holographic telepresence, and multimedia-rich extended reality (XR) services \cite{hu2021vision,vannithamby2017towards,agiwal2016next}. For such systems, not only uninterrupted network connectivity, but other quality of service (QoS) parameters such as guaranteed bitrate and latency become critical. Thus, we now proceed to discuss approaches recently proposed for improving session continuity in mmWave/THz systems.


The blockage events in mmWave/THz systems may lead to two principally different effects. When the signal-to-noise plus interference ratio (SINR) falls below a predefined threshold, the UE suffers from outage conditions. The outage probability depends  on many system parameters including  propagation conditions, receiver sensitivity, utilized power, as well as antenna arrays at both BS and UE. The effect of micromobility is similar in nature and is heavily affected by the type of application utilizing the wireless channel \cite{petrov2018effect}. To ensure uninterrupted service in mmWave/THz deployments, one may utilize multiconnectivity operation standardized by 3GPP \cite{3gpp_MC}. According to it, UE is capable of maintaining multiple connections with the nearest BSs within the same RAT. These connections can be utilized for data transmission simultaneously. Even when only one of these connections may be utilized at a time, the packet flow can be rerouted to the backup connection should the current link experience outage conditions. However, the efficient use of this technique inherently requires dense deployments that may not be available at early phases of mmWave/THz systems rollouts. 


Vendors and standardization bodies also consider the support of multiple RATs at the UE via multi-band multiconnectivity option \cite{holma20205g}. This approach addresses the outage problem by timely rerouting traffic to the backup connection when the currently active link becomes unusable. However, the inherent rate mismatch between RATs utilized in future 5G/6G deployments, especially, between mmWave/THz and LTE interfaces, may render this capability useless in practical use cases. Moreover, recall that the outage events caused by micromobility and blockage by small dynamic objects such as human or vehicle bodies are characterized by a rather short duration and high frequency \cite{raghavan2018spatio,petrov2020capacity,gapeyenko2017temporal}. Thus, the resulting traffic that needs to be supported at lower rate interfaces is expected to be bursty in nature. These occasional traffic spikes caused by high bitrate sessions temporarily offloaded onto lower rate interface may negatively affect the service performance of other sessions and also lead to inefficient resource utilization of the corresponding BSs. As a result, traffic protection strategies might be required.


The blockage may not lead to an outage when the SINR still remains higher than the one associated with the lowest possible MCS. In this case, even though the connection is not lost the amount of resources required to maintain the required bitrate increases drastically. To address this case, the authors in \cite{moltchanov2018improving} suggested the use of the resource reservation technique. According to it, a fraction of BS resources is reserved for sessions that are accepted for service ensuring that there is a surplus of resources for them when their state changes from non-blocked to blocked. This approach is fully localized at the BS and thus can be utilized at the early rollouts of mmWave/THz RATs when the network density is insufficient to efficiently utilize multiconnectivity operation. Furthermore, it does not require maintaining active connections to more than a single BS reducing the complexity of the UE implementation. However, by joining this technique with multiconnectivity one may attain additional gain in terms of session service performance \cite{begishev2021joint}.



\subsubsection{Traffic Coexistence}

Modern cellular systems are being developed having multiple types of traffic in mind with drastically different service requirements. As an example, the NR radio interface is expected to support at least two traffic types \cite{dahlman20205g,holma20205g}. At one extreme, there is eMBB service which is an extension of the broadband access provided in 4G LTE technology having similar service requirements. In industrial automation scenarios, 5G NR is also expected to enable specific services, such as positioning, clock synchronization, joint tasks execution. All these applications are characterized by extreme latency and reliability requirements and need to be supported via URLLC service. This implies that future BSs need to support a mixture of traffic with drastically different service requirements at the air interface. Mechanisms for supporting eMBB or URLLC in isolation are current the focus of ongoing studies, see, e.g., \cite{samuylov2020characterizing,moltchanov2018improving,begishev2019quantifying} for eMBB and \cite{mahmood2018reliability,rao2018packet,mahmood2019resource} for URLLC. Their joint support, however, requires advanced techniques at the BS side. 


There have been two principally different approaches for enabling the coexistence of eMBB and URLLC at the air interface. The first one implies the use of smart scheduling techniques. This approach has been taken in \cite{anand2020joint} to formalize the optimization problem of joint scheduling of these services. While discussing their results, the authors concluded that the major impact is produced by the minimum scheduling interval utilized by the technology in question. To alleviate this limitation, several studies suggested the use of the non-orthogonal multiple access (NOMA) technique, e.g., \cite{kassab2018coexistence,dougan2019noma,kotaba2020urllc}. According to NOMA, properly encoded URLLC transmissions can be superimposed on top of already scheduled eMBB traffic and the message content can be restored at the receiving end. While this approach may indeed drastically reduce the latency, the reliability of communications still remains a challenging problem. On top of this, the intended UEs need to be always awake to receive the intended transmissions.



An alternative approach is based on explicit resource allocation and prioritization techniques. When the minimum scheduling interval allows for satisfying latency constraints, one may utilize, e.g., network slicing techniques at the air interface to explicitly allocate a part of resources to URLLC traffic \cite{sallent2017radio,popovski20185g,koucheryavy2021quantifying}. However, as a load of URLLC traffic may not be known in advance this approach may result in inefficient use of resources. A viable alternative is to utilize priorities between traffic flows as illustrated in \cite{markova2020prioritized,markoval2020priority}. Compared to the resource reservation technique, this approach does not suffer from resource utilization problem but may lead to reduced service performance of eMBB flows.

\subsection{5G/6G Services, Use-Cases and Traffic Specifics}

\subsubsection{5G/6G Services}

The advent of 5G/6G cellular systems could radically improve the customer experience, but it poses new challenges for network operators, device manufacturers, and telecommunications infrastructure: (i) UEs will require redesign or upgrades to ensure that they properly meet the higher data rates and power requirements, (ii) UEs may become more sophisticated to meet consumer expectations for small and miniaturized devices, thus, reliability and miniaturization of component technology will become a necessity, (iii) higher data transfer rates will mean higher device temperatures, which will cause performance and safety issues and will require more efficient power utilization and longer battery lifetime.


5G addresses the challenges of latency, power consumption, hardware complexity, bandwidth, and reliability. For example, the challenges of requiring eMBB and URLLC are being addressed with 5G networking technologies. In contrast, future 6G technology will be developed in a holistic manner to jointly meet extremely demanding network requirements (e.g., ultra-high reliability, capacity, efficiency, and low latency) given the projected economic, social, technological, and environmental context by 2030. To this aim, below, we present the characteristics and anticipated requirements of use cases that, because of their generality and complementarity, are believed to represent future 6G services. For similar outlook of 5G requirements we refer to ITU-R M.2410 \cite{series2017minimum} and also for the assessment of the current evolution of 5G systems provided in recent studies, e.g., \cite{henry20205g,fuentes20205g}.

\begin{figure}[!t]
\vspace{-0mm}
\centering
\includegraphics[width=1.0\columnwidth]{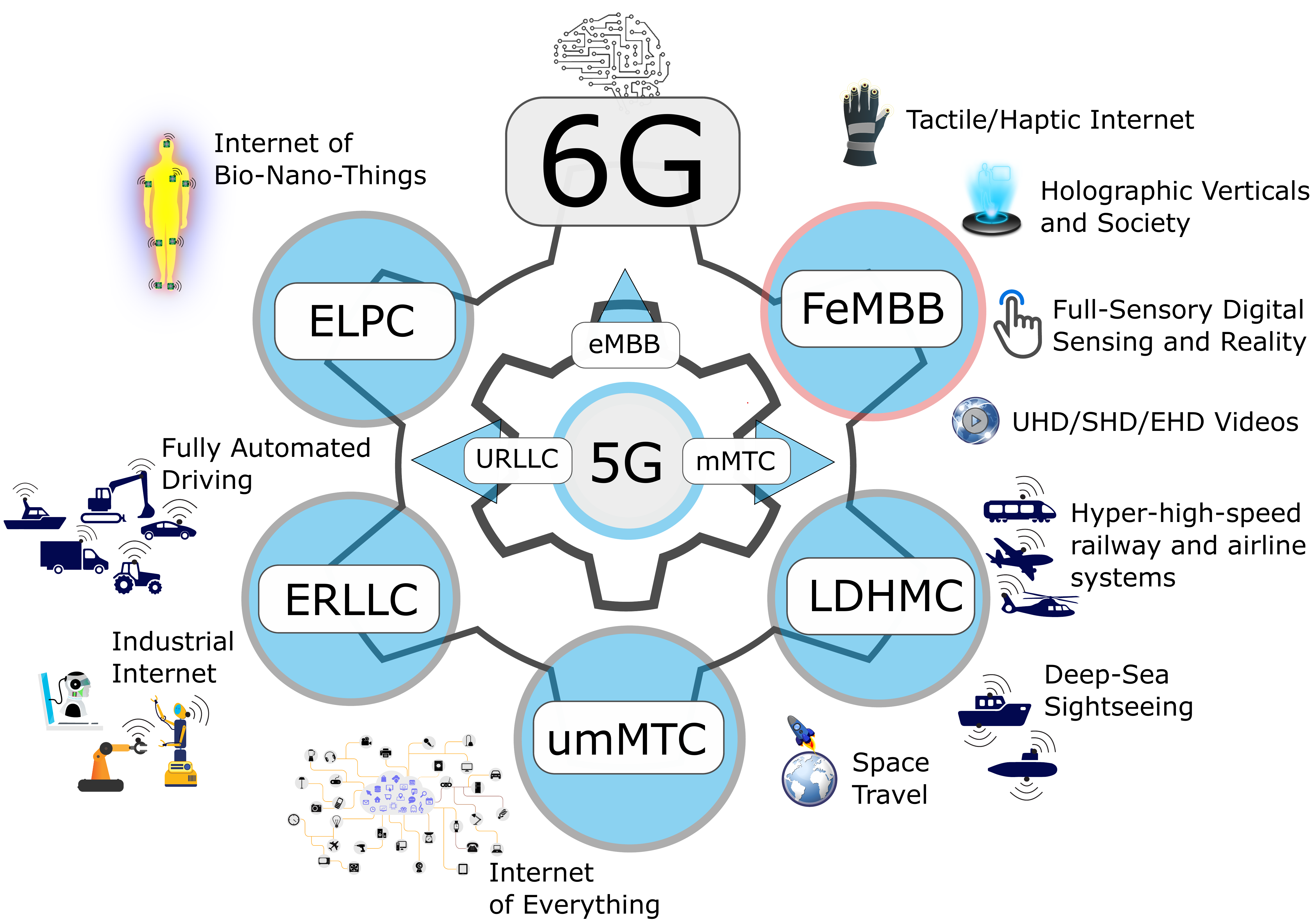}
\caption{Illustration of typical applications in 6G.}
\label{fig:typicalApps}
\vspace{-2mm}
\end{figure}


KPIs for analyzing and evaluating 6G wireless networks include high growth in peak data rates, data rates, traffic throughput, connection density, latency and mobility, as well as the use of additional spectrum and energy efficiency, and presents the following requirements \cite{key_technolog6G}:
\begin{itemize}
	\item{the user peak data rate is not less than 1 $Tbit/s$, which is 100 times higher than that of 5G; for some scenarios, such as backhaul and direct transmission in THz band (x-haul), the peak data rate is expected to reach 10 $Tbit/s$;}
	\item{the user data rate is 1 $Gbps$, which is 10 times faster than 5G; it is also expected to provide data rates of up to 10 $Gbps$ for some scenarios such as indoor access;}
	\item{delay at the air interface is expected to be 10-100 $\mu{}s$ at high mobility speeds reaching 1000 $km/h$ -- this will provide acceptable QoE for scenarios such as hyper-high-speed railway (HSR) and aviation systems;}
	\item{the number of connections is ten times higher than that of 5G -- this will allow achieving up to $10^2$ devices/$km^2$ and zone throughput up to 1 $Gb/s/m^2$ for scenarios such in highly crowded conditions;}
	\item{energy efficiency is expected to be 10-100 times and spectral efficiency -- 5-10 times higher than 5G.}
\end{itemize}


Mobile Internet and Internet of Everything (IoE) are considered to constitute a powerful platform for the development of 6G to support holographic and high-precision communication, tactile applications to provide a full sensory experience (such as sight, hearing, smell, taste and touch). Achieving these factors requires processing very large amounts of data in near real-time with extremely high throughput (approximately Tb/s) and low latency. In addition, 6G wireless networks will provide \cite{key_technolog6G}: (i) support of ultra high definition (SHD) and ultra high definition (EHD) video with ultra high bandwidth, (ii) guaranteed connection with extremely low latency (approximately 10 $\mu{}s$) for the industrial Internet, (iii) support for the IoT of nano-things through intelligent wearable devices provided by implantable nano-devices and ultra-low-power nanosensors (on the order of picowatts, nanowatts and microwatts), (iv) support for space and underwater communications to expand the boundaries of human activity to make space travel and explore hard-to-reach sea depths, (v) setting up uniform service in new scenarios and applications such as ultra high speed rail (HSR), (vi) significantly improved vertical 5G applications such as the massive Internet of Things (IoT) and fully autonomous vehicles.

\begin{table}[t!]
	\vspace{-6mm}
	\renewcommand{\arraystretch}{1.0}
	\centering
	\caption{The network features of 5G and the future 6G} \cite{key_technolog6G}
	\addtolength{\tabcolsep}{-4.5pt}
	\begin{tabular}{l|ll}
	\hline
		\multicolumn{1}{c}{\textbf{Type}} & 
		\multicolumn{1}{c}{\textbf{5G}} &                                                              \multicolumn{1}{c}{\textbf{6G}} \\ 
		\hline\hline
	
	    \begin{tabular}[c]
	    	{@{}l@{}}\textit{Usage}\\
			\textit{Scenarios}
		\end{tabular}  & 
		
		\begin{tabular}[c]
				{@{}l@{}}• eMBB \\• URLLC \\• mMTC~~
		\end{tabular} & 
	    
	    \begin{tabular}[c]
	    	{@{}l@{}}• FeMBB \\• URLLC \\• umMTC\\• LDHMC \\• ELPC~~
		\end{tabular} \\ 
		\hline
		\textit{Applications} & 
		
		\begin{tabular}[c]
			{@{}l@{}}• VR/AR/360° Videos\\• UHD Videos\\• V2X\\• IoT\\• Smart City/Factory/Home\\• Telemedicine\\• Wearable Devices
		\end{tabular}                                                                             & 
	
		\begin{tabular}[c]
			{@{}l@{}}• Holographic Society\\• Tactile/Haptic Internet\\• Sensory Sensing/Reality\\• Fully Automated Driving\\• Industrial Internet\\• Space Travel\\• Deep-Sea Sightseeing\\• Internet of Nano-Things
		\end{tabular}  \\ 
		\hline
		
		\begin{tabular}[c]
			{@{}l@{}}\textit{Network}\\
			\textit{Characteristics}
		\end{tabular}   & 
    	
    	\begin{tabular}[c]
    		{@{}l@{}}• Cloudization\\• Softwarization\\• Virtualization\\• Slicing
    	\end{tabular}  & 
      
        \begin{tabular}[c]{@{}l@{}}• Intelligentization\\• Cloudization\\• Softwarization\\• Virtualization\\• Slicing
        \end{tabular} \\ 
		\hline
		
		\begin{tabular}[c]
			{@{}l@{}}\textit{Peak }\\
			\textit{Data Rate}
		\end{tabular} & 20 Gb/s~~ & ≥1 Tb/s \\ 
		\hline
		\begin{tabular}[c]
			{@{}l@{}}\textit{Experienced}\\
			\textit{Data Rate}
		\end{tabular}  & 0.1 Gb/s & 1 Gb/s \\ 
		\hline
		\begin{tabular}[c]
			{@{}l@{}}\textit{Spectrum}\\
			\textit{Efficiency}\end{tabular}  & 3× that of 4G~~ & 5–10× that of 5G~~ \\ 
		\hline
		\begin{tabular}[c]
			{@{}l@{}}\textit{Network}\\
			\textit{Energy}\\
			\textit{Efficiency}\end{tabular} & 10–100× that of 4G~~ & 10–100× that of 5G~~  \\ 
		\hline
	
		\begin{tabular}[c]
			{@{}l@{}}\textit{Area Traffic}\\
			\textit{Capacity}
		\end{tabular}   & 10 Mb/s/$m^2$~~  & 1 Gb/s/$m^2$~~ \\ 
		\hline
		
		\begin{tabular}[c]
			{@{}l@{}}\textit{Connectivity}\\
			\textit{Density}
		\end{tabular} & $10^6$ Devices/$km^2$~~  & $10^7$ Devices/$km^2$~~ \\ 
		\hline
		\textit{Latency}  & 
	
		\begin{tabular}[c]{@{}l@{}}1 ms \\~
		\end{tabular}   & 10–100 µs\\ 
		\hline
		\textit{Mobility}& 500 km/h~~  & ≥1,000 km/h~~   \\ 
		\hline
		\textit{Technologies}  & 
	
		\begin{tabular}[c]
			{@{}l@{}}• mmWave Communications\\• Massive MIMO\\• LDPC and Polar Codes\\• Flexible Frame Structure\\• Ultradense Networks\\• NOMA\\• Fog/Edge Computing\\• SDN/NFV/Network Slicing~~
		\end{tabular} & 
    
    	\begin{tabular}[c]
		{@{}l@{}}• THz Communications\\• SM-MIMO\\• LIS and HBF\\• OAM Multiplexing\\• Laser and VLC\\• Blockchain Sharing\\• Quantum Comm. \\• AI/Machine Learning~~
       \end{tabular}                                                  \\
		\hline
	\end{tabular}
	\label{tab:key_application}
\end{table}

Therefore, the typical scenarios are shown in Fig. \ref{fig:typicalApps} and in Table \ref{tab:key_application} (ELPC -- extremely low-power communications, FeMBB -- further-enhanced mobile broadband, LDHMC -- long-distance and high-mobility communications, NFV -- network function virtualization, SDN -- software-defined networking, UHD -- ultra-high definition, umMTC -- ultra-massive machine-type communications, V2X -- vehicle to everything; LDPC -- low-density parity check codes) should be supported in 6G networks. 6G technology extends eMBB, URLLC, and mMTC services further to their extremes simultaneously introducing new edges to the baseline 5G service triangle with long distance, extremely high-speed, and ultra low power communications.

\subsubsection{(F)eMBB Use-Cases and Applications}


In this work, we will mainly focus on eMBB services for 5G networks and FeMBB services for 6G systems. These services will have to provide both higher bandwidth as compared to 4G systems in densely populated areas and expanded coverage for those on the move. Next, we will define the main use-cases of (F)eMBB services for current 5G and future 6G networks. Three typical use-cases for 5G eMBB service are:
\begin{enumerate}
  \item{\textit{Enhancing the Fan Experience.} This is an example of public places such as stadiums and arenas, where thousands of fans simultaneously connect to the Internet and share their experiences using apps on their mobile devices. In this case, the demand for connection and more bandwidth grows exponentially. With situations like this, eMBB can handle and provide users with improved broadband access in densely populated areas. Enhanced video capabilities combined with improved network bandwidth can provide fans with real-time mobile access. It will also allow fans to upload videos of their stadium experiences to social media in real time and share them outside of the stadium.}
  \item{\textit{Smart/Safe Cities.} eMMBs can improve security by placing AR/VR-enabled video devices in strategic locations that were previously not possible with legacy 4G technology. Traffic flow monitoring and instant AI analysis views can save the seconds needed to send paramedics to the scene of a car accident. Police services, tow trucks, and other service vehicles, also unmanned aerial vehicles, may be dispatched earlier at the same time. This will allow to quickly eliminate the violation and restore normal traffic flow. Traffic signals can be linked to the flow of cars to eliminate traffic jams on the street.}
  \item{\textit{An Untethered Virtual Experience.} The eMBB will support the key functions of our day-to-day work -- and it will do so wirelessly. From people who use cloud applications while commuting to work, employees who communicate with the office, to an entire smart office where all devices are connected wirelessly, 5G will connect and make our work easier.}
\end{enumerate}


We will now proceed by presenting the overview of the eMMB service application for 6G networks. For such services, we highlight the following main applications (see Fig. \ref{fig:typicalApps} and Table \ref{tab:key_application}): Holographic Verticals and Society, Tactile/Haptic Internet, Full-Sensory Digital Sensing and Reality:
\begin{enumerate}
	\item{\textit{Holographic Verticals and Society.} Holographic communication is the next evolution in multimedia, delivering 3D images from one or more sources to one or more destinations, providing an immersive 3D experience for the end user. The possibility of interactive holography on the web will require a combination of very high data rates and ultra-low latency. The first arises from the fact that the hologram is composed of several three-dimensional images, and the second is based on the fact that parallax is added so that the user can interact with the image, which also changes depending on the position of the viewer. This is essential for providing an immersive 3D user experience \cite{6g_dohler}.\\
	Note that there are significant problems in the implementation of holographic communications, especially, in connection with their widespread use. These problems arise at all stages of holographic video systems and extend from signal generation to display. As far as we know, there are no standards for how to transfer data to a display. Recording digital holograms is another problem, as specialized optical installations may be required. Computer-generated holograms require significant computational resources compared to classical image rendering due to the many-to-many relationship between the source and hologram pixels. The high bit rates required do not take advantage of established compression methods such as the Joint Photographic Experts Group (JPEG)/Moving Image Experts Group (MPEG) since the statistical properties of holographic signals are very different from video images.}
	\item{\textit{Tactile/Haptic Internet.} After using holographic communication to transmit a virtual vision of people close to real species, events, the environment, it will be very convenient to remotely exchange physical interaction via the tactile Internet in real time. Many applications fall into this category. Consider the following examples \cite{6g_dohler}:}
		\begin{itemize}
		\item{\textit{Robotization and Industrial Automation:} We are on the cusp of a revolution in manufacturing, driving networks that facilitate communication between people and between people and machines in Cyber ​​Physical Systems (CPS). This so-called Industry 4.0 vision enables the development of new applications. This approach requires communication between large connected systems without the need for human intervention. Remote production control is based on the management and control of industrial systems in real time. Robotics will need guaranteed real-time control to avoid oscillatory motion. Advanced robotics scenarios in manufacturing require a maximum target latency of 100 $\mu{} s$ and a round trip time of 1 ms. Human operators can control remote machines using virtual reality or holographic communication, and are assisted by tactile sensors, which can also include triggering and control through kinesthetic feedback;}
		\item{\textit{Autonomous driving:} Through communication and coordination between vehicles (V2V) or between vehicles and infrastructure (V2I), autonomous driving can lead to a significant reduction in road accidents and congestion. However, collision avoidance and remote driving will likely require a delay of the order of a few milliseconds.}
	\end{itemize}
\end{enumerate}

\subsubsection{5G/6G Traffic Specifics}

Performance assessment of 4G and older cellular systems is conventionally assessed by utilizing the stochastic geometry \cite{haenggi2012stochastic,haenggi2009stochastic,blaszczyszyn2018stochastic}. The rationale is that those systems mainly target adaptive applications with appropriate transport layer (via TCP rate feedback) or application layer (via application layer rate control) adaptivity to constantly changing the wireless channel and available resources conditions. Contrarily, current 5G and future 6G systems primarily target non-elastic/adaptive traffic rate-greedy with limited adaptation capabilities such as VR/AR, tactile Internet, holographic telepresence, etc, \cite{hu2021vision,vannithamby2017towards,agiwal2016next}.

In our paper we mainly concentrate on non-elastic/adaptive applications that may not tolerate outage and rate degradation. The principal difference compared to elastic/adaptive traffic is that these sessions require time-varying amount of resources in spontaneously changing channel conditions, especially, under blockage and micromobility impairments. As a result, new approaches that are capable of accounting for channel conditions and resource allocations need to be developed.

\section{Modeling Principles of THz/mmWave Systems}\label{sect:approaches}

In this section, we briefly overview the basic principles of stochastic geometry analysis of cellular systems that is still utilized as one of the tools for radio part characterization in joint new frameworks suitable for mmWave/THz systems. Then, we proceed by assessing the current state-of-the-art in queuing models providing the abstraction of dynamic resource allocation at mmWave/THz BSs. Finally, we briefly sketch the basic principles of the new frameworks required for performance assessment of non-elastic/adaptive applications.

\subsection{Stochastic Geometry Approach}

\subsubsection{Random Variable Transformation Technique}

Stochastic geometry operates with random events in two- or three-dimensional space representing locations of BSs and UEs. To this aim, it utilizes spatial point processes (Poisson, Matern, etc.), as well as random graphs and spatial tessellations (for example, Dirichlet tessellations leading to Voronoi cells) to simulate BS service areas. The main probability theory tool the stochastic geometry is based upon is the functional transformation of random variables (RV). Recall, that given the joint probability density function (pdf) $w_n(x_1,\dots,x_n)$ of RVs $\{\psi_1,\dots,\psi_n\}$ and Jacobian of functional transformation from $\{\psi_1,\dots,\psi_n\}$ to $\{\nu_1,\dots,\nu_n\}$,
\begin{align}
J=\frac{\partial{}(x_1,\dots,x_n)}{\partial{}(y_1,\dots,y_n)}=
\left|\begin{matrix}
\frac{\partial{}x_1}{\partial{}y_1}&\dots&\frac{\partial{}x_1}{\partial{}y_n}\\
\vdots&\dots&\vdots\\
\frac{\partial{}x_n}{\partial{}y_1}&\dots&\frac{\partial{}x_n}{\partial{}y_n}.\\
\end{matrix}\right|,
\end{align}
then the sought joint pdf is given by \cite{ross,samuylov2015random}
\begin{align}\label{eqn:meth_RVtrans}
W_n(y_1,\dots,y_n)=\sum_{k}w_n(x_{1k},\dots,x_{nk})|J|,
\end{align}
where $x_{ik}=(y_1,\dots,y_n)$, $i=1,2,\dots,n$, is $k$-th branch of inverse function.

When in the original set of RVs we have $m<n$, i.e., $\{\nu_1,\dots,\nu_m \} $, then the system (\ref{eqn:meth_RVtrans}) must be supplemented with $m-n$ RVs $\nu_j=\psi_j$, $j=m+1,\dots,n$. Then, the joint pdf is obtained by integrating over the variables $y_{m + 1},\dots,y_n$ as
\begin{align}
W_n(y_1,\dots,y_m)&=\int\dots\int\sum_{k}w_n(x_{1k},\dots,x_{nk})\times{}\nonumber\\
&\times{}|J|d_{y_{m+1}}\dots{}d_{y_n}.
\end{align}

For example, the distance from a random point to the $i$-th neighbor in the Poisson field of BSs obeys the Gamma distribution with a pdf \cite{moltchanov2012distance}
\begin{align}\label{eqn:distancePoissonProcess}
f_{i}(x)=\frac{2(\pi\lambda)^{i}}{(i-1)!}x^{2i-1}e^{-\pi\lambda{}x^{2}},\,x>0,\,i=1,2,\dots.
\end{align}

Now, one can determine the received power from the nearest UE by utilizing the model of the $i$-th BS at the receiver by applying the propagation model in the form $P(x)=Ax^{-\zeta}$, where $x$ is the distance between the receiver and the transmitter, $\zeta$ is the attenuation constant, $A$ -- a constant that depends on the properties of the transmitter, receiver and carrier frequency. Specifically, the inverse function and the modulus of its derivative are
\begin{align}
G(y)=P^{-1}(x)=\left(\frac{y}{A}\right)^{-1/\zeta },\,G'(y)=\frac{\left(\frac{y}{A}\right)^{-\frac{1}{\zeta }-1}}{A \zeta }.
\end{align}

Thus, by applying (\ref{eqn:meth_RVtrans}) the received power from the $i$-th BS can be written as
\begin{align}
f_{i}(y)&=\frac{2(\pi\lambda)^{i}}{(i-1)!}\left(\frac{y}{A}\right)^{-(2i-1)/\zeta }e^{-\pi\lambda{}\left(\frac{y}{A}\right)^{-2/\zeta }}\times{}\nonumber\\
&\times{}\frac{\left(\frac{y}{A}\right)^{-\frac{1}{\zeta }-1}}{A \zeta },\,y>0,\,i=1,2,\dots.
\end{align}

For fairly simple scenarios, for example, the triangle scenario with one receiver-transmitter pair and one source of interference, considered in, e.g., \cite{samuylov2015random}, or for a part of cellular network with six sources of interference, considered in \cite{etezov2016distribution}, or in indoor premises, also with a limited number of interference sources, see \cite{samuylov2017analytical}, the use of this method allows you to obtain pdfs of basic characteristics of communication channels, such as signal-to-interference ratio (SIR), SINR, spectral efficiency, and data rate. However, such simple stochastic geometry models become more complicated in the case of UEs movement as shown e.g., \cite{orlov2017time,fedorov2017sir} or when the number of interference sources is RV itself, \cite{petrov2017interference,kovalchukov2018evaluating,kovalchukov2018analyzing}.

\subsubsection{Random Field of Interferers}

In those cases, where the number of interference sources is a RV, the use of the transformation of RVs does not allow obtaining closed-form expressions for the metric of interest. Consider the example where BSs operating at the same frequency are distributed according to the Poisson process over the plane with some intensity $\lambda$. The UEs associated with the BSs are located at a fixed distance from the receivers, $r_0$. Consider some randomly selected receiver and some metric, e.g., SIR, that can be written as $S(x)=\frac{C_0}{C_1+I}$, where $C_0$ is a constant that determines the level of the received power from BS located at a fixed distance $r_0$ from the receiver, $C_1$ is a constant that takes into account transmission and reception losses, $I$ is the total interference from the field of interference sources,
\begin{align}\label{eqn:metrics02}
I=C_2\sum_{i=1}^{\infty}R_{i}^{-\zeta},
\end{align}
where $C_2$ is a constant that takes into account the transmit and receive gains, $\zeta$ is a path loss exponent, $R_i$ is the distance from the interfering BS to UE.

In the considered scenario, $R_i$ are RVs whose distributions obey the generalized Gamma distribution in (\ref{eqn:distancePoissonProcess}). Using the RV transformation technique one may determine pdf of the interference from any $i$-th BS. However, the aggregate interference is a random sum of RVs, where each component has its own distributions. In this case, one may utilize the Campbell's theorem \cite{chiu2013stochastic}, to obtain the moments of interference as
\begin{align}\label{eqn:meth_campbell}
E[I^{n}]=\int_{r_{B}}^{D}(C_2r^{-\zeta})^{n}2\lambda\pi{}rdr,
\end{align}
where $r_B$ is the minimum distance from the UE, where interfering BS might be located, $D$ is the maximum distance, where interference is non-negligible, $2\lambda\pi{}r$ is the probability that there is interfering BS at the $dr$ increment of circumference.

The result in (\ref{eqn:meth_campbell}) can be extended to accommodate LoS blockages, antenna directivity, 3D deployments, random BS and UE heights, and non-Poisson distribution of BSs, etc. as shown in \cite{petrov2017interference,kovalchukov2018evaluating,kovalchukov2018analyzing}.

Further, to obtain estimates of the moments of the SIR, SINR, spectral efficiency, and data rate, one may again utilize RV transformation technique. Consider as an example SIR function. In the absence of pdf of aggregate interference, one can use the expansion of the sought functions in a Taylor series around the mean of $I$, as shown in, e.g., \cite{petrov2017interference}. Particularly, for the mean value of the RV $Y=g(X)$, where $ X $ is a RV with mean value and variance $\mu_{X}$ and $\sigma^{2}_X$, we have
\begin{align}\label{eqn:saddle_04}
&E[Y]=g(\mu_{X})+\frac{g''(\mu_{X})}{2}\sigma^{2}_X,\nonumber\\
&\sigma^{2}[Y]=[g'(\mu_{X})]^{2}\sigma^{2}_X-\frac{1}{4}[f''(\mu_{X})\sigma^{2}_X]^2.
\end{align}

Note that the obtained moments can be further utilized to construct probabilistic bounds on the considered metrics of interest by applying, e.g., Markov, Chebyshev, and Hoeffding inequalities \cite{wasserman2004inequalities}. As an alternative, one may also utilize Laplace-Stieltjes transform technique to get transforms of SIR, SINR, and rate metrics. However, these transforms can rarely be converted back to the RV domain.

\subsubsection{Rate Approximations}

Stochastic geometry can also be used to estimate the rate provided to UEs in the presence of competing users. However, for such an assessment, one needs to make additional assumptions about the resource allocation between UEs. Assume, for example, that in the BS service area of ​​radius $R$ there the number of active UEs are distributed according to Poisson's distribution with the parameter $\lambda\pi{}R^2$, where $\lambda$ is the density of users per squared meter. Let us also assume that the location of each user is uniformly distributed in the BS coverage area, and the available frequency band, $B$, is evenly distributed among  UEs. This scenario corresponds to the deployment of BSs in the Poisson field of active UEs with the intensity $\lambda$. In this deployment, the bandwidth available to a UE additionally ``thrown'' into the service area of BS is provided by $B/i$ with Poisson probability
\begin{align}\label{eqn:meth_poissonPr}
p_{i}=\frac{(\lambda\pi{}R^2)^{i}}{i!}e^{-\lambda\pi{}R^2},\,i=1,2,\dots,
\end{align}
while the achieved rate in the presence of $i$ competing users can be determined by transforming a RV describing the distance from the user to the BS, $f_X(x)=2x/R^2$, $x>0$, according to the Shannon rate, i.e.,
\begin{align}
C_i=\frac{B}{i}\log_{2}[1+S(x)],\,i=1,2,\dots,
\end{align}
where $S(x)$ is the SNR at a distance $x$ from the BS. The final result can be obtained by summing up the rates corresponding to $i$ users in the service area, weighted by the coefficients (\ref{eqn:meth_poissonPr}).

\subsubsection{Limitations of Stochastic Geometry}

Note that the considered scenario can be extended to the case of resource allocations between competing UEs. For example, max-min resource sharing, which equalizes user speeds, or proportional sharing, which gives priority to users who are in more favorable conditions, can be obtained by introducing additional weighting factors, see, e.g., \cite{moltchanov2014optimal,gerasimenko2015cooperative}. It is also possible to extend it further to the $\alpha$-weighted priority, where the operator has the ability to control the sharing of speeds between users, as shown in \cite{gerasimenko2016adaptive}. Although in our work we mainly concentrate on non-elastic/adaptive traffic, we sketch the analysis for such adaptive type of traffic with mmWave/THz specifics in Section \ref{sect:adaptive}, while systems with elastic traffic are briefly outlined in Section \ref{sec:servcoefs}.

The stochastic geometry approach is still inherently limited to the case of elastic full-buffer traffic, where UE always have an infinite amount of data to transmit and may adapt its transmission rate to the current wireless channel and system conditions (i.e., number of active UEs). In the context of mmWave/THz systems, this approach is also not suitable for capturing the performance of applications sensitive to outage conditions that can be caused by either blockage or micromobility.

\subsection{Queuing-Theoretic Approach}

The session service process in systems with stochastic arrival processes, service times, and a limited amount of resources is traditionally modeled by the methods of queuing theory. To capture the resource allocation in the context of mmWave/THz systems, in addition to these specifics, the queuing-theoretical models need to be supplemented with the randomness of the amount of requested resources caused by the location of the UEs in the coverage area of BSs. While the stochastic geometry alone cannot capture resource allocation dynamics, it may complement queuing-theoretic models by abstracting the radio part characteristics to the form suitable for queuing theory. However, the methods of the queuing theory itself need to be expanded to include models capturing randomness of the amount of requested resources. Below, we briefly review the current state-of-the-art and recent developments in this direction, concentrating on the so-called resource queuing systems.

\subsubsection{Conventional Service Models}

The roots of the resource queuing systems go back to the celebrated Erlang-B and Erlang-C formulas describing the call drop probability in $M/M/C/C$ queuing system and waiting for probability in $M/M/C$ queuing system, respectively. Since then, Erlang's results have been widely used for performance analysis of communications systems. Moreover, they have been extended and generalized following technological advances.

\begin{figure}[!t]
	\centering
	\includegraphics[width=0.4\columnwidth]{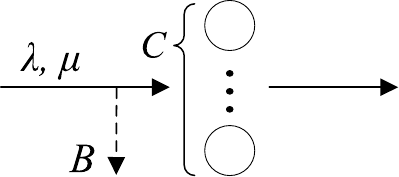}
	\caption{Erlang's loss model -- $M/M/C/C$ queuing system: exponential arrival and service processes with intensities $\lambda$ and $\mu$, finite number of servers $C$ representing currently served sessions.}
	\label{fig:rels_1}
	\vspace{-4mm}
\end{figure}

It started with trunking -- a way to provide network access to multiple users by making them share a group of circuits. In terms of queuing theory, the process can be described by a loss service system depicted in Fig. \ref{fig:rels_1}. The system consists of $C$ servers, each of which is available to an arriving customer whenever it is not busy. Customers arrive according to a Poisson process of rate $\lambda $, i.e., interarrival times are independent and have exponential distribution with a mean $1/\lambda $. Service times are independent and have exponential distribution with a mean $1/\mu $. An arriving customer is blocked and lost if it finds all servers busy. By studying this system in 1917, Erlang derived the famous relation now known as the Erlang loss formula, or Erlang-B, which provides the probability that a call is lost
\begin{align}\label{eqn:rels_1}
	E_{C} (\rho )=B=\frac{\rho ^{!} }{C!} \left(\sum _{k=0}^{!}\frac{\rho ^{k} }{k!}  \right)^{-1} ,  
\end{align}
where $\rho =\lambda /\mu $ is the mean number of arrivals during the mean service time, referred to as traffic intensity. This expression has seen several extensions, for example, the multi-class arrivals. As a result, generalized loss systems are suitable for modeling multi-rate circuit-switched communications and have proved useful, among other applications, for performance analysis of Time Division Multiplexing (TDM) systems, where multiple time slots can be allocated to reduce delay.

The Erlang formula received further development in application to multiservice broadband networks, which has resulted in a class of models named loss networks, or multiservice loss networks [Kelly, 1991; Ross, 1995]. A classical loss network is a general model of a circuit-switched network carrying multi-rate traffic. The model is well-suited for bidirectional traffic flows, because the reverse traffic for a given pair of nodes may have different bandwidth requirements. This evolved model further enabled the analysis of the first wireless general purpose telecommunication systems like GSM, which has played a pivotal role in the telecommunications revolution \cite{rappaport1996wireless}. It is a traditional circuit-switched telephony system, and that is exactly why the focus is on the call blocking probabilities as one of the key performance metrics of GSM systems.


Further technology development brings us the core technology of the fourth generation (4G) cellular networks -- LTE, with its predecessor of the third generation (3G), UMTS technology. It is substantially different from the radio access technologies of previous generations as LTE  is completely packet switched with the smallest unit of resource assignment being a single physical resource block (PRB), providing highly efficient resource management for multiple users. Here, PRBs are allocated to user sessions by schedulers based upon the signal characteristics between the transmitter and receiver. Therefore, the amounts of resources (the number of PRBs) allocated to sessions vary and can be considered as random variables (more precisely, as discrete random variables). In subsequent radio access technologies, such as 5G New Radio (NR) systems operating in a millimeter-wave (mmWave) band, the resource requirement of a session can be represented by a continuous random variable. This is why performance analysis of such systems required another major modification of loss models, giving rise to loss systems with random resource requirements (ReLS). Loss systems with random resource requirements constitute the new evolution of the Erlang formula.

\subsubsection{Baseline Resource Queuing System Formalization}\label{sect:RQSformal}

\begin{figure}[!t]
	\centering
	\includegraphics[width=1.0\columnwidth]{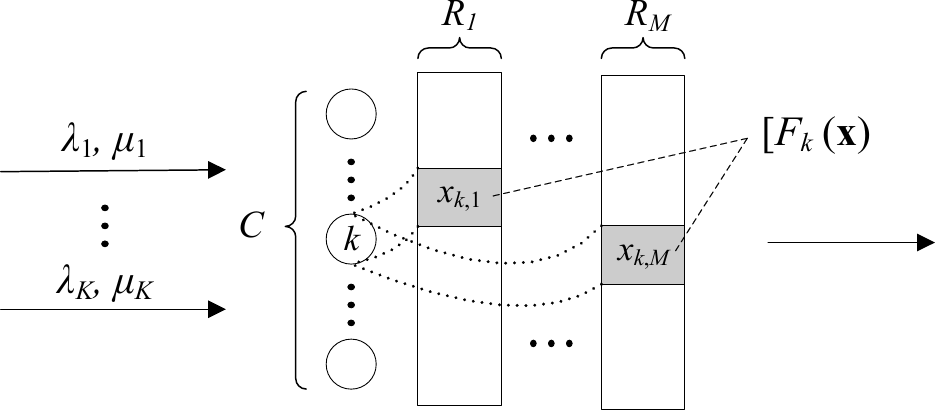}
	\caption{Multi-resource loss system with random resource requirements: multiple flows of sessions with arrival and service intensities $\lambda_i$, $\mu_i$, finite number of servers $C$ representing currently served sessions, and multiple resource pools modeling quantities associated with active sessions, e.g., amount of requested resources \cite{naumov2021}.}
	\label{fig:rels_2}
	\vspace{-4mm}
\end{figure}

The main difference between resource queuing systems and classic ones is that in the former a session upon arrival, in addition to a server, demands a random amount of limited resources, that are occupied for the duration of its service time. The resulting models can thus reflect the fact that user session requirements for the radio resources of a base station vary due to random user positioning, and, consequently, to random spectral efficiency of the transmission channel. A multi-resource ReLS, depicted in Fig. \ref{fig:rels_2}, relies on the following assumptions. Let a $C$-server loss system have $M$ types of resources of capacities given by a vector $R=(R_{_{1} } ,...,R_{M} )$. Class $k$ sessions arrive according to a Poisson process of rate $\lambda _{k} $ and have a mean service time of $1/\mu _{k} $, $\rho _{k} =\lambda _{k} /\mu _{k} $, $k=1,\ldots ,K$.

The $i$-th customer of class $k$ requires an amount $r_{km} (i)$ of resources of type $m,$which is a real-valued random variable. Resource requirements ${\it r}_{k} (i)=(r_{k_{1} } (i),...,r_{k_{M} } (i))$, $i=1,2,...$, of class $k$ customers are nonnegative random vectors with a cumulative distribution function (CDF) $F_{k} (x).$An arriving customer receives service if it finds at least one server free and the required resource amounts available. Then, the required resource amounts are allocated, and the customer holds them, along with a server, for the duration of its service time. Upon departure, the allocated resources are released.

The analysis of these complicated systems can be substantially simplified by using what can be called ``pseudo-lists''. In this approach, we deal with an otherwise same service system, but the resource amounts released upon a departure are assumed random and may differ from the amounts allocated to the departing customer upon its arrival. We assume that given the totals of allocated resources -- i.e., for each resource type, the resource amount held by all customers combined -- and the number of customers in service, the resource amounts released upon a departure are independent of the system's behavior prior to the departure instant and have an easily calculable CDF. Stochastic processes representing the behavior of such simplified systems are easier to study since there is no need to remember the resource amounts held by each customer: the totals of allocated resources suffice. This fact will be crucial for more complex systems. The reason is that for more complex systems, usually, one fails to derive an analytical expression for the stationary probabilities and the equilibrium equations have to be solved numerically. The proposed short-cut approach hence permits obtaining the performance measures of the system numerically in a reasonable time.

Note, that for more complex systems, it still has not been proved that the aggregated states' probabilities of the original process equal the stationary probabilities of the corresponding simplified process. Nevertheless, we can reasonably believe that such a simplification provides a good approximation, with an acceptable error. When applied in practice, however, this assumption is to be verified, for instance through simulation.

\section{Models of Individual Radio Components}\label{sect:system}

In this section, we survey analytically tractable models of various components that can be utilized as building blocks for specifying the prospective mmWave and THz deployment scenarios. These include BS and UE locations, propagation, antenna, blockage, micromobility, beamsearching models. For each model, we discuss the level of abstraction, parameterization and comment on its applicability to considered RATs. 

We would like to note that from the system-level performance modeling and characterization point of view there are no principal differences between mmWave and THz specifics as both technologies will be subject to blockage and micromobility impairments (albeit with quantitatively different impact) as well as utilize highly directional antenna arrays at both BSs and UEs sides. This implies that the same models can be utilized for both considered bands. Furthermore, these specifics are unique to mmWave/THz, making them very unreliable for prospective services as compared to legacy microwave systems such as LTE, and also constitute the main challenge for developing new service models of sessions and their mathematical models for these systems. The main difference is in quantitative values and also in propagation models that we will highlight explicitly in the discussion below.


\subsection{Deployment Models}

The choice of the deployment model, i.e., indoor/outdoor as well as BS and UE locations, for mmWave and THz systems is a more critical question as compared to microwave ones. The rationale is that now scenario geometry that involves not only deployment premises but the type of the surrounding objects that may cause blockage starts to play an important role in systems performance. Particularly, in addition to the effects of indoor and outdoor propagation the relative positioning of these objects affects the tractability of considered deployments.
Furthermore, the directionality of antenna radiation patterns coupled with inherent multi-path propagation of mmWave/THz bands forces researchers to switch from conventional two-dimensional (2D) models to more comprehensive three-dimensional (3D) ones. Finally, in addition to UE and BS locations, one should also need to specify blocker types, locations, and/or their mobility.

\begin{figure}[!t]
	\centering
	\subfigure[{Outdoor random deployment}]
	{
		\includegraphics[width=0.75\columnwidth]{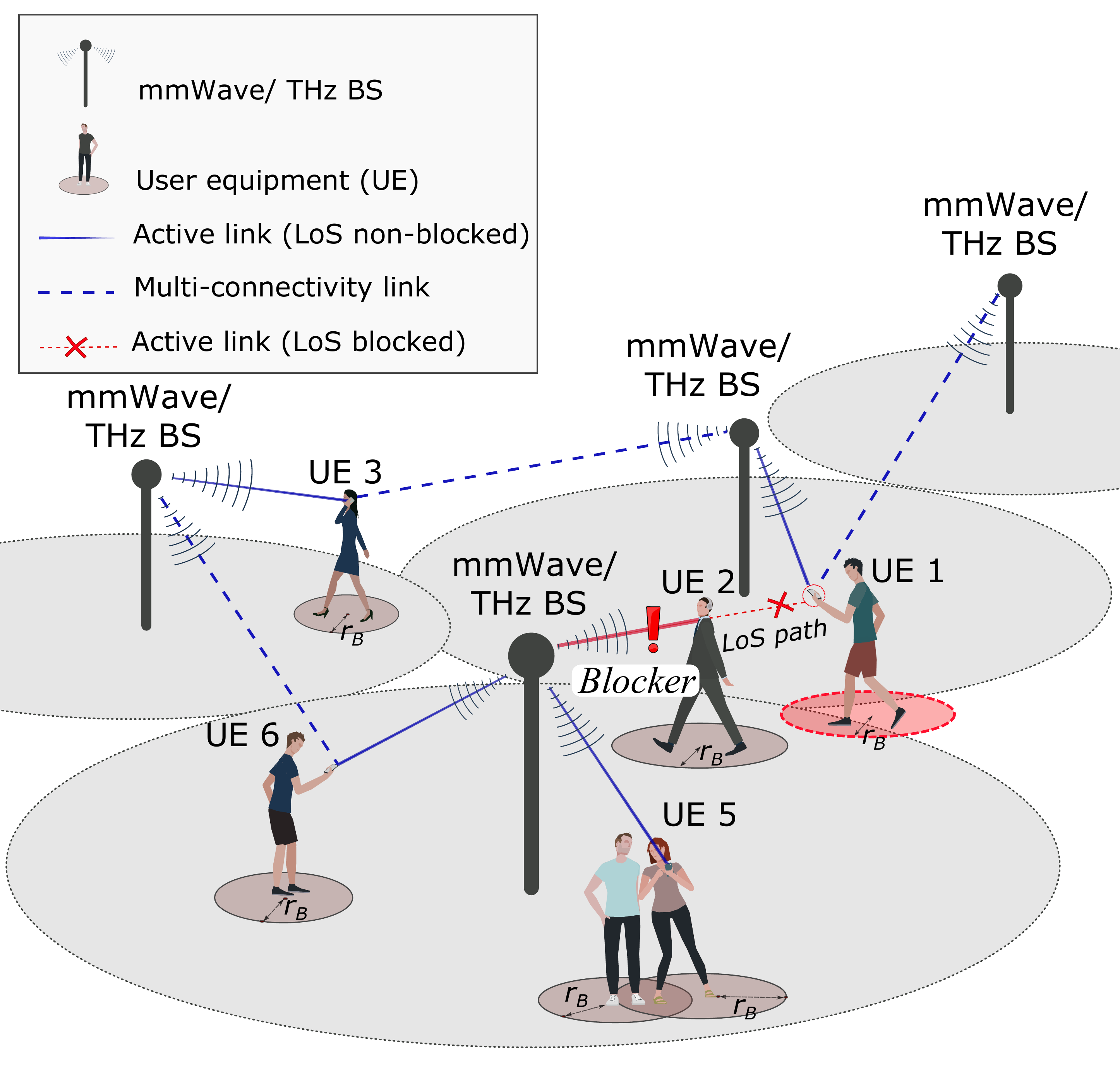}
		\label{fig:randomDep}
	}\\
	\subfigure[{Outdoor semi-regular deployment}]
	{
		\includegraphics[width=0.75\columnwidth]{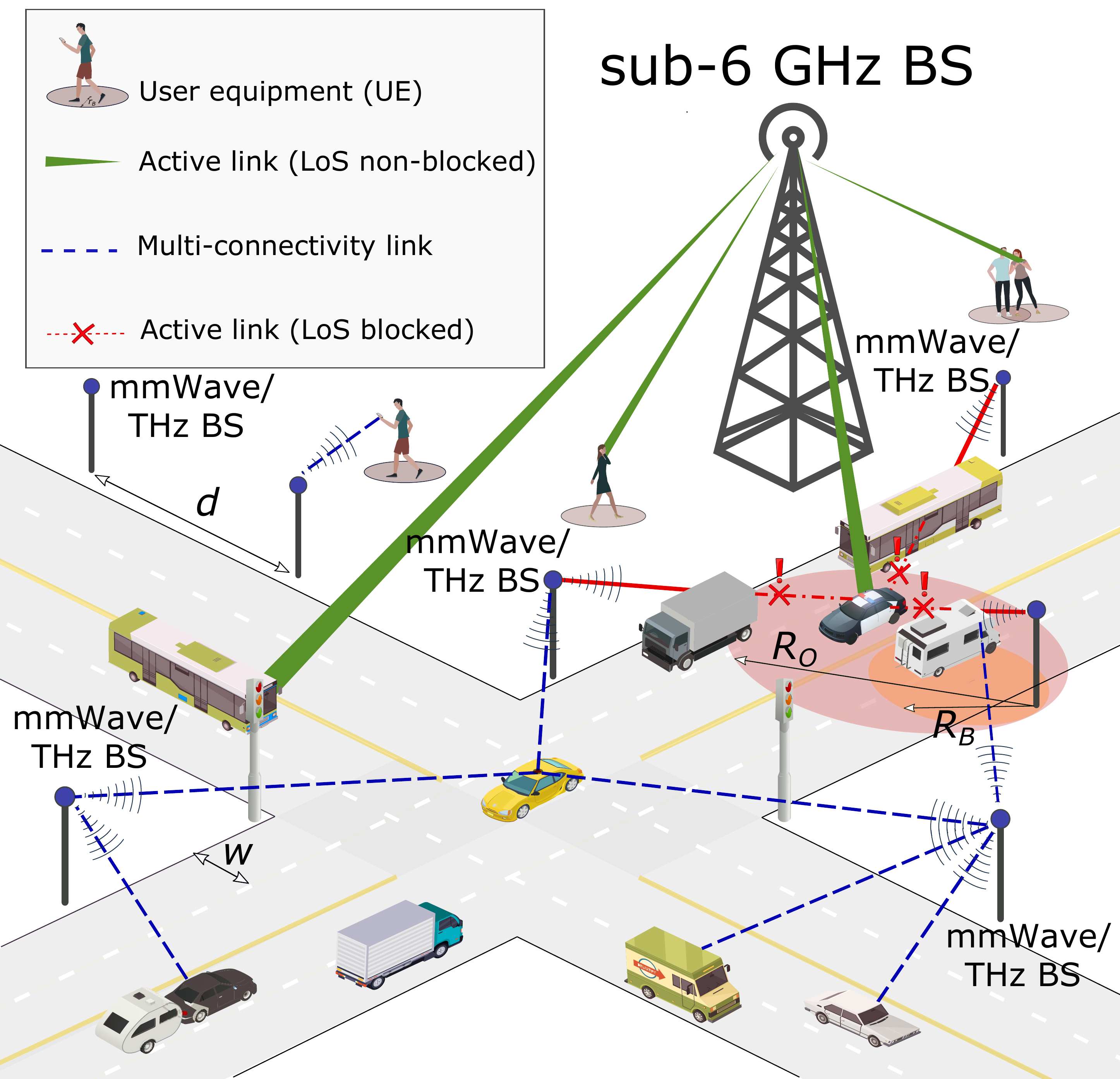}
		\label{fig:regularDep}
	}\\
	\subfigure[{Indoor regular deployment}]
	{
		\includegraphics[width=0.75\columnwidth]{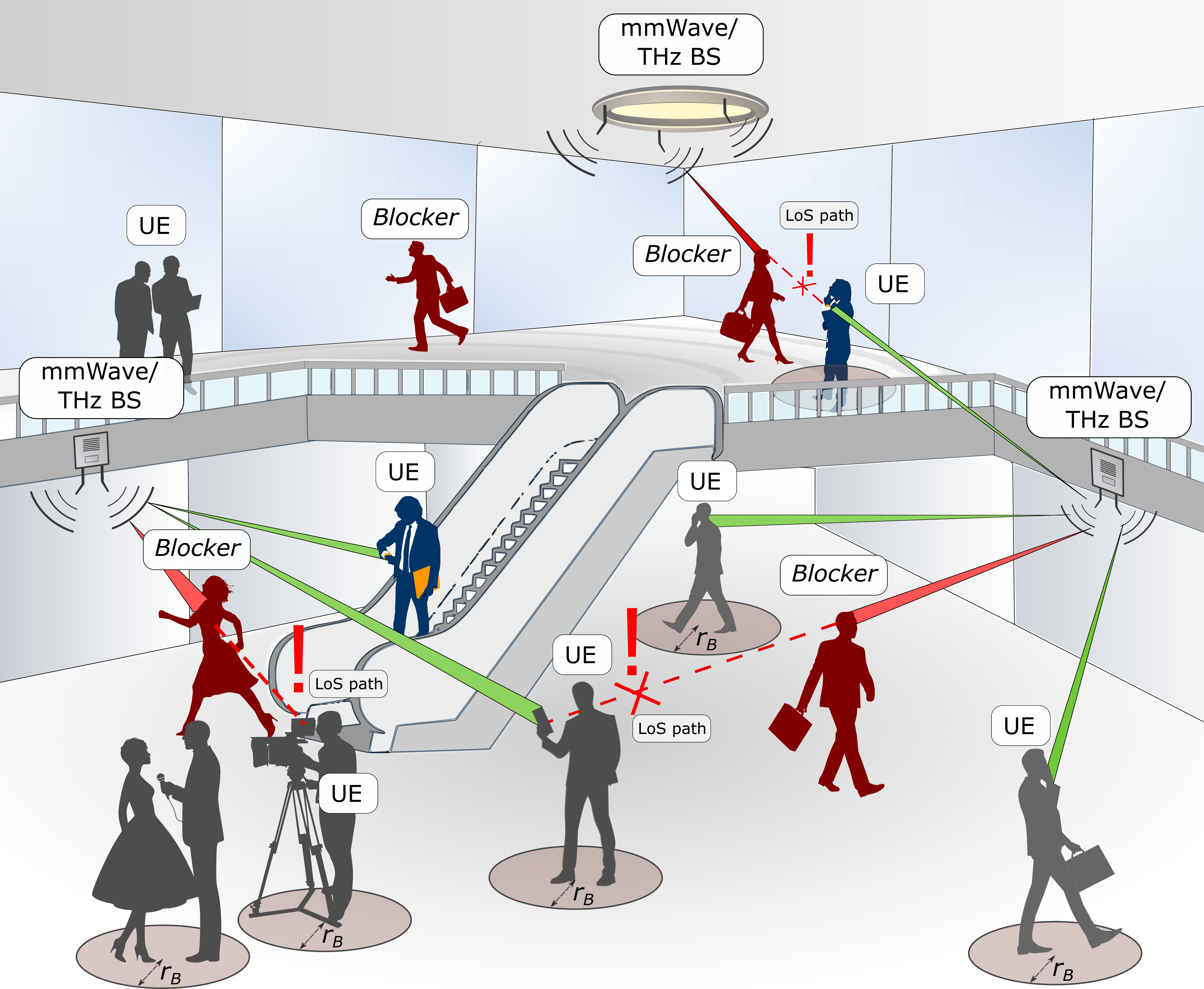}
		\label{fig:regularDep}
	}
	\caption{Illustration of different deployment types.}
	\label{fig:deployments}
	\vspace{-4mm}
\end{figure}

\subsubsection{Outdoor Purely Random Deployments}


Following the studies of microwave LTE systems \cite{elsawy2013stochastic,elsawy2016modeling}, the early works on mmWave and THz systems performance, e.g., \cite{maksymyuk2015stochastic,di2015stochastic,petrov2017interference}, assumed purely stochastic 2D deployments, where BSs, UEs, and blockers are assumed to follow certain stochastic processes in $\Re^2$, see Fig. \ref{fig:randomDep}. Here, active UEs are often characterized as a separate process or included (as a fraction of) in the specification of blockers representing humans. Often, a homogeneous Poisson point process (PPP) is utilized for tractability reasons. Indeed, the geometric distances between points in PPP are readily available \cite{moltchanov2012distance}, the coverage area of a BS can be roughly approximated by Voronoi diagrams \cite{haenggi2009stochastic}, while the distance between BS and UE can be found assuming circle approximation of Voronoi cells.


The rationale behind the use of purely stochastic deployments is that the locations of BSs may not follow the regularity assumption (i.e., cellular deployments) and possess a certain degree of randomness as discussed in \cite{li2015statistical}. Furthermore, UE may indeed be naturally randomly distributed within the cell coverage area. Finally, the use of 2D planar deployments is warranted when antenna radiation patterns are not extremely directional in the vertical dimension. Such purely random 2D deployments may still be relevant for mmWave systems in outdoor open space conditions, where the coverage of a single BS is relatively large reaching few hundreds of meters, such as squares, parks, suburbs, etc. 


Recently, extensions of these deployment models to 3D open space environments have been provided \cite{kovalchukov2018analyzing,kovalchukov2019evaluating,yi2019unified,yi2020clustered} in context of unmanned aerial vehicle (UAV) communications. Here, in addition to the planar deployment assumption, one has to characterize vertical dimension by supplying BS, UE and blocker height distributions. Assuming analytically convenient distributions such as exponential or uniform mathematically tractable models can be provided. 

\subsubsection{Outdoor Semi-Regular Deployments}

As mmWave and THz BSs are characterized by much smaller coverage areas compared to sub-6 GHz BS, the impact of scenario geometry and type of the objects in the channel is expected to be profound. This concerns both outdoor and indoor environments. First, in urban street deployments, locations of BSs are no longer stochastic and likely follow regular deployment along the street, e.g., BSs are mounted on lampposts or building walls \cite{he2019channel,kumari2020optimization,begishev2018connectivity}, UE are also naturally located along the sidewalks and crossroads. In addition to humans blocking the propagation paths, vehicles moving along straight trajectories may contribute to the blockage process in mmWave and THz systems \cite{boban2019multi,park2017millimeter,petrov2020measurements,eckhardt2021channel}, see Fig. \ref{fig:regularDep}. In city squares, BSs may also be installed along the perimeter \cite{begishev2021joint,kovalchukov2018improved}. Finally, the dimensions of blockers, e.g., vehicles, can be comparable to the communications distances and thus one may need to take into account detailed blockers geometry. The latter is also critical for indoor THz system deployments \cite{chen2021channel,xing2021millimeter}. 

The abovementioned specifics naturally lead to semi-regular environments with stochastic factors interrelated with deterministic ones. Particularly, the conventional Poisson process may not be applicable for modeling vehicle and pedestrian locations due to potential overlapping between adjacent objects requiring the utilization of hardcore processes. The potential specific locations of BS, i.e., building corners, walls, lampposts, crossroad centers lead to different performance results \cite{petrov2019analysis}. The analysis is further complicated by 3D specifics including heights of blockers, BS and UEs. As a result, the use of stochastic geometry for addressing radio part performance, although feasible, is usually more complicated and tedious as compared to purely stochastic deployments as discussed in  \cite{petrov2019analysis,wang2017blockage,wang2018mmwave}.

\subsection{Indoor Regular Deployments}

Indoor deployments present the most complex environment for modeling purposes. The main rationale is very complex propagation environment that differs from building to building making impossible to infer general laws of BS locations, see Fig. X. For this reason, studies assessing performance of mmWave/THz BS indoor deployments are forced to deal with regular fixed locations almost prohibiting mathematical system-level performance assessment and forcing investigators to reduce the scenario to single-room environment \cite{moldovan2017coverage} or utilize computer simulations or both \cite{bai2014coverage,petrov2018last}. Thus, to still obtain qualitative insights on performance measures of interest, similarly to outdoor deployment either semi-regular deployment or even fully stochastic one are often assumed with appropriate propagation models, see, e.g., \cite{bai2014coverage2,wu2020interference,wu2019interference,wang2020interference}.


\subsection{Propagation Models}

Generally, there are two types of propagation models for mmWave and THz communications system: (i) ray-tracing models and (ii) models based on fitting of measurements data. Although the former may precisely account for details of the propagation environment leading to very accurate models, they, by design, cannot be utilized in the mathematical modeling of communications systems. However, these models as well as empirical experiments allow to formulate empirical  models. These models are based on the fitting experimental data to mathematical expressions and can further be classified into two large groups: (i) ``averaged'' models and (ii) 3D multi-path cluster-based models.

\subsubsection{Averaged Models}

The SINR at the UE is written as
\begin{align}\label{eqn:genericProp}
S(y)=\frac{P_AG_AG_UL(y)}{N_0 + I},
\end{align}
where $y$ is the 3D distance between BS and UE, $P_{A}$ is the transmit power at BS, $G_A$ and $G_{U}$ are the antenna gains at the BS and the UE, respectively, $N_0$ is the thermal noise at the UE, $L(y)$ is the path loss, and $I$ is the interference.

	




Note that propagation $L(y)$ and interference $I$ components in (\ref{eqn:genericProp}) are often RVs. In (\ref{eqn:genericProp}), one may also account for fast fading and shadow fading phenomena via additional RVs with exponential and Normal distributions, respectively, \cite{umi}. To simplify the model, the shadow fading effect is often accounted for by margins, $M_{S,1}$ and $M_{S,2}$ for the LoS non-blocked. These margins are specified in \cite{umi}. 

To define the path loss, $L(y)$, for mmWave systems one may utilize the models defined in \cite{umi}. Particularly, the urban-micro (UMi) path loss is dB scale is readily given by
\begin{align}\label{eqn:plDb}
\hspace{-2mm}L_{dB}(y)=
\begin{cases}
32.4 + 21\log_{10}y + 20\log_{10}{f_c},\,\text{non-bl.},\\
32.4 + 31.9\log_{10}y + 20\log_{10}{f_c},\,\text{blocked},\\
\end{cases}\hspace{-1mm}
\end{align}
where $f_c$ is the carrier measured in GHz. The UMi path loss model for the $140$ GHz band has been introduced in \cite{xing2021propagation} while \cite{xing2021millimeter} reports an indoor-hall (InH) model for the same frequency. 

The path loss defined in (\ref{eqn:plDb}) can also be converted to the linear scale by utilizing the generic representation $A_iy^{-\zeta_i}$, where $A_i$, $\zeta_i$, $i=1,2$, are the propagation coefficients corresponding to LoS non-blocked ($i=1$) and blocked ($i=2$) conditions, i.e,
\begin{align}\label{eqn:prop_coeff}
A=A_1=A_2=10^{2\log_{10}{f_c}+3.24},\,\zeta_1=2.1,\,\zeta_2=3.19.
\end{align}
	
Now, the SINR at the UE can then be rewritten as
\begin{align}\label{eqn:prop}
S(y)=\frac{P_{A}G_{A}G_{U}}{A(N_0+I)}[{y}^{-\zeta_1}[1-p_B(y)]+{y}^{-\zeta_2}p_B(y)],
\end{align}
where $p_B(y)$ is the blockage probability \cite{gapeyenko2016analysis}
\begin{align}\label{eqn:blockage}
p_{B}(y)=1-\exp^{-2\lambda_Br_B\left[\sqrt{y^2-(h_A-h_U)^2}\frac{h_{B}-h_{U}}{h_{A}-h_{U}}+r_B\right]},
\end{align}
where $\lambda_B$ is the blockers density, $h_B$ and $r_B$ are the blockers' height and radius, $h_U$ is the UE height, $h_U\geq{}h_B$, $h_A$ is the BS height, $y$ is the 3D distance between UE and BS.
	
Finally, introducing the coefficient
\begin{align}\label{eqn:Ci}
C=P_AG_AG_U/A,
\end{align}
the SINR at UE can be compactly written as
\begin{align}\label{eqn:finalPropModel}
S(y)=\frac{C}{N_0+I} [{y}^{-\zeta_1}[1-p_{B}(y)]+ {y}^{-\zeta_2}p_{B}(y)].
\end{align}

\subsubsection{Absorption Losses in mmWave and THz Bands}


The unique property of the mmWave and THz channels is the atmospheric (molecular) absorption \cite{jornet2011channel,jornet2014femtosecond}. In the mmWave band, absorption is mostly due to the oxygen molecules while in the THz band it is mainly caused by the atmospheric water vapor \cite{hitran}. These losses may induce frequency selectivity in the channel characteristics. With absorption losses accounted for, the SINR at the UE takes the following form
\begin{equation}\label{eq:s_rx1}
S(y)=\frac{P_AG_AG_UL(y)L_A(y)}{N_0 + I},
\end{equation}
where the additional factor $L_A(y)$ represents absorption losses. Following \cite{jornet2011channel}, the absorption loss is defined as 
\begin{equation}\label{eq:l_a}
L_{A}(y) = 1/\tau(y),
\end{equation}
where $\tau(y)$ is the medium transmittance described by the Beer-Lambert-Bouguer law. The latter is related to the frequency dependent absorption coefficient $K(f)$ as $\tau(y) \approx e^{-K(f)y}$. The values of $K(f)$ can be obtained from \cite{hitran} as described in detail in \cite{jornet2011channel,jornet2014femtosecond}.

Note that in mmWave band absorption losses are only non-negligible in the 60 GHz band that is utilized for WLAN systems. In the THz band,  there are so-called transparency windows \cite{boronin2014capacity}, where the impact of these losses is negligible as well. Finally, the absorption phenomena may also lead to the molecular noise theoretically predicted in \cite{jornet2011channel}. The theoretical model for molecular noise has been proposed in \cite{boronin2015molecular}. However, recent measurements \cite{kokkoniemi2016discussion} did not reveal any noticeable impact of the molecular noise phenomenon.

\begin{figure}[!t]
\centering
\includegraphics[width=1.0\columnwidth]{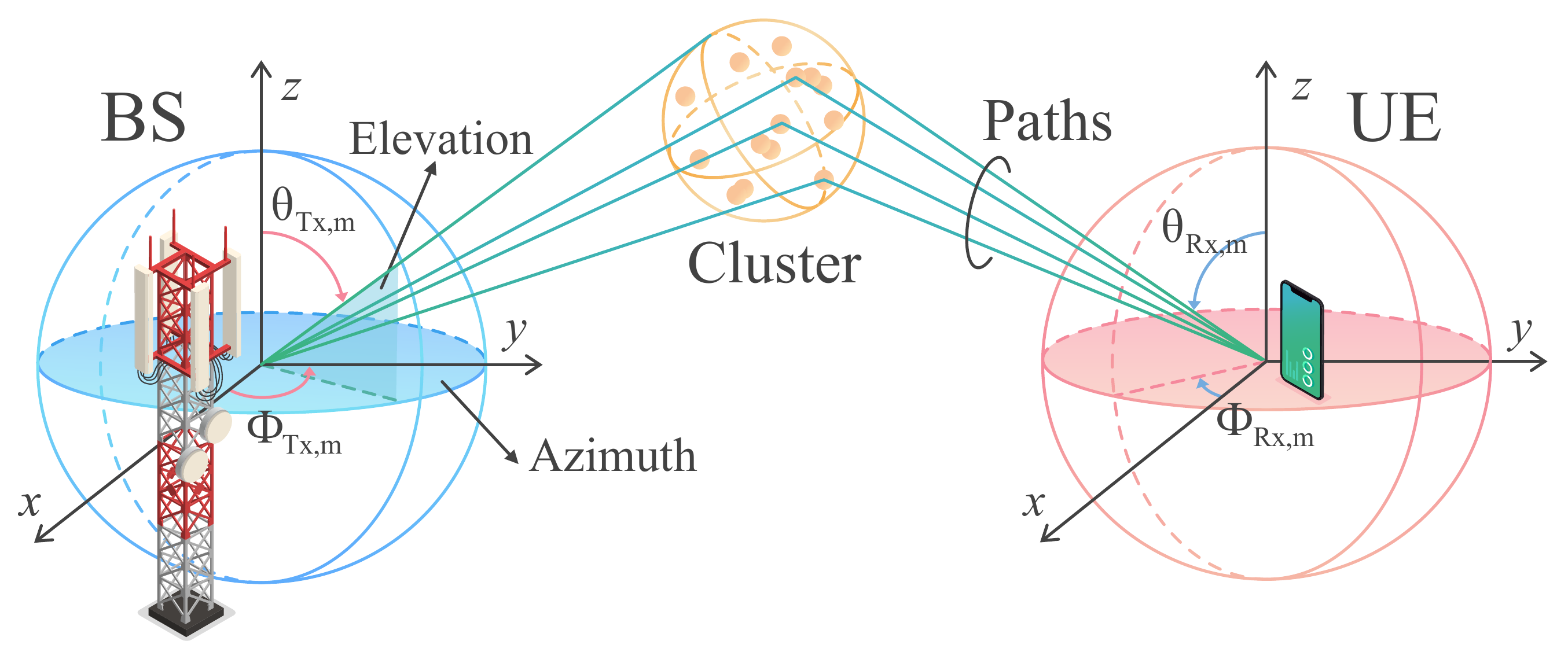}
\caption{Illustration of the 3D cluster-based propagation model: a nuber of paths between BS and UE are available each characterized with its own delay and power profiles \cite{umi}.}
\label{fig:cluster}
\vspace{-2mm}
\end{figure}

\subsubsection{Cluster-based Models}

MmWave and THz channels are inherently characterized by multi-path nature. To this aim, 3GPP standardized 3D channel model for $6-100$ GHz band in~\cite{umi}. The structure of the model, see Fig. \ref{fig:cluster}, is essentially similar to the LTE 3D model specified in \cite{standard_15_lte}, with enhancements inherent to mmWave propagation such as a blockage. According to it, the received power consists of energy coming from LoS and reflected paths. Here, the term cluster is interpreted as a surface potentially leading to the reflection of propagating rays. Taking the number of clusters in the range of $5-20$ as an input, this model relates the specifics of the propagation environment to a set of parameters that include: (i) zenith of departure and arrival (ZoD/ZoA), (ii) azimuth of departure and arrival (AoD/AoD), (iii) number of rays (paths) in a cluster, (iv) cluster delay, (v) power fraction of a cluster, and (vi) other parameters such as Ricean K-factor.


The 3GPP 3D cluster model has an explicit algorithmic structure providing no closed-form expressions for the abovementioned parameters. However, the authors in \cite{gapeyenko2018analytical} provided analytical approximations for selected parameters required for the use of this model in the mathematical analysis of mmWave systems. Particularly, they have shown that the power of a cluster follows Log-Normal distribution while AoA and ZoA can be approximated by Laplace distribution, i.e.,
\begin{align}\label{eqn:theta_Ai}
&f_{\theta_i}(y;x)=\frac{1}{2a_{i,2}(x)}e^{-\frac{|y-a_{i,1}(x)|}{a_{i,2}(x)}},\,i=1,2,\dots,W,\nonumber\\
&f_{P_{S,i}}(y;x)=\frac{1}{ya_{i,4}\sqrt{2\pi}} e^{-\frac{(\ln y - a_{i,3})^2}{2a_{i,4}^{2}}},\, i=1,2,\dots,
\end{align}
where $a_{i,1}(x)$, $a_{i,2}(x)$, $a_{i,3}(x)$, and $a_{i,4}(x)$, $i=1,2,\dots,W$, are the parameters estimated from statistical data, see \cite{gapeyenko2018analytical} for details. It has also been shown in \cite{gapeyenko2018analytical} that $a_{i,1}(x)$ are independent of the cluster number $i$ and only depend on the separation distance $x$. Further, the mean of ZoA for all clusters coincides with the constant ZoA of the LoS cluster. In its turn, $a_{i,2}(x)$, $a_{i,3}(x)$, and $a_{i,4}(x)$ are independent of the distance and only depend on the cluster number, see \cite{gapeyenko2018analytical}. The use of the model is demonstrated in \cite{petrov2019exploiting,gapeyenko2018analytical,begishev2021joint}.


As of now, 3D cluster model parameters are only available for mmWave bands for outdoor deployment conditions \cite{umi}. For other environments, such as vehicle-to-vehicle (V2V) and vehicle-to-infrastructure (V2I) communications exhaustive data for parameterization have been recently reported \cite{boban2019multi,he2019propagation}. Despite similar structure is expected to be retained by THz models as well, only a few comprehensive measurements studies are available to date that can be used to infer channel parameters, see e.g. \cite{eckhardt2021channel} for V2V, 
\cite{guan2019measurement} for the train to infrastructure links, as well \cite{he2017stochastic} for a static kiosk application.


\subsubsection{Other Types of Impairments}

In addition to the path loss as well as blockage considered in the subsequent sections, meteorological conditions such as rain \cite{sun2017novel,ghosh2014millimeter,khan2011mmwave}, fog \cite{frey1999effects,liebe1989millimeter}, snow \cite{foessel1999short} and foliage \cite{nashashibi2004millimeter,schwering1988millimeter} may provide additional impairments on mmWave and THz propagation summarized in Table \ref{tab:effects}. As one may observe, foliage produces the most significant impact resulting in up to $2$ dB/m of additional degradation. On the other hand, the impact of snow, fog, and cloud is insignificant, i.e., less than $1$ dB/km. Finally, up to 10 dB/km of signal degradation is induced by rain. The impact of these environmental conditions is expected to be higher for the THz band. These values can be utilized as an additional constant in path loss models defined above.


\subsection{Static and Dynamic Blockage Models}

\begin{figure}[!t]
\vspace{-0mm}
\centering
\includegraphics[width=1.0\columnwidth]{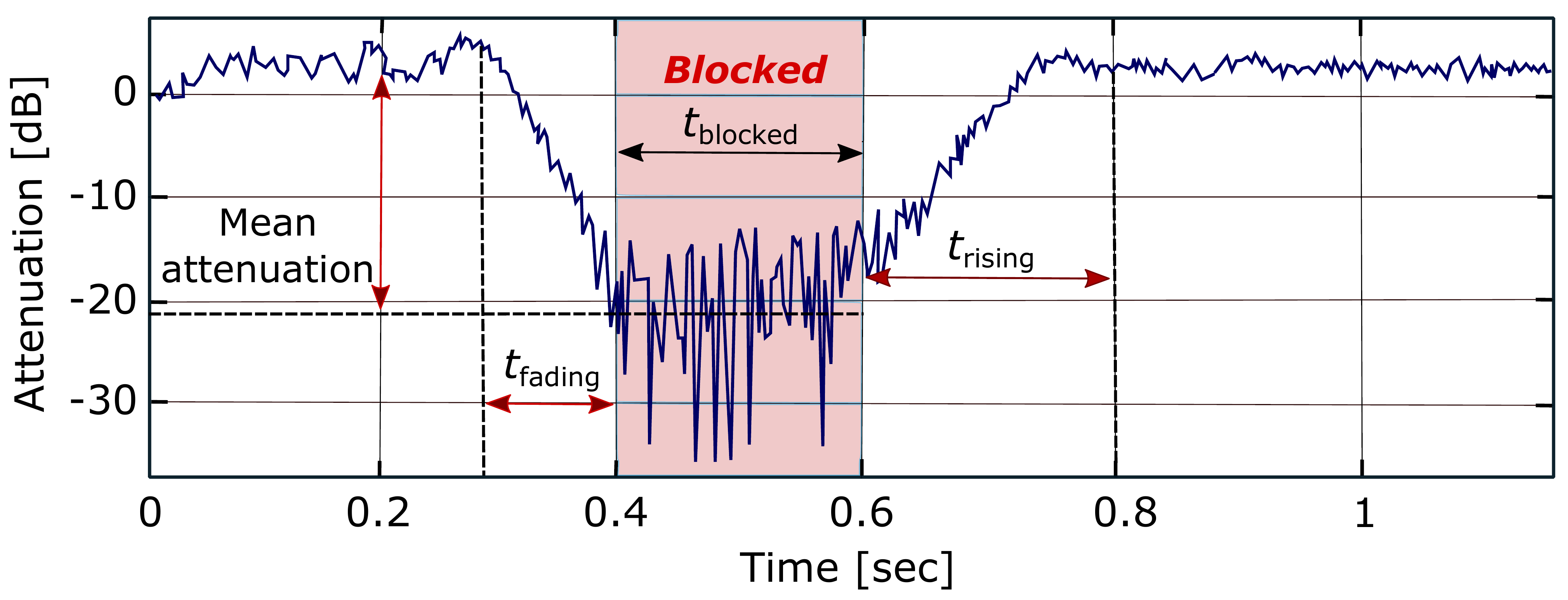}
\caption{Typical blockage profile at 60 GHz: mean attenuation characterizes the signal drop while fade, blockage, and rise intervals describe time-dependent metrics \cite{slezak2018empirical}.}
\label{fig:pathGain}
\vspace{-2mm}
\end{figure}

Blockage caused by dynamic objects in the channel is an inherent property of mmWave and THz systems. Depending on the induced attenuation on top of the path loss, the blockage may or may not lead to outage conditions. In the latter case, the communications may still be possible at much reduced modulation and coding scheme (MCS). In the former case, depending on outage duration, application layer connectivity may or may not be interrupted. Thus, to comprehensively characterize it for performance evaluation studies one needs to provide: (i) attenuation values induced by different objects in the channel and (ii) blockage intervals under different types of UE and blockers and their mobilities.

The dynamic human body blockage introduces additional uncertainty in the channel resulting in drastic fluctuations in the received signal. An illustration of the typical measured path loss experienced as a result of human body blockage by UE in the 60 GHz band is shown in Fig.~\ref{fig:pathGain}. The absolute values of blockage induced attenuation heavily depend on the type of blockers. For human body blockage in mmWave band values of losses in the range of 15--25 dB have been reported \cite{weiler2016environment,maccartney2016millimeter,maccartney2017rapid}.  For THz band these losses are expected to reach $40$ dB \cite{bilgin2019human}. Blockage by vehicles at 300 GHz heavily depends on the vehicle type and geometry and reported to be from 20 dB at the front-shield glass level up to 50 dB at the engine level. They are considerably higher than those for mmWave band. Particularly, at $60$ GHz, 5 dB--30 dB blockage losses have been reported. Note that the values of blockage losses also depend on the vehicle size and the number of them between communicating entities \cite{yamamoto2008path}. For 28 GHz, the authors in \cite{boban2019multi,park2017millimeter,park2018vehicle} also report the following height-dependent vehicle blockage losses: 11 dB--12.2 dB for 1.7 m, 13.3 dB for 1.5 m, and 30 dB--40 dB for 0.6 m.

The fading and recovery phases highlighted in Fig.~\ref{fig:pathGain} are reported to be on the scale from tens to a couple of hundreds of milliseconds \cite{slezak2018empirical,maccartney2016millimeter,maccartney2017rapid}. Similar observations have been made recently for THz links \cite{bilgin2019human}. In mathematical modeling, these phases are often omitted assuming that fading and recovery phases are negligibly small compared to the blockage duration. The blockage models utilized in performance evaluation frameworks attempt to predict the probability of blockage and blockage duration that can be caused by multiple blockers occluding the propagation paths between UE and BS.

\begin{figure}[!t]
\vspace{-0mm}
\centering
\includegraphics[width=0.8\columnwidth]{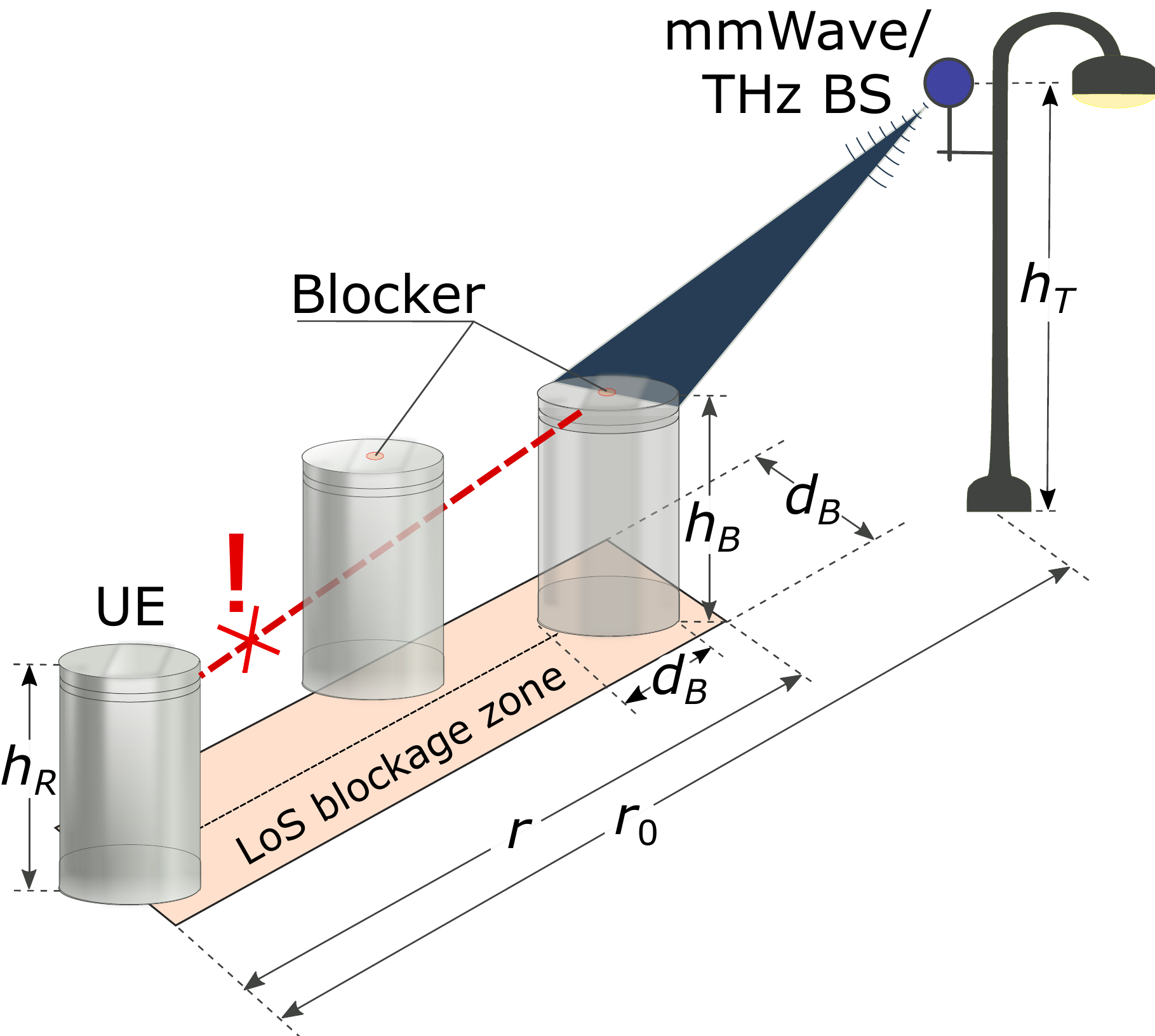}
\caption{Illustration of the LoS blockage zone: the length is defined by the distance between UE and BS while the width is specified by the blocker's ``radius'' \cite{gapeyenko2016analysis}.}
\label{fig:meth_staticBlockage}
\vspace{-2mm}
\end{figure}

\subsubsection{Static Human Body Blockage Models} 

Most of the static blockage models proposed in the literature, e.g., \cite{bai2014analysis,gapeyenko2016analysis}, assume the following model. Consider the static case of stationary UE in $\Re^2$ located in stationary homogeneous PPP of blockers with intensity $\lambda_B$ considered in \cite{gapeyenko2016analysis}. Assume that UE is located at a 2D distance $r_0$ from the BS. UE and BS are located at heights $h_T$ and $h_R$, respectively. Human bodies are represented by cylinders with constant height $h_B$ and base diameter $d_B=2r_B$. In the considered scenario, one may introduce the so-called ``LoS blockage zone'' of rectangular shape with sides $2r_B$ and $r$ as shown in Fig. \ref{fig:meth_staticBlockage}. Observe that the LoS is blocked when a center of at least one blocker is located inside this zone. Utilizing the void probability of the Poisson process the LoS blockage probability immediately follows
\begin{align}\label{eqn:meth_probNbl}
p_{B}(x)=1-e^{-2\lambda_Br_B\left[r_0\frac{h_{B}-h_{R}}{h_{T}-h_{R}}+r_B\right]}.
\end{align}

Analyzing (\ref{eqn:meth_probNbl}) one may deduce that the blockage probability increases exponentially as a function of blockers density. The heights of UE and BS also affect the final values.  Note that this model can be utilized in those cases when UE and blockers are stationary or to predict the time-averaged behavior of mobile blockers in the mobile field of blockers. However, in performance evaluation frameworks addressing session continuity parameters, one needs to characterize blocked and non-blocked times explicitly as discussed below.

\subsubsection{Mobile UE and Static Blockers}



\begin{figure}[t!]
\vspace{-0mm}
\centering
\includegraphics[width=0.8\columnwidth]{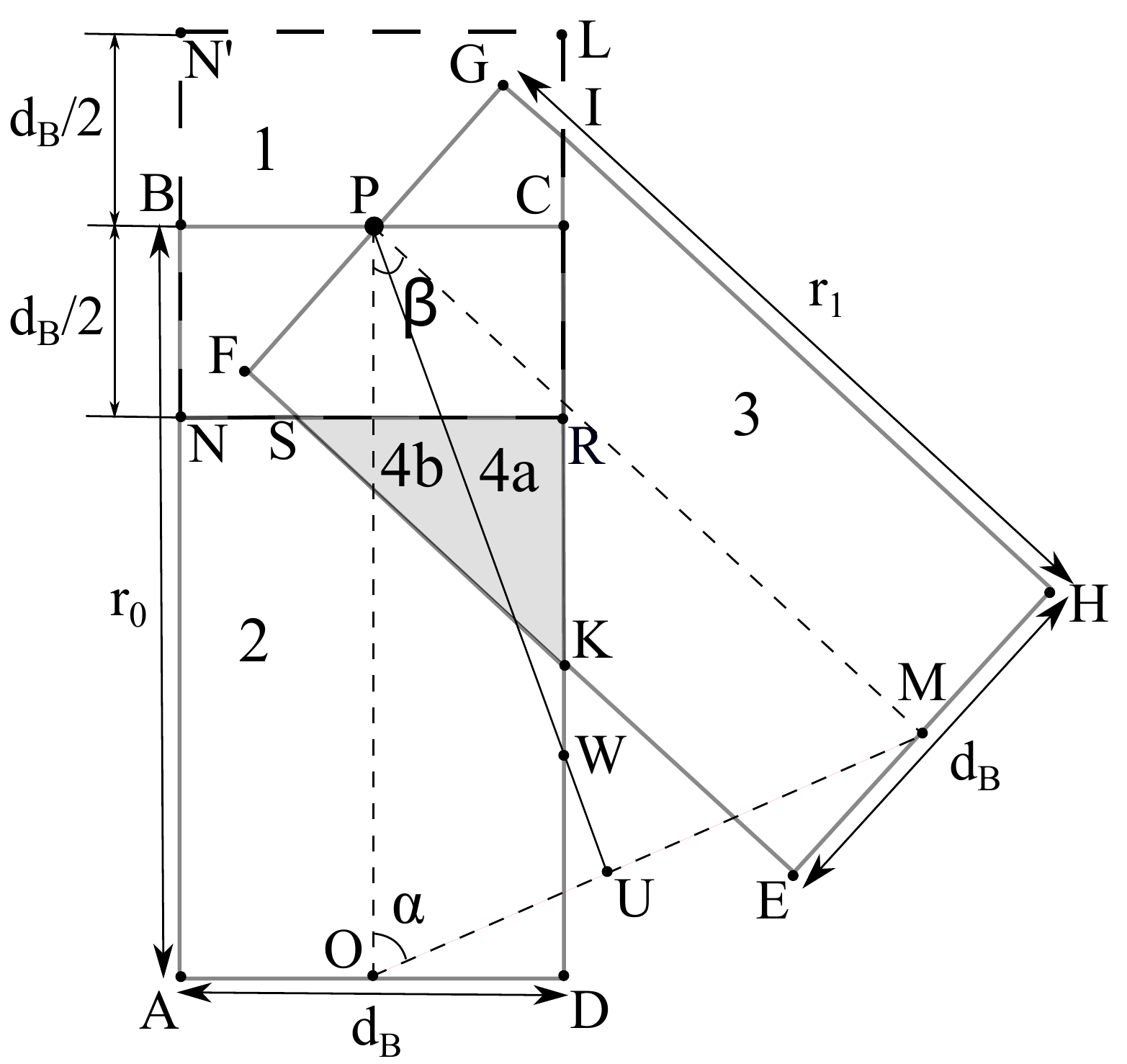}
\caption{Illustration of the LoS blockage zones for the case with moving UE: UE moves from point $O$ to point $M$, BS is located at point $P$; zones $4a$ and $4b$ define dependences between LoS states at $O$ and $M$ \cite{samuylov2016characterizing}.}
\label{fig:meth_MobileUE}
\vspace{-2mm}
\end{figure}

Consider the case when UE moves according to the uniform rectilinear pattern in a static homogeneous PPP of blockers considered in \cite{samuylov2016characterizing}. In this case, in addition to the fraction of time UE spends in blockage, one is also interested in conditional blockage probabilities of UE at $M$ given a certain state at $O$, see Fig. \ref{fig:meth_MobileUE}. To capture dependence between links states one may utilized conditional probabilities defined as $p_{ij}$ P\{at M LoS blocked/non-blocked given that LoS at O is blocked/non-blocked\}. These probabilities can be formed in a matrix as
\begin{align}\label{eqn:1}
P=
\begin{pmatrix}
p_{00} & p_{01} \\
p_{10} & p_{11} \\
\end{pmatrix}
\end{align}
where states $0$ and $1$ reflect the non-blocked and blocked states. In general, these probabilities are a function of (i) distance from $O$ to $P$, $r_0$, (ii) distance from $O$ to $M$, $d$, (iii) angle $\angle$POM, see Fig. \ref{fig:meth_MobileUE}, (iv) blockers density $\lambda_A$, (v) heights of UE and BS at $P$, $O$, and $M$, $h_C$. Observing that
\begin{align}\label{eqn:1m}
p_{00}=1-p_{01},\quad p_{10}=1-p_{11},
\end{align}
one needs $p_{00}$ and $p_{11}$ to fully parameterize (\ref{eqn:1}).

The illustration of the model is shown in Fig. \ref{fig:meth_MobileUE}, where rectangles represent the areas affecting the blockage of $PO$ and $PM$ links. Their width and length are determined by the blockers diameter $d_B=2r_B$ and the link lengths $PO$ and $PM$. The intersection area of these rectangles visually represents the correlation between link states. Particularly, this zone can be further divided into sub-zones that affect it differently. More specifically, zone $4$ is further split into two smaller zones, $4a$ and $4b$, representing the area on the right and left sides, respectively, along with the $PU$ line of intersection of two planes. Observe that zone $1$ can be excluded as it is fairly small for any potential geometry. Zones $2$ and $3$ affect LoS state at $O$ and $M$, respectively. Furthermore, LoS state at both $O$ and $P$ is simultaneously affected by zones $4$a and $4$b. By using these observations, the probability $p_{00}$ is obtained as
\begin{align}\label{eqn:p00_0}
p_{00}  = P\left[\text{nB at M}|\text{nB at O}\right] = \frac{P\left[\text{nB at M}\cap\text{nB at O}\right]}{P\left[\text{nB at O} \right]},
\end{align}
where $nB$ and $B$ denote LoS non-blocked and blocked states. Once $p_{ij}$ are determined, conditional and unconditional cumulative distribution function (CDF) and pdf of time spent in blocked and non-blocked states can be determined as shown in \cite{gapeyenko2019spatially}.

\subsubsection{Static UE and Mobile Blockers}


Consider the case when a static UE is surrounded by blockers moving according to random direction model (RDM, \cite{nain2005properties}) considered in \cite{gapeyenko2017temporal}. Here, the main emphasis is on the UE state evolution in time. In addition to the metrics related to the time UE spends in the blocked and non-blocked states, one is also interested in the conditional probability of blockage at time $t_1$, given that the state at $t_0$, $t_0<t_1$ is either blocked or non-blocked. The latter can be further utilized to obtain CDF and pdf of time in blocked and non-blocked states. 

\begin{figure}[b!]
\vspace{-2mm}
\centering
\includegraphics[width=0.9\columnwidth]{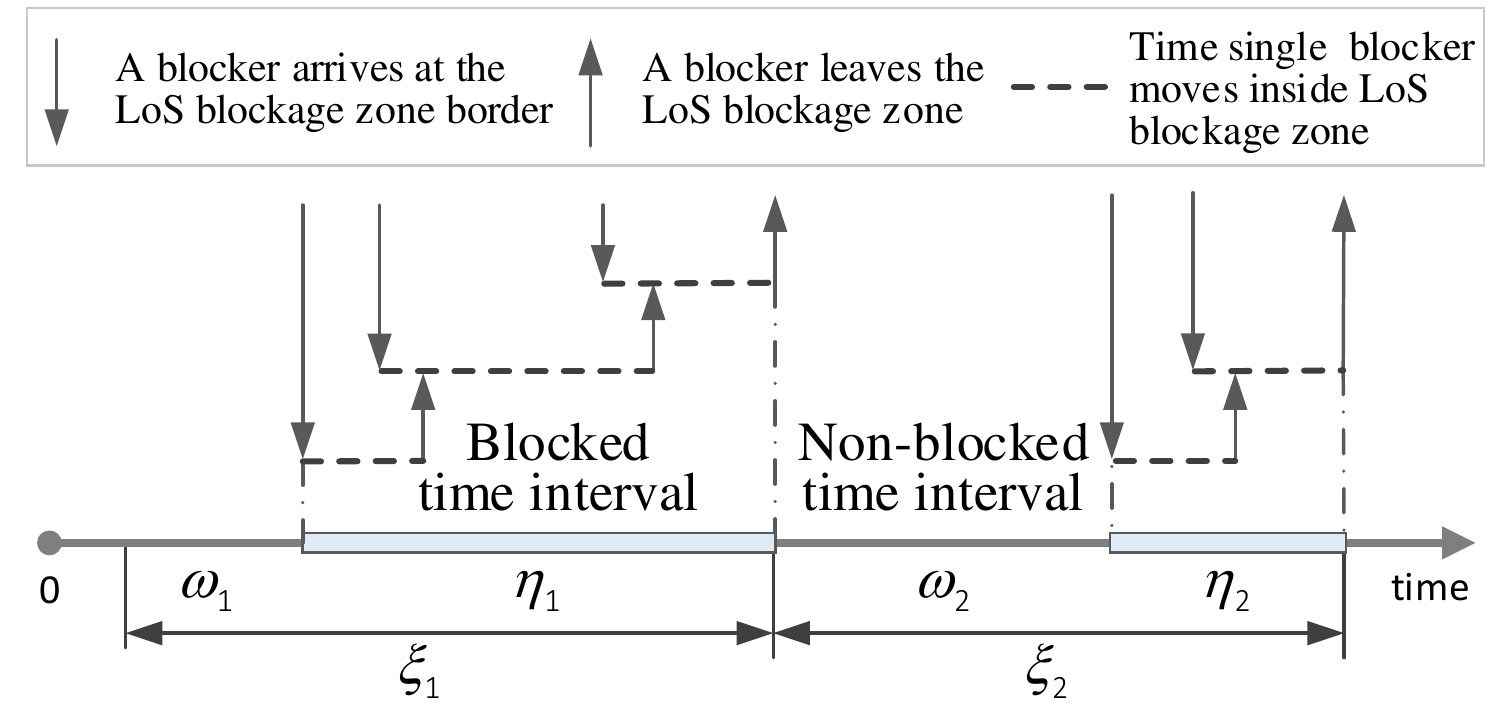}
\caption{Renewal process associated with the LoS blockage: blocked period is initiated by the arrival of a blocker to the LoS blockage zone and terminated when the last blocker leaves the LoS blockage zone empty; then, the non-blocked period is initiated forming a renewal process \cite{gapeyenko2017temporal}.}
\label{fig:renewal_process}
\vspace{-0mm}
\end{figure}

To determine these characteristics, the authors in \cite{gapeyenko2017temporal} rely upon the renewal theory. Particularly, the time UE state evolution in time is illustrated in Fig. \ref{fig:renewal_process}. The arrival process of blockers to the LoS blockage zone is assumed to be homogeneous Poisson. The time a single blocker spends in the LoS blockage zone follows the general distribution obtained in \cite{gapeyenko2017temporal}. Alternatively, to simplify the model, observe that the length of LoS blockage zone is often much larger than the width \cite{gapeyenko2016analysis}, implying that one may assume that the blockers enter the blockage zone at the right angle to the long side leading to the constant zone residence time. Further, by utilizing the analogy with queuing theory, the alternating blocked and non-blocked periods coincide with the free and busy periods in equivalent M/G/$\infty$ queuing system, where the service time is the time spent by a single blocker in the LoS blockage zone, leading to the structure represented in Fig. \ref{fig:renewal_process}. Although the busy interval is readily available from \cite{daley1998idle}, the structure of its CDF is rather complicated. Alternatively, one may replace M/G/$\infty$ model with a simpler equivalent M/M/$\infty$ queue having exponential LoS zone residence time with the same mean.

The approach utilized to determine the sought conditional probabilities is based on convolutions in the time domain. Particularly, the following relation is established
\begin{align}\label{eqn:meth_sum}
p_{00}(\Delta{t}) = \hspace{-1mm}\sum_{i=0}^{\infty}P\{A_{i}(\Delta{t})\},p_{01}(\Delta{t}) = \hspace{-1mm}\sum_{i=1}^{\infty}P\{B_{i}(\Delta{t})\},
\end{align}
where the events $A_{i}(t)$ describe the cases of being initiated in the non-blocked state at $t_{0}$ and ending up in the non-blocked state in $\Delta{t}=t_{1}-t_{0}$, while having exactly $i$, $i=0,1,\dots$, blocked periods during $\Delta{t}$. Similarly, $B_{i}(t)$ are the events corresponding to starting in the non-blocked interval at $t_{0}$ and ending up in the blocked interval, while having exactly $i$, $i=1,2,\dots$, non-blocked periods during $\Delta{t}$. Observe that the absolute values of $A_i(t)$ and $B_i(t)$ diminishes quickly making computation of $p_{00}(\Delta{t})$ and $p_{01}(\Delta{t})$ feasible.

\subsubsection{Mobile UE and Mobile Blockers}


The generic model when blockers and UEs are both mobile is much more difficult to handle. The only study that addressed this case the authors are aware of is \cite{moltchanov2019analytical}, where a Markov model has been developed to capture the mean duration of blocked and non-blocked periods. Several mobility models have been considered. For uniform rectilinear movement of UE in a field of blockers moving according to RDM, the infinitesimal generator of the Markov blockage models takes the following form
\begin{align}\label{eqn:ctmc_two}
\Lambda(g(0))=\left[
\begin{matrix}
-\alpha(g(0))&\alpha(g(0))\\
\beta(g(0))&-\beta(g(0))\\
\end{matrix}\right],
\end{align}
where $\alpha(g (0))=1/E[\Theta (g (0))]$, $\beta(g (0)) = 1 /E[\Omega (g (0))]$ are the mean duration of intervals when UE is in blocked and non-blocked conditions. These parameters depend on the UE and blockers mobility models and are derived in \cite{moltchanov2019analytical}. Since the model presumes exponentially distributed blocked and non-blocked intervals, it is approximate in nature capturing the dynamics of the means of these intervals \cite{gapeyenko2017temporal}.

\subsubsection{Spatial and Temporal Consistencies}




There are two critical requirements propagation models for THz and mmWave bands need to satisfy to represent real propagation conditions evolving in time. These are spatial and temporal consistencies ensuring that the model provides consistent prediction in time and space. The correlated channel behavior is caused by both macro objects such as buildings and micro objects including the human crowd or vehicles around the UE of interest. Ensuring spatial and temporal consistencies is complicated by the presence of dynamic blockers in the channel.

The 3GPP 3D cluster-based model has been extended to capture spatial correlation caused by large static objects in the channel such as buildings. Particularly, the standard in \cite{umi} suggests three different methods. According to the first one, known as \textit{the method of spatially-consistent RVs}, the correlation is artificially introduced to clusters by altering RV utilized for their generation. In the second alternative, called \textit{the geometric stochastic approach}, a grid is added to the plane, where UEs are located. The large-scale channel model parameters are generated in a correlated manner for all UEs located within the same grid cell. As a result, in this model, two nearby UEs may still have completely independent conditions. Furthermore, the grid dimensions are chosen rather arbitrarily. Finally, according to the last method, \textit{method of geometrical cluster locations}, small-scale 3D cluster model parameters such as cluster delays, AoD and AoA, ZoD and ZoA are generated in a correlated manner. It is critical that all the considered models the parameters affecting the correlation distance are chosen arbitrarily without direction relation to the blockers density and geometry.


There have been only a few attempts other than those standardized by 3GPP to capture spatial correlation caused by mobile objects including vehicles and human crowds. The study in \cite{MacCartney_GC_cor} measured the UE blockage probability to multiple BSs and reported correlation states between multiple links. The authors in \cite{Aditya} proposed a mathematical model to derive blockage probability to multiple BS simultaneously. Note that their model is principally similar to the one described above for the case of mobile UE and static blockers. Based on the results of \cite{samuylov2016characterizing} the authors in \cite{gapeyenko2019spatially} provided a model ensuring spatial consistency that does not require the correlation distance to be known in advance. They revealed that the correlation distance in mmWave band for a wide range of blockers density is limited to just a few meters.

\begin{figure}[!b]
\vspace{-2mm}
\centering	
\includegraphics[width=0.8\columnwidth]{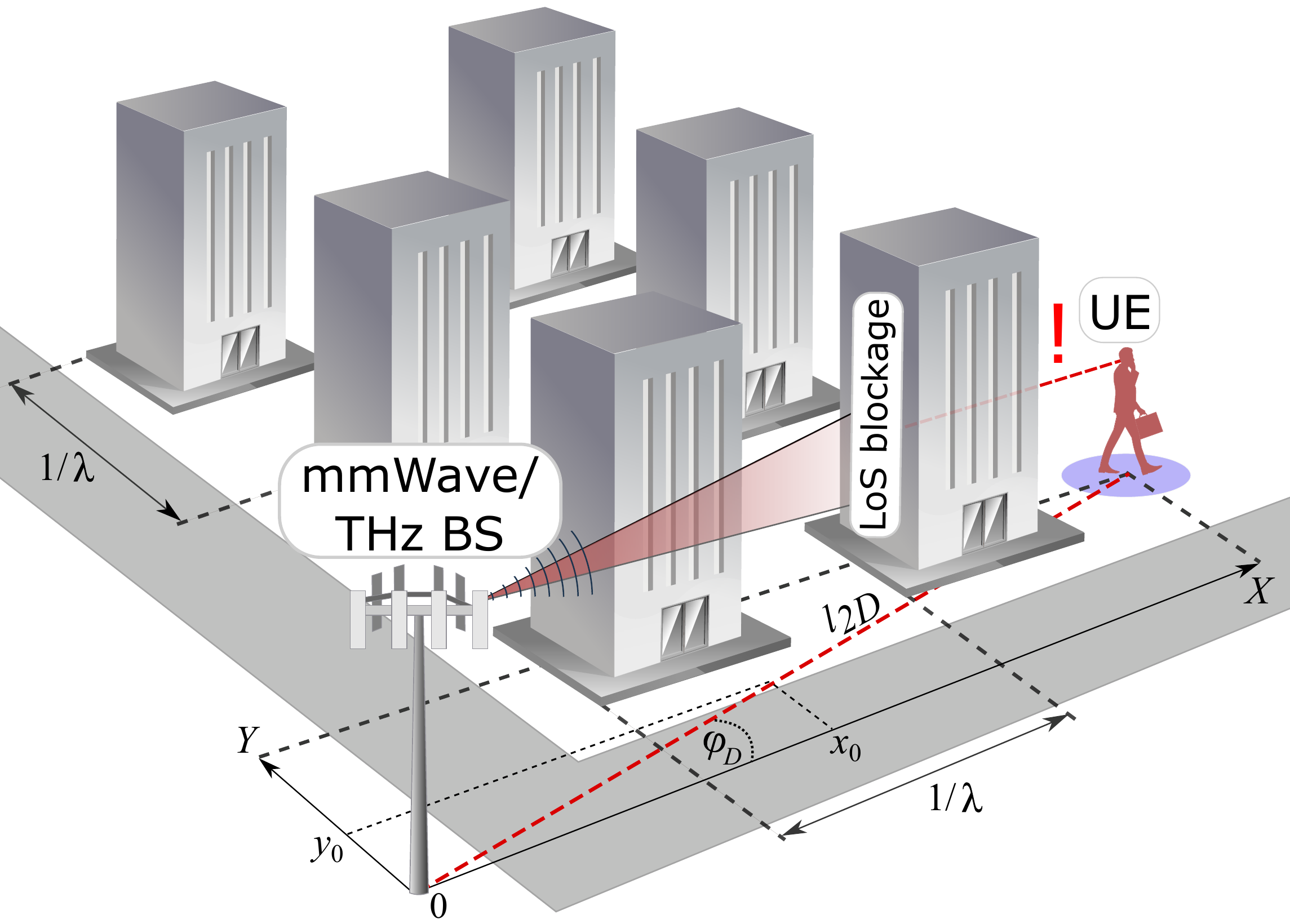}
\caption{LoS blockage probability by buildings: LoS path can be occluded by a series of buildings \cite{gapeyenko2021line}.}
\label{fig:block_building}
\vspace{-0mm}
\end{figure}

\subsection{Blockage by Large Static Objects}


In addition to blockage by small dynamic objects such as humans or vehicles, propagation paths in mmWave and THz systems can be blocked by large static objects such as buildings \cite{heath2,gapeyenko2021,Baccelli_manhattan,Heath_LoS_ball}, see Fig. \ref{fig:block_building}. To distinguish from dynamic blockage, in 3GPP terminology, blockage by large static objects is referred to as non-LoS conditions. 

There have been several models proposed in the past to model building blockage phenomena. The critical part of these models affecting their analytical tractability is the choice of the city deployment. Particularly, in \cite{heath2} the authors propose a random shape theory to model irregular deployment of buildings. The buildings are represented as rectangles with random sizes having their centers forming a homogeneous PPP. The proposed approach suits well for those deployments, where building locations are purely stochastic. The authors in \cite{Heath_LoS_ball} formalized the so-called ``ball'' model, where there is always LoS in a certain circularly shaped area around UE. In addition to the stochastic city deployments, studies addressing the regular city grids have also been performed. In \cite{Baccelli_manhattan} the city grid has been captured using the Manhattan Poisson line process (MPLP), where the lines represent the streets. In general, analysis of LoS probability in semi-regular deployments is more involved compared to stochastic ones.


The regular city grid deployment has also been considered by standardization bodies including both ITU-R and 3GPP. Specifically, ITU-R P.1410 \cite{ituBlock} considers the frequency band $20-50$ GHz and presents LoS probability in the following form
\begin{align}
\hspace{-2mm}P_{LoS}^{ITU} = \prod_{n=0}^{m}\left[1-\exp\left[\frac{[h_T - \frac{(n+\frac{1}{2})(h_T - h_R)}{m+1}]^2}{2\gamma^{2}}\right]\right],
\end{align}
where $m=\lfloor{}r\sqrt{(\alpha \beta)}\rfloor{} - 1$ is the mean number of buildings in between UE and BS, $r$ is the 2D distance measured in kilometers, $h_T$, and $h_U$ and are BS and UE heights. The parameters $\alpha$, $\beta$, and $\gamma$ are the input model parameters representing the deployment specifics including building dimensions, density, and height. One of the limitations of the model is that it does not capture 2D spatial locations of buildings concentrating on buildings locations on a LoS projection between UAV and BS to $\Re^2$. Although this model accounts for different mean heights of BS and UE, it does not capture their height distributions as well as LoS AoD. 



In TR 36.777 3GPP has also proposed a large-scale LoS blockage model. This model differentiates between deployment types providing separate solutions for them. The structure of the model is also unique for different heights of BS and UE. For UMi environment with appropriate heights of BS and UE, the LoS blockage probability is
\begin{align}
\hspace{-2mm}P_{LoS}^{3GPP} = 
\begin{cases}
1, &l_{2D}\leq d, \\
\frac{d}{l_{2D}} + \left[1-\frac{d}{l_{2D}}\right]\exp\left[\frac{-l_{2D}}{p_{1}}\right], &l_{2D}> d,\\
\end{cases}
\end{align}
where the variables $p_1$ and $d$ are defined as
\begin{align}
&p_{1} = 233.98 \log_{10}(h_R) - 0.95\nonumber\\
&d = \max (294.05 \log_{10} (h_R) - 432.94,\,\,18).
\end{align}


The generic structure of the 3GPP model makes it complicated to apply it for the specific city deployment with unique dimensions. In general, the 3GPP model also requires a careful choice of parameters to assess the performance related to the specific deployment.


The ITU-R model has been extended to 3D deployments by the authors in \cite{gapeyenko2021}. Although the model has been originally developed for UAV LoS blockage it can also be applied to the case of BS to terrestrial UE communications. Following their study, the LoS probability in Manhattan Poisson line process (MPLP \cite{morlot2012population}) is provided by, see Fig. \ref{fig:block_building},
\begin{align}\label{eqn:prop1}
&P_{\text{LoS}}(l_{2D}, \phi_{D}) = \nonumber\\
&=F_{H_B}(h_{m}^{0})\exp\Bigg(-\lambda \int\limits_{x_0}^{l_x}\Big[1-F_{H_B}\big(h_{m}^{x}(x)\big)\Big]\text{d}x-\nonumber\\ &-\lambda \int\limits_{y_0}^{l_y}\Big[1-F_{H_B}\big(h_{m}^{y}(y)\big)\Big]\text{d}y\Bigg),
\end{align}
where $F_{H_B}(x)$ is the CDF of building block height, $F_{H_B}(h_{m}^{0})$ is the probability that the height of the first building along the LoS path is lower than LoS height, $h_{m}^{0}$, $F_{H_B}\big(h_{m}^{x}(x)\big)$ and $F_{H_B}\big(h_{m}^{y}(y)\big)$ are the probabilities that the sides perpendicular to the $Ox$ and $Oy$ axes are lower than the LoS heights, $h_{m}^{x}(x)$ and $h_{m}^{y}(y)$, at the point of their intersection, respectively. The authors  have also derived closed-form solutions for uniformly, exponentially and Rayleigh distributed building heights.


The large-scale static and small-scale dynamic blockage models need to be utilized together for performance assessment of mmWave and THz deployments. This can be done similarly to \cite{margarita2018multicon}, where the authors determined zones corresponding to exhaustive superposition of nLoS and blockage states, i.e., (LoS,blocked), (LoS,non-blocked), (nLoS,blocked), (nLoS, non-blocked). Further, the performance of UEs located in these zones can be analyzed separately and then combined by weighting with probabilities of UE being located in these zones. We also note that due to inherent limitations on the communications range, blockage by buildings may not be relevant for THz cellular systems.




\begin{figure*}[!t]
\vspace{-0mm}
\begin{align}\label{eqn:overallPDF}
f_{T_A}(t)&=\frac{\frac{e^{-\frac{(\log (t)-\mu_x)^2}{2 \sigma_x^2}}}{\sigma_x} \left[2-\text{erfc}\left(\frac{\mu_y-\log (t)}{\sqrt{2} \sigma_y}\right)\right]+\frac{e^{-\frac{(\log (t)-\mu_y)^2}{2 \sigma_y^2}}}{\sigma_y} \left[2-\text{erfc}\left(\frac{\mu_x-\log (t)}{\sqrt{2} \sigma_x}\right)\right]}{ 4   \sqrt{2 \pi } t       \left[1-\frac{1}{2} \text{erfc}\left(\frac{\mu_\phi-\log (t)}{\sqrt{2} \sigma_\phi}\right)+\frac{1}{2} \text{erfc}\left(\frac{\mu_\theta-\log (t)}{\sqrt{2} \sigma_\theta}\right)\right]^{-1}}
+\nonumber\\
&\qquad{}\qquad{}\qquad{}\qquad{}\qquad{}\frac{\frac{e^{-\frac{(\log (t)-\mu_\phi)^2}{2 \sigma_\phi^2}}}{\sigma_\phi} \left[2-\text{erfc}\left(\frac{\mu_\theta-\log (t)}{\sqrt{2} \sigma_\theta}\right)\right]+\frac{e^{-\frac{(\log (t)-\mu_\theta)^2}{2 \sigma_\theta^2}}}{\sigma_\theta} \left[2-\text{erfc}\left(\frac{\mu_\phi-\log (t)}{\sqrt{2} \sigma_\phi}\right)\right]}{ 4   \sqrt{2 \pi } t       \left[1-\frac{1}{2} \text{erfc}\left(\frac{\mu_x-\log (t)}{\sqrt{2} \sigma_x}\right)+\frac{1}{2} \text{erfc}\left(\frac{\mu_y-\log (t)}{\sqrt{2} \sigma_y}\right)\right]^{-1}}.
\end{align}
\normalsize
\hrulefill
\vspace{-0mm}
\end{figure*}

\begin{figure}[b!]
\vspace{-2mm}
\centering
\includegraphics[width=0.63\columnwidth]{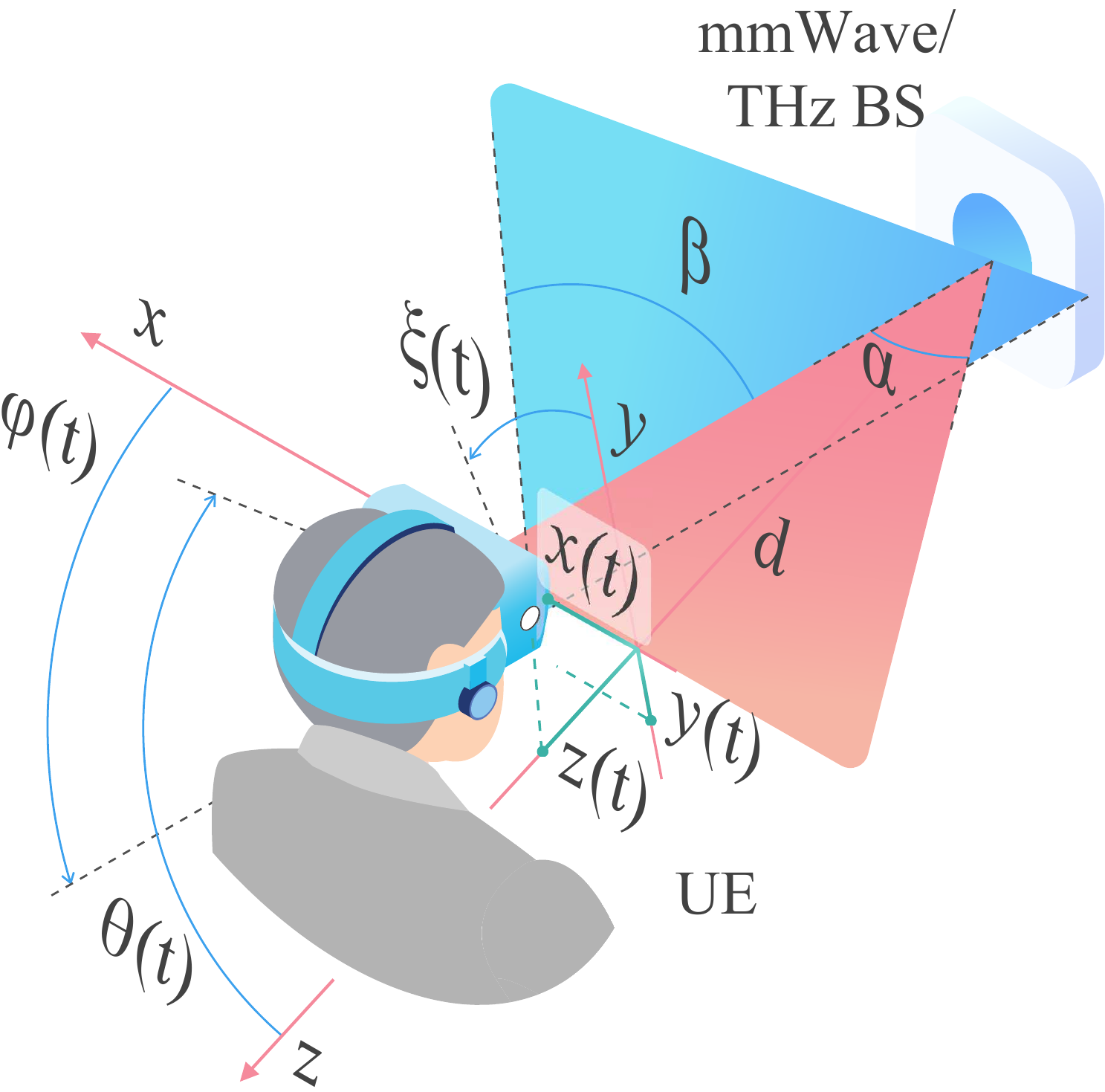}
\caption{Illustration of the UE micromobility process: motions over Cartesian $Ox$, $Oy$, and $Oz$ axes represented by $x(t)$, $y(t)$, and $z(t)$ processes; rotations over vertical (yaw), transverse (pitch), and roll (longitudinal) axes represented by $\phi(t)$, $\theta(t)$ and $\zeta(t)$, respectively, \cite{petrov2020capacity}.}
\label{fig:micromobility}
\vspace{-0mm}
\end{figure}

\subsection{UE Micromobility Models}

In addition to static and dynamic blockage affecting the performance of users, for mmWave/THz systems micromobility, manifesting itself in fast UE displacements and rotations, have to be considered \cite{hur_dynamic_mmWave,petrov2018effect,kokkoniemi2020impact}, see Fig. \ref{fig:micromobility}. This phenomenon happens during the communications process may cause frequent misalignments of the highly-directional THz beams, resulting in fluctuations of the channel capacity and outage events \cite{peng_dynamic_thz}. The preliminary studies indicate that the outage in THz systems might happen at much smaller timescales compared to blockage, i.e., on the order of few tens of milliseconds \cite{petrov2018effect,petrov2020capacity}. This property challenges the development of efficient beamtracking algorithms that are critical for mmWave and THz communications.

To date, only a few mocromobility models have been proposed. The authors in \cite{petrov2020capacity} first demonstrated that small displacements over $Oz$ axis as well as roll (longitudinal axis) motion do not affect link performance. They further observed that movements over $Oz$ and rotation over roll (longitudinal) axes does not affect beamalignment and proceeded with the decomposition technique by modeling motion over Cartesian $Ox$ and $Oy$ axes as well as rotations over vertical (yaw) and transverse (pitch) axes, $\phi(t)$ and $\theta(t)$, see Fig.~\ref{fig:micromobility}, by mutually independent Brownian motions. The authors revealed that the pdf of time to outage due to beam misalignment $f_{T_A}(t)$ follows (\ref{eqn:overallPDF}), where $\text{erfc}(\cdot)$ is the complementary error function, $\mu_{(\cdot)}$ and $\sigma_{(\cdot)}$ are the parameters of the corresponding displacement and rotation components that can be estimated from the empirical data. 

We specifically note that the model in (\ref{eqn:overallPDF}) represents only the essentials of the UE micromobility process as it does not capture potential dependence between movements, distance-dependent velocity and drift to the origin. Observing that micromobility may lead to outage conditions, for performance evaluation purposes one needs to jointly represent the blockage and micromobility processes. This can be done by, e.g., superposing blockage process on top of micromobility process or abstracting both processes by a certain stochastic process with the joint intensity of link interruptions. Additional attempts have been done recently by the authors in \cite{stepanov2021accuracy,stepanov2021statistical} who performed detailed measurements campaigns for different applications and demonstrated that two-dimensional Markov models can accurately capture micromobility when application does not directly control user's behavior, e.g., video/VR viewing.


\begin{figure}[t!]
\vspace{-0mm}
\centering
\includegraphics[width=0.8\columnwidth]{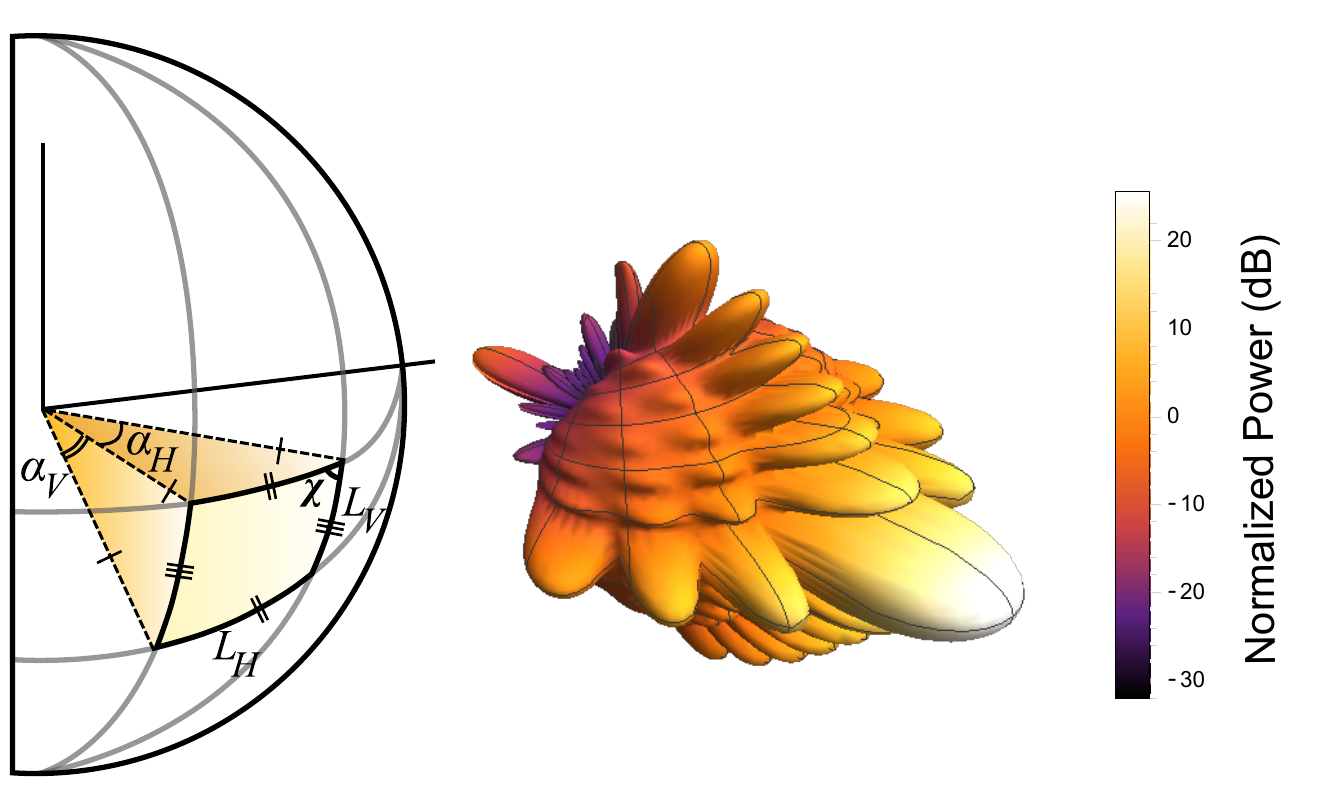}
\caption{Illustration of the 3D antenna radiation pattern models: right -- realistic $15\times{}4$ pattern obtained using the algorithm in 3GPP TR 37.977; left -- 3D HPBW approximation by following \cite{kovalchukov2018evaluating}.}
\label{fig:coneModel}
\vspace{-2mm}
\end{figure}

\subsection{Antenna Array Models}

The ability to accurately capture essential features of BS and UE antenna radiation patterns is critical for performance analysis of mmWave and THz communications systems. A very detailed model based on correlated superposition of individual antenna elements can be produced by utilizing the 3GPP procedure specified in 3GPP TR 37.977, see Fig. \ref{fig:coneModel}. However, this model is not analytically tractable. As a result, various approximations are utilized in the literature. Below, we consider typical models utilized for these purposes.

\subsubsection{2D Cone and Cone-Plus-Sphere Models}

\begin{figure}[!b]
\vspace{-2mm}
\centering
    \subfigure[2D cone antenna model]
    {
        \includegraphics[width=0.75\columnwidth]{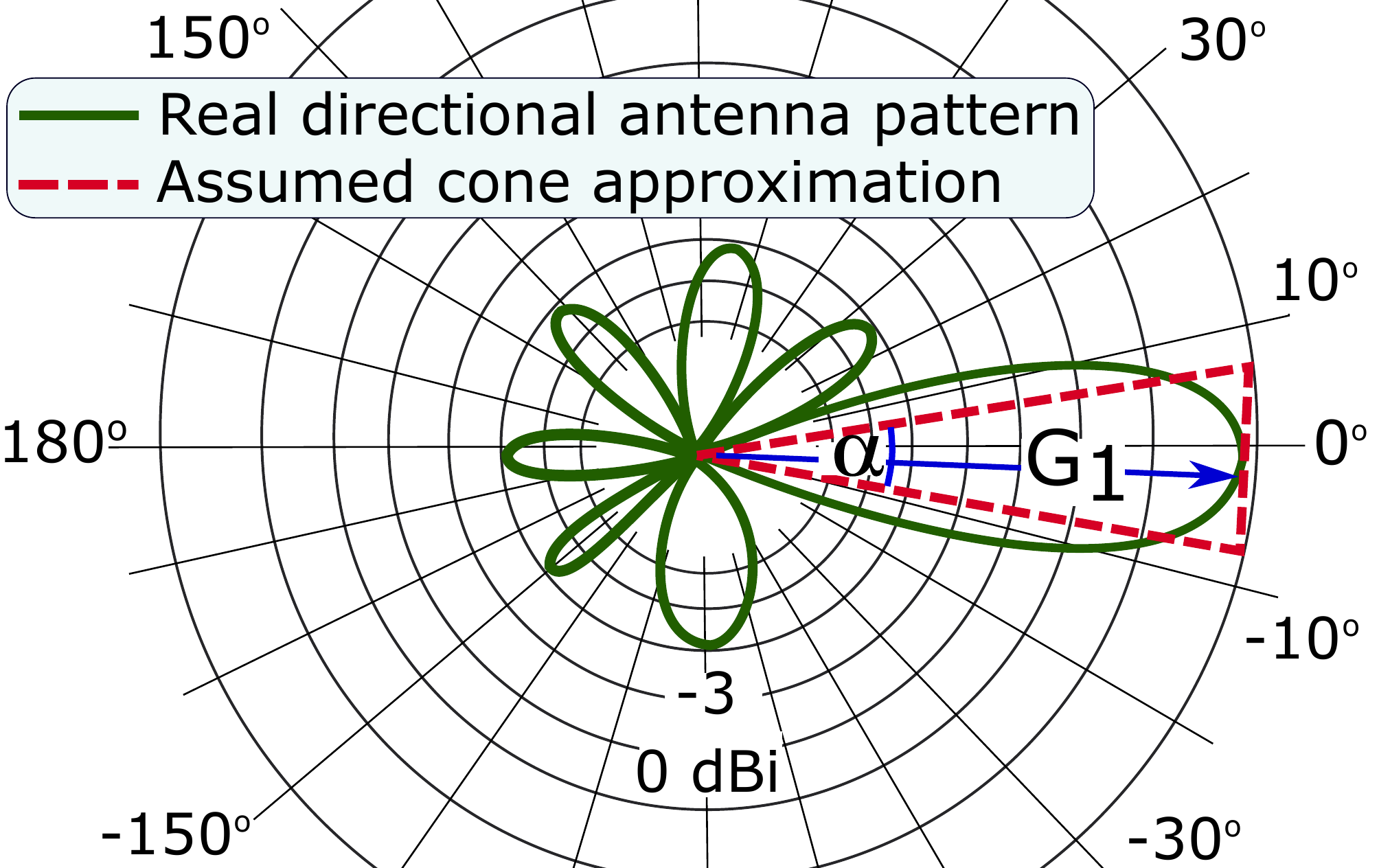}
        \label{fig:antenna_a}
    }\vspace{+3mm}\\
    \subfigure[2D cone-plus-sphere model]
    {
        \includegraphics[width=0.75\columnwidth]{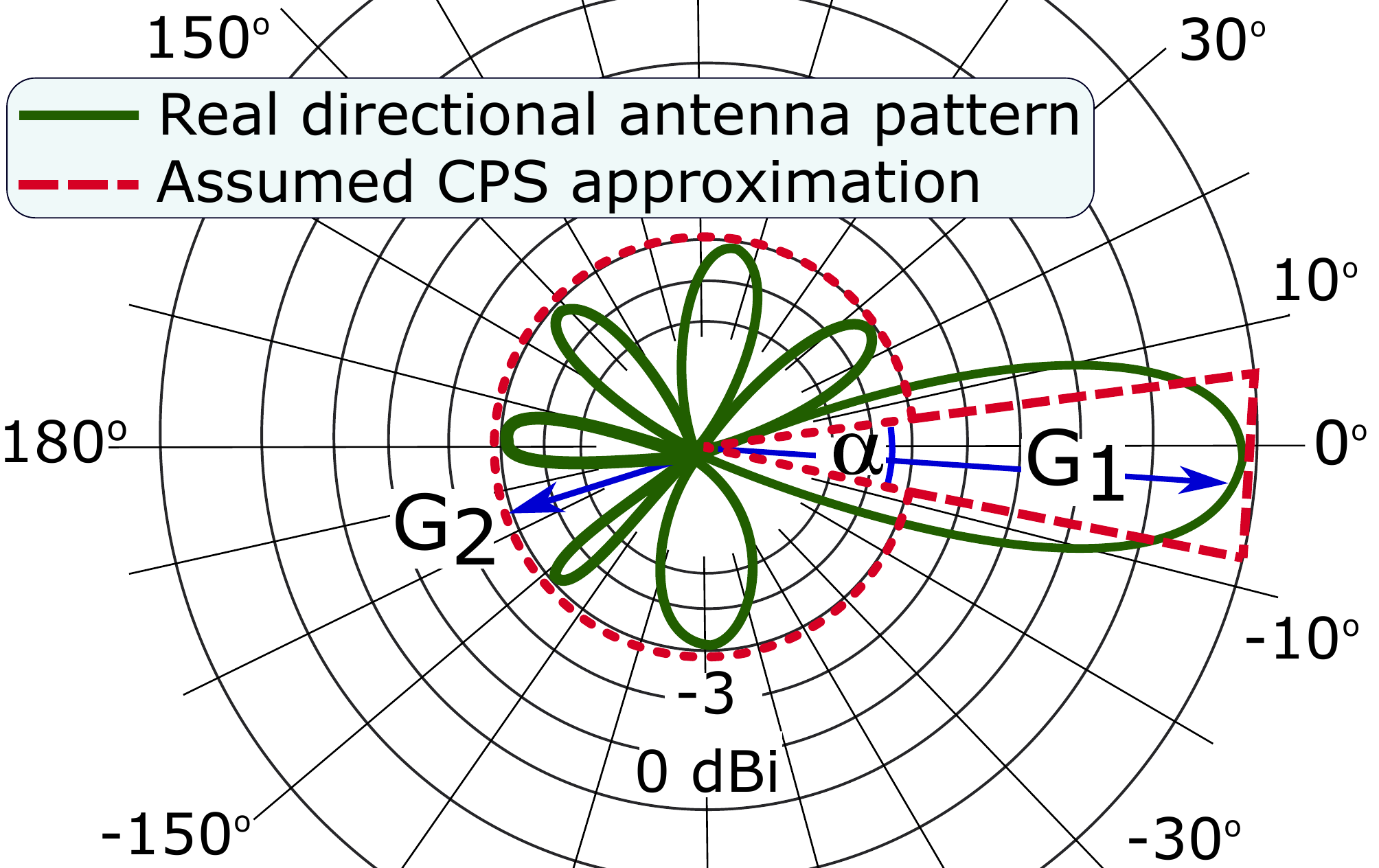}
        \label{fig:antenna_b}
    }
    \caption{Illustration of the 2D approximate antenna models: $G_1$ -- gain over the main lobe, $G_2$ -- approximate gains of side and back lobes \cite{petrov2017interference}.}
    \label{fig:antennas}
\end{figure}

The simplest models capturing the essential feature of mmWave and THz systems is the so-called 2D cone model and the extended 2D cone-plus-sphere model, see Fig.~\ref{fig:antennas}, considered in many early studies of mmWave and THz systems \cite{petrov2017interference,singh2011interference,park2013incremental,petrovGlobecom,biason2019multicast,zhang2012wireless}. According to the first model, illustrated in Fig.~\ref{fig:antenna_a},
a cone-shaped pattern parameterized by a single parameter $\alpha$ -- HPBW of the main lobe -- is utilized for representing the antenna radiation pattern. The second model, shown in Fig.~\ref{fig:antenna_b}, attempts to account for imperfections of antenna design by modeling back and side lobes as a sphere around UE. To filly parameterize the former model one needs to determine the antenna gain $G$ as a function of the directivity angle $\alpha$. The second model requires gain values corresponding to the main and side lobes, $G_1$ and $G_1$, associated with a chosen directivity angle $\alpha$ and the power loss parameter to back and side lobes $k$.


The power spectral density (PSD) $P_{R}$ at a distance $r$ is
\begin{align}
P_R=\frac{P_{A}}{S_A}=\frac{P_{A}}{2\pi r h},
\end{align}
where $\alpha$ is the antenna directivity, $P_A$ is the emitted power, and $S_A$ is the area of the wavefront surface. The latter is known to be $h=r[1-\cos(\alpha/2)]$. By applying the free-space propagation model (FSPL), the power spectral density at the wavefront surface $P_{R}$ is given by
\begin{align}
P_{R}=\frac{P_{A}}{S_A}=P_{A} \frac{G}{4\pi r^{2}},
\end{align}
leading to the following for the main lobe gain
\begin{align}
G = \frac{2}{1-\cos(\alpha/2)}.
\end{align}



Consider now the 2D cone-plus-sphere model. Let $k_1$ and $k_2$ denote the fractions of power split between the main lobe and back and side lobes, i.e., $k_1+k_2=1$. By applying the same logic as in the case of the cone model we may write
\begin{align}
\begin{cases}
P_{R,1} 2\pi r^2 [1-\cos(\alpha/2)] = k_{1}P_{A}\\
P_{R,2} 2\pi r^2 [1+\cos(\alpha/2)] = k_{2}P_{A}\\
k_{1}+k_{2} = 1
\end{cases},
\end{align}
where, according to the FSPL, we have
\begin{align}
\begin{cases}
P_{R,1} = G_{1} P_{A}/4\pi r^2\\
P_{R,2} = G_{2} P_{A}/4\pi r^2
\end{cases}.
\end{align}

Thus, there exists the following relationship 
\begin{align}
G_{1}[1-\cos(\alpha/2)] + G_{2}[1+\cos(\alpha/2)] = 2.
\end{align}

Using $k=k_{1}/k_{2}$ we see that $G_{2}=kG_{1}$ and $G_{1}$, $G_{2}$ are
\begin{align}
\begin{cases}
G_{1} = 2[(1-\cos(\alpha/2))+k(1+\cos(\alpha/2))]^{-1}\\
G_{2} = k G_{1}
\end{cases}.
\end{align}


\subsubsection{3D Models}

One of the critical issues of the considered models is their 2D nature implying that they cannot be utilized for mmWave and, especially, THz systems featuring antennas forming radiation patterns in horizontal and vertical dimensions, simultaneously. Following \cite{kovalchukov2019evaluating,kovalchukov2018analyzing} we consider an antenna pattern approximated by a pyramidal zone, see Fig.~\ref{fig:coneModel}. This model is fully defined by angles, $\alpha_{V}$ and $\alpha_{H}$.


To obtain the gain $G$ as a function of $\alpha_{V}$, and $\alpha_{H}$, we first notice that the wavefront surface area is provided by the area of the spherical rectangle, see Fig.~\ref{fig:coneModel}. Applying the law of cosines~\cite{Petrera2014}, we obtain $\cos\chi$ as
\begin{align}\nonumber
	\cos\chi&=\frac{\cos\left(\frac{\pi}{2}-\frac{L_H}{2}\right)-\cos\left(\frac{\pi}{2}-\frac{L_H}{2}\right)\cos\left(L_V\right)}{\sin\left(\frac{\pi}{2}-\frac{L_H}{2}\right)\sin\left(L_V\right)}=\\
	&=
	\frac{\sin\left(\frac{L_H}{2}\right)}{cos\left(\frac{L_H}{2}\right)}\frac{1-\cos\left(L_V\right)}{\sin\left(L_V\right)}\nonumber=\\
	&=\tan\left(\frac{L_H}{2}\right)\tan\left(\frac{L_V}{2}\right).
\end{align}

The quarter of the spherical excess is $(\rho-\pi/2)$ leading to
\begin{align}
	\cos\left(\rho-\frac{\pi}{2}\right)=\tan\left(\frac{L_H}{2}\right)\tan\left(\frac{L_V}{2}\right),
\end{align} 
where $L_H$ and $L_V$ are spherical geodesics corresponding $\alpha_{H}$ and $\alpha_{V}$. Hence the sought area is
\begin{align}
	S_A=4\arcsin\left(\tan\frac{\alpha_V}{2}\tan\frac{\alpha_H}{2}\right).
\end{align} 

Since the wavefront psd is provided as $P_{R}=A r^{-\zeta}$, the gain of the main lobe is obtained as follows
\begin{align}
	G(\alpha_V,\alpha_H) =\frac{4 \pi}{S_A}=\frac{\pi}{\arcsin\left(\tan\frac{\alpha_V}{2}\tan\frac{\alpha_H}{2}\right)}.
\end{align}    

\subsubsection{Antenna Parameterization}

The introduced antenna arrays model requires directivity angles as the input, $\alpha$ for 2D antenna models and $\alpha_V$ and $\alpha_H$ for 3D antenna models. To determine them, recall that HPBW of the antenna array is determined by the number of antenna elements in the considered plane. By utilizing the array maximum $\theta_m$ and the 3-dB point, $\theta_{3dbh}$, the following relation holds for the HPBW of the linear array \cite{constantine2005antenna}
\begin{align}
\alpha = 2|\theta_m - \theta_{3db}|,
\end{align}
where $\theta_m$ is related to the phase excitation difference, $\beta$ as
\begin{align}
\theta_m=\arccos(-\beta/\pi),
\end{align}
leading to $\theta_m=\pi/2$ for $\beta=0$.

The upper and lower $3$-dB points are given by\cite{constantine2005antenna}
\begin{align}
\theta_{3db}^{\pm}=\arccos[-2.782/(N\pi)].
\end{align}

Table~\ref{tab:array_comp} shows HPBWs and their approximation via empirical law $102/N$ for linear arrays, where $N$ specifies the number of array elements. We also note that instead of calculating the antenna gains for considered models analytically based on wavefront area one may utilize \cite{constantine2005antenna}
\begin{align}
G = \frac{1}{\theta_{3db}^+-\theta_{3db}^{-}}\int_{\theta_{3db}^-}^{\theta_{3db}^+} \frac{\sin(N\pi\cos(\theta)/2)}{\sin(\pi\cos(\theta)/2)}d\theta.
\end{align}

The antenna gains are summarized in Table~\ref{tab:gain}.

\begin{table}[t!]
\centering
\caption{Antenna HPBW and its approximation, reproduced from \cite{gerasimenko2018multicon}}
\label{tab:array_comp}
\begin{tabular}{p{2cm}p{3cm}p{2cm}}
\hline
\textbf{Array}&\textbf{Value, direct calculation} &\textbf{Approximation}\\
\hline\hline
64x1 & 1.585 &  1.594\\
\hline
32x1& 3.171 & 3.188\\
\hline
16x1& 6.345 & 6.375\\
\hline
8x1& 12.71 & 12.75\\
\hline
\end{tabular}
\vspace{-2mm}
\end{table}

\begin{table}[b!]
\vspace{-2mm}
\centering
\caption{Antenna array gains, reproduced from \cite{gerasimenko2018multicon}}
\label{tab:gain}
\begin{tabular}{p{2cm}p{3cm}p{2cm}}
\hline
\textbf{Array}&\textbf{Gain, linear} &\textbf{Gain, dB}\\
\hline\hline
64x1 & 57.51 &  17.59\\
\hline
32x1& 28.76 & 14.58\\
\hline
16x1& 14.38 & 11.57\\
\hline
8x1& 7.20 & 8.57\\
\hline
4x1& 3.61 & 5.57\\
\hline
\end{tabular}
\end{table}

\subsection{Beamsearching Algorithms}

The beam misalignment caused by micromobility and blockage affects the already strict time budget for beamtracking in mmWave and THz communications. Inherently requiring high gains at both BS and UE sides, these systems would require antenna arrays with tens (mmWave band) or even hundreds (THz band) of array elements to form "pencil-wide" beams \cite{akyildiz2016realizing}. This would lead to extremely large beamforming codebooks at both communications sides drastically increasing the beamsearching time. This time can be reduced by minimizing the arrays switching time. However, given the expected order of magnitude increase in the number of antenna array elements, the array switching time has to be decreased down to nanoseconds from the current state-of-the-art few microseconds that may not be feasible for modern arrays design options \cite{piazza2008design,kim201828}. 

\begin{table*}[!t]\footnotesize
\centering
\caption{Details of the considered deployment and radio part models}
\begin{tabular}{l|l|l|l}
\hline
Model&Type&Specifics and parameterization&Complexity\\
\hline\hline
\multirow{2}{*}{Deployment} 	& Indoor & Regular & High \\
\cline{2-4}
								 	& Outdoor, open area & PPP & Low \\
\cline{2-4}
								 	& Outdoor, street grids & Semi-regular & High \\
\hline
\multirow{2}{*}{Blockage} 		& By large buildings 	& Empirical 3GPP \cite{umi}, analytical ITU-R model \cite{ituBlock}, MLNP-based \cite{gapeyenko2021} & Low \\
\cline{2-4}
									& Human blockage	& Mobile/static UE and blockers, RDM mobility & Low-high\\
\hline
\multirow{2}{*}{Antenna array}	& Analytical & Cone, cone-plus-sphere, parameterized via Tables \ref{tab:array_comp} and \ref{tab:gain}  & Low \\
\cline{2-4}
									& Empirical & Fitted to measurements & Low \\
\hline
\multirow{2}{*}{Micromobility} 	& Analytical & Brownian motion, explicit pdf of link active time & Low \\
\cline{2-4}
									& Empirical & Fitted to measurements in \cite{stepanov2021statistical}, 1D/2D discrete Markov chains & High \\
\hline
\multirow{3}{*}{Propagation} 	& Averaged & 3GPP UMi/UMa from \cite{umi} & Low \\
\cline{2-4}
									& Cluster-based & Fitted to the model in \cite{umi}, e.g., \cite{gapeyenko2018analytical} & High \\
\cline{2-4}
									& THz specific & Averaged with exponential path loss component, $e^{-Kx}$ & Low \\
\hline
\multirow{2}{*}{Beamalignment} & Type of search & Full/iterative search, parameterized by the array switching time/number of elements \cite{gerasimenko2018multicon} & High \\
\cline{2-4}
									& Design type & Cellular periodic, WLAN-style on-demand, see \cite{petrov2020capacity} & High\\
\hline
\end{tabular}
\label{tab:radioCookbook}
\vspace{-0.0cm}
\end{table*}


The algorithmic beamtracking improvements are nowadays considered as the main option for efficient utilization of mmWave and THz resources. The approaches originally proposed for beamtracking design in mmWave systems, in addition to hierarchical iterative mechanisms putting one of the sides in omnidirectional regime similarly to IEEE 802.11ad/ay technologies \cite{wigig2012,wei2017exhaustive} and various algorithmic improvements \cite{va2016beam,lim2019beam}, may utilize: (i) external localization information provided by e.g., GPS or 5G/6G positioning services \cite{shim2014application,alrabeiah2020viwi,brambilla2020sensor} or radar information \cite{petrov2019unified}, (ii) lower-band RATs for provisioning of direction towards BS/UE \cite{gui2021network,liu2018millimeter}, (iii) information available in the past when the connection has been up \cite{kutty2015beamforming,aykin2020mamba,jeong2020online}.


To abstract specifics of the beamsearching procedure, models for performance evaluation purposes need to specify two parameters as a function of beamsearching algorithm and antenna array: (i) beamsearching time and (ii) time instance when beamsearching is initiated. Consider exhaustive search as an example. Here, to establish a direction to the BS, both UE and BS need to attempt all feasible configurations leading to the beamsearching time of $T_S=N_UN_A\delta$, where $N_U$ and $N_A$ are the number of UE and BS configurations and $\delta$ is the array switching time. For hierarchical search, UE and BS perform beamsearching by switching the other side to omnidirectional mode. Here, the beamsearching time is $T_S=(N_U +N_A)\delta$. By analogy, the beamsearching time of any algorithm can be defined as a function (possibly probabilistic) of $N_U$, $N_A$ and $\delta$. As one may deduce, the array switching time, $\delta$, is a crucial parameter for future mmWave and THz systems. In general, it heavily depends on array implementation and may vary on the timescale from microseconds to milliseconds. As an example, IEEE 802.11ad recommends utilizing arrays with $\delta=1$ ms leading to $4$ ms and $0.41$ ms of beamsearching time for exhaustive and hierarchical search, respectively, and $64$ and $4$ antenna elements at BS and UE sides.



Addressing the time, when beamsearching is initiated we distinguish between two system design options, cellular and WLANs \cite{petrov2020capacity}. In the former case, beamsearching is performed periodically with interval $T_P$ . Note that the frame duration in mmWave NR and THz systems (1 ms for mmWave NR \cite{nrmcs} and might get smaller for THz radio interface) is smaller than the time to the outage, caused by the micromobility reported in \cite{petrov2020capacity}. Thus, in this case, one may assume that only blockage leads to outage situations. According to WLAN design \cite{wigig2012}, beamsearching is performed once the connection is lost. Thus, not only blockage but micromobility may cause an outage.



\subsection{Summary and Usage of Models}

The choice of deployment and radio part models heavily depends on the desired accuracy of the resulting performance evaluation framework and also lead to different levels of complexity when combining them to abstract the radio channel specifics (see Section \ref{sect:param} for examples) in the models of session service process considered further in Section \ref{sect:perf}. To this aim, Table \ref{tab:radioCookbook} provides a concise summary of these models and also verbally comments on the complexity of the resulting combined  framework.

First of all, we would like to note that all of these models provided in Table \ref{tab:radioCookbook} are inherently analytically tractable, that is, they can be utilized to build appropriate performance evaluation frameworks for mmWave/THz systems. In terms of deployment, the simplest for analysis is the PPP distribution of both BSs and UEs while the most complex are those models presuming semi-regular deployments, i.e., those utilized for outdoor city center deployment options. To provide an abstraction of radio part specifics these models may require additional link-level or system-level simulations, see \cite{begishev2018connectivity,moltchanov2021performance}. Building indoor models is the most complicated from the radio part point of view. Here, not only deployment models need to be assumed to be semi-regular but the propagation model needs to be more detailed than the average ones, preferably, based on ray-tracing or 3D cluster-based model. This significantly complicates radio channel abstraction particularly, estimation of the coverage range of BS and associated resource request distribution of sessions.


The human body blockage model capturing both mobility of UEs and blockers simultaneously is scarce and generally capture the first moment of blockage statistics only. Forcing either blockers or UEs to stationary random locations, one may potentially obtain detailed knowledge of blockage phenomena but looses in the model's ability to capture real-life users behavior. As compared to human body blockage, the nature of micromobility is still not fully studied with just a few recently published studies revealing Markov nature of micromobility for applications that does not control user behavior rejecting this hypothesis for other applications \cite{stepanov2021statistical,stepanov2021accuracy}. For the former class, inherently Markov in nature Brownian motion or discrete two-dimensional empirically fitted models reported in \cite{stepanov2021statistical} can be utilized without compromising accuracy.

\begin{table*}[!t]\footnotesize
\centering
\caption{Traffic types, corresponding service models and metrics of interest}
\begin{tabular}{p{0.225\textwidth}|p{0.225\textwidth}|p{0.225\textwidth}|p{0.225\textwidth}}
\hline
Traffic type&Non-elastic/adaptive traffic &Elastic traffic &Adaptive traffic\\
\hline\hline
No blockage/micromobility & Baseline RQS, Section \ref{sect:RQSformal}, \ref{sec4-1} & PS models with service coefficients, Section \ref{sec:servcoefs}  & Stochastic geometry \cite{haenggi2012stochastic,elsawy2013stochastic,elsawy2016modeling} \\
\hline
Blockage/micromobility & RQS with events causing service interruptions, Section V-D & Not available & Stochastic geometry \cite{moltchanov2021ergodic,shafie2020coverage} \\
\hline
Reservation functionality & RQS with partial resource availability, Section \ref{sec4-3} & Not available & Not available, Section \ref{sect:adaptive} \\
\hline
Priorities & RQS with priorities, Section \ref{sect:priorities} & Not available & Stochastic geometry \cite{haenggi2012stochastic,elsawy2013stochastic,elsawy2016modeling} \\
\hline
Multiconnectivity & RQS network, Section \ref{sec4-4} & Not available & Stochastic geometry \cite{shafie2021coverage,moltchanov2021ergodic}\\
\hline
Reservation/mulitconnectivity & RQS network with partial availability, Section \ref{sec4-4} & Not available & Not available\\
\hline
\multicolumn{4}{c}{Typical performance metrics}\\
\hline
User-centric & New and ongoing session loss probabilities  & Session service time & UE spectral efficiency/rate, fraction of time in outage\\
\hline
Operator-centric & System resource utilization  & Fairness & Fairness\\
\hline
\end{tabular}
\label{tab:serviceCookbook}
\vspace{-0.0cm}
\end{table*}



The most well-understood sub-models in terms of usage are propagation models for mmWave and THz bands. Here, the incremental accuracy is usually expected by going from averaged models to 3D cluster-based, to ray-traced empirical models. For the first two models, a completely mathematical abstraction of the radio part can be utilized as demonstrated in, e.g., \cite{begishev2021joint,begishev2019quantifying}, while for the latter one ray-tracing simulations are required for a given scenario. Also, some of the studies indicate that system-level performance metrics are not heavily affected by the choice of the propagation model. Particularly, the authors in \cite{gapeyenko2018analytical} demonstrated that the outage probability computed analytically for averaged and 3D cluster-based models coincide implying that the latter one does not provide any substantial accuracy improvement. Furthermore, while all these models are already available for mmWave band only a few steps are made towards standardization of these classes of model for THz band.


The most frequently utilized model for directional antenna arrays is the simple cone model. However, the use of this model is known to provide biased results for highly directional antenna radiation patterns as shown in \cite{petrov2017interference} making it not suitable for system-level analysis of THz systems. Cone-plus-sphere model allows to characterize interference more precisely. Finally, we note that the radio abstraction part provided in Section \ref{sect:param} may also accept the measured antenna radiation pattern without a significant increase in modeling complexity. Finally, we note that beamalignment time and, generally, beamsearching processes are rarely accounted for in mathematical system-level performance evaluation studies. The rationale is that this process is deterministic making it difficult to account for in stochastic models. Nevertheless, it is still feasible as shown in \cite{moltchanov2021ergodic,gerasimenko2018multicon}. The difference between different types of beamsearching procedure is mainly related to the time it takes to locate the beam and the time instant when beamsearching is initialed. In the former case, the performance metrics are often more difficult to obtain \cite{petrov2020capacity}.


\section{Radio Abstraction and Parameterization}\label{sect:param}

In this section, we outline the basic techniques for parameterizing the models introduced in the previous section. Recall that the queuing part  accepts the following parameters as the input: the pmf of the amount of requested resources, $\{p_j\}_{j\geq0}$, and the temporal intensity of the UE stage changes, $\alpha$. These parameters abstract the propagation and antenna specifics, UE locations, blockage, and micromobility and represent them in the form suitable for queuing analysis. Below, we first specify the parameterization procedure in detail for the baseline model having PPP deployment of mmWave/THz BSs and UEs randomly located in the BS coverage area. Whenever possible we also sketch the extension for other deployment cases.

\begin{figure}[t!]
	\centering\hspace{-5mm}
	\includegraphics[width=0.8\columnwidth]{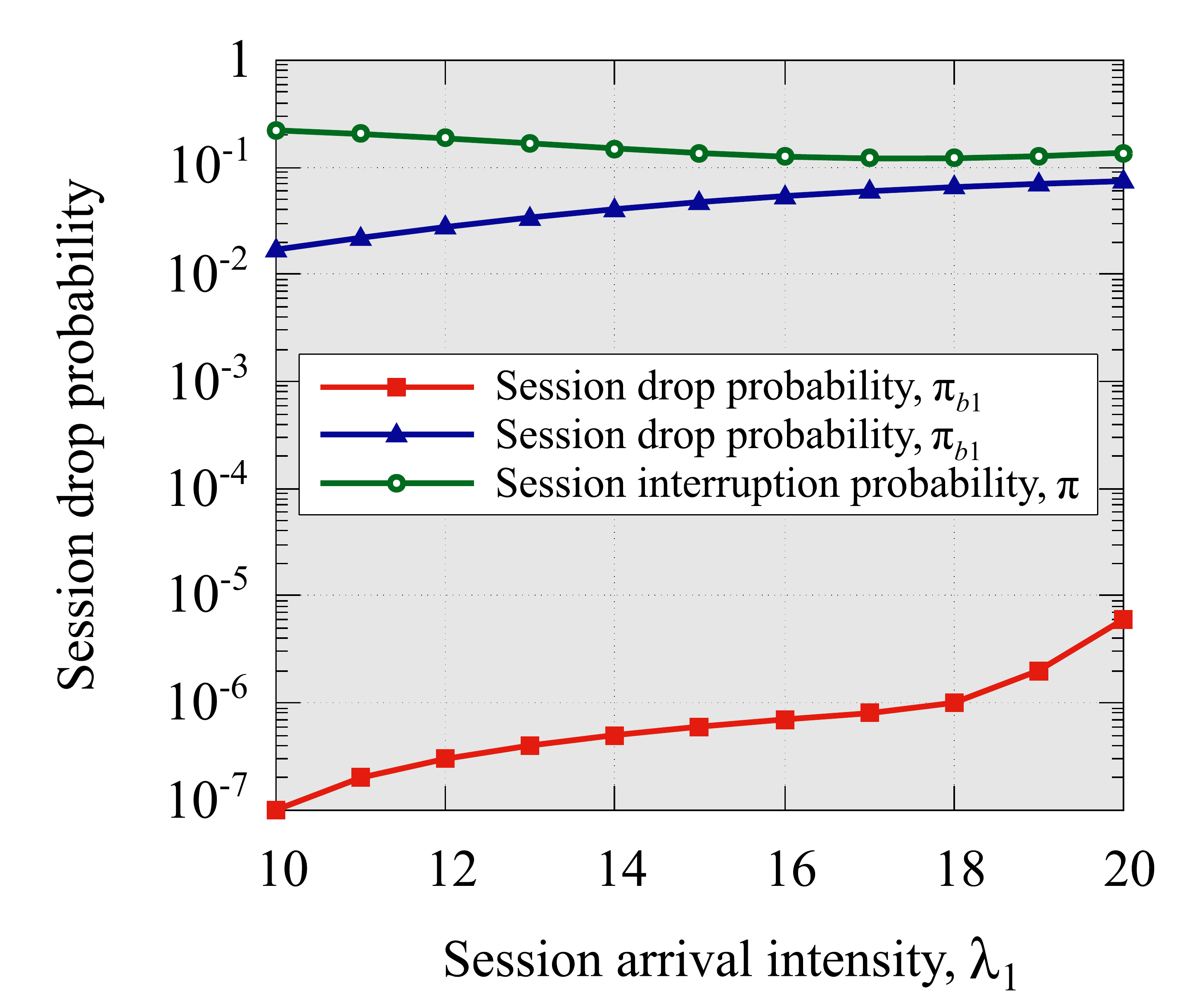}
	\caption{Performance metrics for prioritized service at BSs: enforcing pre-emptive priority at the mmWave/NR air interface efficiently isolates higher priority traffic, e.g., URLLC, from the lower priority one, e.g., eMBB.}
	\label{fig:interup}
	\vspace{-2mm}
\end{figure}

\subsection{Resource Request Characterization}


For certainty, consider the propagation model defined in Section \ref{sect:system} excluding the exponential component responsible for atmospheric absorption. To derive the pmf of the amount of requested resources, we start with SINR at the UE located at the distance of $y$ from the mmWave/THz along the propagation path is provided by
\begin{align}\label{eqn:sinr_derive}
	S(y)=\frac{C}{N_0+I}[{y}^{-\zeta_1}[1-p_{B}(y)]+{y}^{-\zeta_2}p_{B}(y)],
\end{align}
where $C=P_AG_AG_U/A$, $P_{A}$ is the BS transmit power, $G_A$ and $G_{U}$ are the antenna array gains at the BS and UE sides, respectively, $N_0$ is the thermal noise, $I$ is the interference, $A$, $\zeta_1$ and $\zeta_2$ are the propagation coefficients provided in (\ref{eqn:prop_coeff}).


There is a single unknown to determine in (\ref{eqn:sinr_derive}), the interference, $I$. Observe that this is a RV that depends on the plethora of deployment factors including BS deployment model, utilized antenna arrays, emitted power, the density of blockers, etc. For a given deployment scenario, interference can be estimated by utilizing conventional stochastic geometry approaches, e.g., \cite{kovalchukov2019evaluating,kovalchukov2018analyzing,chen2021coverage,petrov2017interference,shafie2021coverage,shafie2021coverage}. Finally, if one wants to account for more detailed propagation specifics, e.g. fast and shadow fading, these can also be added to the numerator of (\ref{eqn:sinr_derive}). For example, shadow fading distribution is known to follow Normal distribution with zero mean and standard deviation that can be found in \cite{umi}. 


Analyzing the structure of (\ref{eqn:sinr_derive}) one may observe that coefficients $P_A$, $G_A$, $G_U$, $N_0$ are all constants. The blockage probability $p_B(y)$ can be derived utilizing the  blockage models specified in (\ref{eqn:blockage}). On the other hand, the distance between UE and BS, $y$, interference, $I$ as well as additional components such as shadow fading and fast fading are all RVs. The presence of multiple RVs drastically complicates the derivation of pmf of resources requested by a session as shown in \cite{samuylov2015random,samuylov2017analytical,kovalchukov2019evaluating}. Thus, to simplify derivations, interference is often captured by an interference margin corresponding to the mean value in a given deployment of interest. In what follows, we illustrate how to derive the mean value of interference when the fast fading component is neglected while the shadow fading is captured by the shadow fading margin.

\subsubsection{Interference Characterization}


For a considered deployment, $I$, the interference can be written as
\begin{align}\label{eqn:interference}
	I=\sum_{i=1}^{N}C(y_i^{-\zeta_1}[1-p_{B}(y_i)]+y_i^{-\zeta_2}p_{B}(y_i)),
\end{align}
where $Y_i$ are the distances to interfering BS.


Observe, that interference in (\ref{eqn:interference}) is a random function of RVs. Unfortunately, there are no effective methods to derive its distribution. For dense deployments, however, one may approximate it utilizing the Normal distribution. Alternatively, when approximating interference in (\ref{eqn:sinr_derive}), one may utilize the mean values of interference as an interference margin. The latter can be obtained by applying the Campbell theorem expressing the moments of aggregated interference at a randomly selected (tagged) UE as follows \cite{chiu2013stochastic}
\begin{align}\label{eqn:campbell}
	E[I^{n}]=&\int_{0}^{{r}_I}C[x^{-\zeta_1}(1 - p_B(x)) + x^{-\zeta_2}p_B (x)]^n\times\nonumber\\
	\times &p_{C}(x)2\xi\pi{}xdx,
\end{align}
where $2\xi\pi{}xdx$ is the probability that there is BS in the radius increment of $dx$, $p_{C}(x)$ is the so-called exposure probability, that is, the probability that the antennas of interfering BSs are oriented towards the considered UE, and $p_{B}(x)$ is the blockage probability at the distance $x$, $r_I$ is the maximum radius, where BSs contribute non-negligible interference at the tagged UE.



The only unknown component in (\ref{eqn:campbell}) is the exposure probability, $p_C(x)$. For 2D deployments or when the vertical directivity is much smaller compared to the horizontal one, one may determine it following \cite{petrov2017interference} as
\begin{align}
	p_C(x)=p_C=\alpha_A\alpha_U/4\pi^2,
\end{align}
where $\alpha_A$ and $\alpha_U$ are HPBW angles of antennas at BS and UE sides. These parameters can be estimated as a function of the number of antenna elements and the type of utilized antenna model as discussed in Section \ref{sect:system}. For 3D radiation patterns the exposure probability $p_C$ has a more complex structure but can still be obtained in closed-form as shown in \cite{kovalchukov2019evaluating}. 

\subsubsection{SINR Distribution}


Recall that the distance from BS to UE uniformly distributed over the circularly shaped area is given by pdf in the form of $f_{Y}(y)=2y/r_E^2$, $0<y<r_E$, where $r_E$ is the BS coverage radius. Observe, that depending on the density of BS in the environment the effective coverage radius, $r_E$, can be limited by SINR (sparse deployments) or by cell boundaries (dense deployments). This, in turn, affects the distribution of UE locations in the cell and, the received signal strength, and SINR. Thus, the coverage radius, $r_E$, is determined by two distances -- inter-BS distance, $r_{E,V}$, and the maximum coverage of BSs, $r_{E,S}$, i.e., $r_{E}=\min(r_{E,V},r_{E,S})$. The latter can be obtained by determining the maximum distance between the UE and the BS, such that the UE in the LoS blocked conditions is not in the outage state. By utilizing the propagation model defined in Section \ref{sect:system} the sought 2D distance $r_{E,S}$ can be written as
\begin{align}\label{eqn:dist_00}
	S= C \left(r_{E,S}^2+(h_A-h_U)^2\right)^{-\frac{\zeta_2}{2}}=S_{th},
\end{align}
where $S_{th}$ is the SINR threshold associated with the worst possible MCS defined for a given technology, e.g., see \cite{nrmcs} for mmWave NR, $h_A$ is the BS height, $h_U$ is the UE height.

Solving for $r_{E,S}$, we obtain
\begin{align}\label{eqn:11}
	r_{E,S} = \sqrt{\left(C/S_{th}\right)^{\frac{2}{\zeta_2}}-(h_A-h_U)^2}.
\end{align}


The second component -- inter-BS distance, $r_{E,V}$ heavily depends on the type of deployment. For hexagonal cellular deployments with tri-sector antenna BS it is specified by as $D=3 R$~\cite{3gpp2018rel15}, where $R$ is the radius of a cell. For semi-regular deployments considered in Section \ref{sect:system} it is defined by the scenario geometry. Finally, for random deployments, one may obtain the sought radius by determining the half distance between typical BS locations. For the PPP field of BSs, the BS coverage area is known to form Voronoi cells. Thus, one may approximate it by the circle with the corresponding mean area. Since there is no analytical expression for the area of a Voronoi cell, one needs to resort to computer simulations or utilize approximations available in, e.g., \cite{kumar1992properties,tanemura2003statistical}.


Once the pdf of the distance to the tagged UE, $f_{Y}(y)=2y/r_E^2$, $0<y<r_E$, as well as interference margin (\ref{eqn:campbell}) are both determined, one can proceed with characterizing SINR distribution. The impact of the shadow fading can also be accounted by utilizing the constant shadow fading margins, $M_{S,i}$ for the LoS non-blocked and blocked states, i.e.,
\begin{align}
	M_{S,i}=\sqrt{2}\sigma_{S,i},
\end{align}
where $\sigma_{S,i}$ is the standard deviations of the shadow fading distribution in LoS non-blocked and blocked states provided in \cite{umi}. Then, the SINR CDF can be obtained using RV transformation technique as a function of a single RV -- distance from UE to BS as described in \cite{ross}. 


To illustrate the derivations, let $S_{nB}$ be a RV denoting the SINR in non-blocked state and $F_{S_{nB}}(x)$, $x>0$, be its CDF. Since UEs are assumed to be uniformly distributed in the circularly shaped BS coverage, the CDF of the distance between UE and BS can be obtained as
\begin{align}
	F_Y(y)=(y^2-(h_A-h_U)^2)/r_{E}^2,
\end{align}
where $y\in(|h_A-h_U|,\sqrt{r_{E}^2+(h_A-h_U)^2})$.


Observe that SINR is a monotonously decreasing function of distance $y$. Thus, the CDF of SINR can be expressed as
\begin{align}
	F_{S_{nB}}(y)=1 - F_Y(C/(N_0+I)y^{\zeta_1/2}).
\end{align}

\begin{figure*}[!!t]
	\vspace{-0mm}
	\begin{align}\label{eqn:pmfShadow}
		F_{S^{dB}_{nB}}(y)&=\frac{1}{2 r_E^2}\Bigg[A^{2/\zeta_1 } 10^{-\frac{y}{5 \zeta_1 }} e^{\frac{\sigma_{S,nB} ^2 \log ^2(10)}{50 \zeta_1^2}} \Bigg[\text{erf}\left(\frac{50 \zeta_1  \log (A)-25 \zeta_1 ^2 \log \left(r_E^2+(h_A-h_U)^2\right)+\sigma_{S,nB}^2 \log ^2(10)-5 \zeta_1  y \log (10)}{5 \sqrt{2} \zeta_1  \sigma_{S,nB}  \log (10)}\right)-\nonumber\\
		&-\text{erf}\left(\frac{50 \zeta_1  (\log (A)-\zeta_1  \log (h_A-h_U))+\sigma_{S,nB}^2 \log ^2(10)-5 \zeta_1  y \log (10)}{5 \sqrt{2} \zeta_1  \sigma_{S,nB}  \log (10)}\right)\Bigg]+\left(r_E^2+(h_A-h_U)^2\right)\times{}\nonumber\\ 
		&\times{}\text{erf}\left(\frac{-10 \log (A)+5 \zeta_1  \log \left(r_E^2+(h_A-h_U)^2\right)+y \log (10)}{\sqrt{2} \sigma_{S,nB}  \log (10)}\right)-(h_A-h_U)^2\times{}\nonumber\\
		&\times{}\text{erf}\left(\frac{\sqrt{2} (-10 \log (A)+10 \zeta_1  \log (h_A-h_U)+y \log (10))}{\sigma_{S,nB}  \log (100)}\right)+r_E^2\Bigg].
	\end{align}
	\hrulefill
	\normalsize
	\vspace{-2mm}
\end{figure*}


One may utilize the described technique to obtain SINR distribution for the blocked state. Further, if one wants to include the effect of the shadow fading distribution the following step needs to be performed. Specifically, let $F_{S_{nB}}^{dB}(y)=F_{S_{nB}}(10^{y/10})$ be CDF of SINR in dB scale. Recall that shadow fading is known to follow Log-Normal distribution in linear scale resulting in Normal distribution in dB scale. By utilizing these observations, the SINR distribution in non-blocked state can be written as
\begin{align}
	S_{nB,S}^{dB}=S_{nB}^{dB}+N(0,\sigma_{S,nB}),
\end{align}
where $N(0,\sigma_{S,nB})$ is the Normal distribution with zero mean and standard deviation $\sigma_{S,nB}$ specifying shadow fading.

Finally, the SINR CDF accounting for both path loss and shadow fading  can be obtained as a sum of two RVs, that is, by convolving $F_{S_{nB}}^{dB}(y)$ and $N(0,\sigma_S)$, immediately leading to
\begin{align}\label{eqn:shadowFading}
	F_{S_{nB,S}}^{dB}(y)=\int_{-\infty}^{\infty}F_{S_{nB}}^{dB}(y+u)\frac{e^{-u^2/2\sigma_S^2}}{\sqrt{2\pi}\sigma_S}du.
\end{align}

Unfortunately, no closed-form expression for \ref{eqn:shadowFading} can be obtained by utilizing the conventional RV transformation technique \cite{ross}. However, one may still obtain the final results in terms of an error function, $\text{erf}(\cdot)$, as provided in (\ref{eqn:pmfShadow}), where
\begin{align}
	A=\frac{P_A10^{G_AG_U/10}}{f_c^2 10^{3.24+L_B/10} 10^{\frac{1}{10} (M_{S,1})} N_0},
\end{align}
$M_{S,1}$ is the interference margin, $L_B$ is the additional losses induced by the blockage. By weighting SINR CDFs corresponding to LoS blocked and non-blocked states with probabilities $p_B$ and $(1-p_B)$ one obtains the final result. Finally, we note that $p_B$ -- the averaged blockage probability, can be estimated as follows
\begin{align}
	\pi_B=\int_{0}^{r_E}p_B(x)\frac{2x}{r_E^2}dx,
\end{align}
where $p_B(x)$ is available from (\ref{eqn:blockage}). The authors in \cite{kovalchukov2019accurate} demonstrated that the resulting SINR CDF can be well approximated by Normal distribution.

\subsubsection{Resource Request Distributions}


To determine resource request distribution we now introduce SINR thresholds corresponding to MCS schemes, $S_j$, $j=1,2,\dots,J$. Further, we define the probability that UE assigned MCS $j$ by $\epsilon_j$. By discretizing SINR CDF $F_{S}(s)$, we obtain
\begin{align}\hspace{-1mm}
	\begin{cases}
		\epsilon_0=F_{S}(S_1),\\
		\epsilon_j=F_{S}(S_{j+1})-F_{S}(S_j),\, j=1,2,\dots,J-1,  \\
		\epsilon_J=1-F_{S}(S_J).\\
	\end{cases}\hspace{-2mm}
\end{align}

The probability $\epsilon_j$ that a session requests $r_j$ RBs can now be used to determine the resource requirements of sessions characterized by a certain requested rate. Note that the latter can be fixed or random depending on the considered scenario.

\subsubsection{Usage of 3D Cluster-Based Propagation Model}

One may characterize the resource request distribution more precisely by utilizing 3D cluster-based propagation model as discussed in \cite{begishev2021multi}. Here, the most complex part is the derivation of the pdf of the received power at a distance $x$ from BS. Assuming that UE always utilizes the strongest cluster the sought pdf is provided by
\begin{align}\label{eqn:strongestCluster}
	f_{P_R}(z;x)=\hspace{-1mm}\sum_{i=1}^{W}\left[(1-p_{B,i}(x))\prod_{j=1}^{i-1}p_{B,j}(x)\right]\hspace{-1mm}f_{P_{n}}(z;x),
\end{align}
$p_{B,i}(x)$ is the cluster blockage probability at distance $x$, $f_{P_{n}}(z;x)$ is the pdf of power of cluster $n$ at the distance $x$. The latter parameters is obtained by utilizing RV transformation technique as follows \cite{ross}
\begin{align}\label{eqn:clusterpower}
	f_{P_{n}}(z;x) = \frac{f_{P_{s,n}}\Big(\frac{P_{n}(x)}{10^{[P_{A} - L_{dB}(x)]/10}}\Big)}{10^{[P_{A}-L_{dB}(x)]/10}},
\end{align}
where $L_{dB}(x)$ is the pass loss in dB, $f_{P_{S,n}}$ is provided in (\ref{eqn:theta_Ai}).

Integrating (\ref{eqn:strongestCluster}) over the cell radius one obtains
\begin{align}
	f_{P_R}(z)=\int_{0}^{r_{E}}f_{P_R}(z;x)\frac{2x}{r_{E}^2}dx.
\end{align}

The other unknowns in (\ref{eqn:strongestCluster}) are the cluster blockage probabilities at distance $x$, $p_{B,i}(x)$. Here, the LoS cluster blockage is obtained similarly to (\ref{eqn:blockage}). However, to estimate probabilities $p_{B,i}$, $i=2,3,\dots$ one needs ZoA pdf, $f_{\theta_i}(y;x)$, provided in (\ref{eqn:theta_Ai}). Particularly, following \cite{gapeyenko2018analytical}, the cluster $i$, $i=2,3,\dots,N$, blockage probability conditioned on  ZoA $y_i$ is given by
\begin{align}\label{eqn:p_Aiy}
	p_{B,i}(y;x)= 1 - e^{-2\lambda_B r_B(\tan y_i (h_B - h_U) + r_B)}.
\end{align}
leading to the cluster blockage probability in the form of
\begin{align}
	p_{B,i}(x)=\int_{-\pi}^{\pi}f_{\theta_i}(y;x)p_{B,i}(y;x)dy,
\end{align}
which can be estimated by numerical integration.

By using  $p_{B,i}$, $i=1,2,\dots,N$ one may also obtain outage probability required in those deployments, where blockage leads to outage conditions as shown in \cite{gapeyenko2018analytical}
\begin{align}\label{eqn:outage1}
	p_{O}(x)=\prod_{i=1}^{N}p_{B,i}(x)+\int_{0}^{S_T}f_{P_R}(z;x)dz,
\end{align}
where $S_T$ is the UE sensitivity threshold.


For complex scenario geometries and/or service rules involving a field of BS and multiconnectivity one may further extend the baseline parameterization models. Particularly, the case of multiconnectivity in square mmWave BS deployments with multiconnectivity operation is considered in \cite{begishev2021joint}.

\subsection{UE State Changes}

There are two critical impairments that may affect session continuity of applications served by mmWave or THz systems -- micromobility and blockage. In the presented framework both can be abstracted using the external Poisson process of ``signals'' associated with sessions currently served at BSs. Below, we first characterize UE state changes caused by micromobility process and then proceed describing state changes induced by dynamic blockage phenomenon. We also specifically note that the queuing framework assumes only one type of impairment to be taken into account. However, it can be extended to the case of more than a single external process associated with sessions. Finally, notice that the Poisson nature of state changes also limits the modeling capabilities to mean values of the state holding times. Extending the framework to more complex external signal patterns is more complicated.

\subsubsection{State Changes due to Micromobility}

Consider first the process of UE state changes caused by micromobility of UE. Here, UE transitions between ``connectivity'' and outage states. The duration of both states heavily depends on the considered   beamsearching strategy, i.e., on the time instant, when beamsearching is initiated. Let $T_C$ and $T_O$ be RVs denoting connectivity and outage time, respectively. In the simplest ``on-demand'' beamsearching scheme, where beamsearching is invoked when the connection is lost the pdf of the connectivity time $f_{T_C}(t)$ coincides with the time to outage $f_{T_A}(t)$ provided in \cite{petrov2020capacity} and given in (\ref{eqn:overallPDF}) while the outage time, $T_O$ coincides with the duration of the beamsearching procedure, $T_B$, discussed in Section \ref{sect:system}. Note that the latter is affected by the utilized beamsearching algorithm and, generally, depends on the number of antenna elements and array switching time.

Another type of the beamalignment strategy is periodic alignment. Here, the beam realignment procedure runs regularly with a period $T_{U}$. Note that this scheme reflects the cellular-style system design with centralized control, while the former ``on-demand'' one is characteristic for WLANs. Here, the connectivity time can be determined as the minimum of $T_A$ provided in (\ref{eqn:overallPDF}) and constant $T_U$, i.e., 
\begin{align}
	f_{T_C}(t)=
	\begin{cases}
		f_{T_A}(t),&T_A<T_U\\
		f_{T_A}(t)\big/\int_{0}^{T_U}f_{T_A}(x)dx,&T_A\geq{}T_U\\
	\end{cases},
\end{align}
while the outage time again coincides with $T_B$.

\begin{figure}[t!]
	\vspace{-0mm}
	\centering
	\includegraphics[width=0.6\columnwidth]{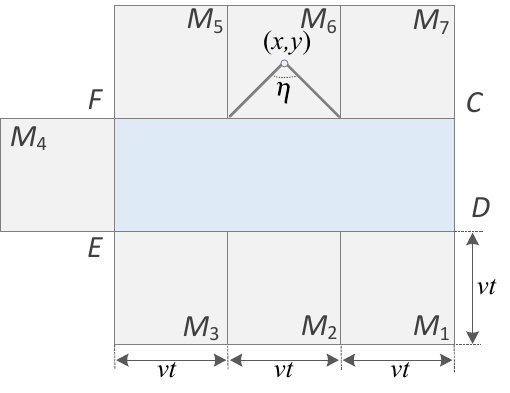}
	\caption{Translation of spatial blocker intensity into temporal domain: shaded areas around the LoS blockage zone reprsent potential locations of blockers that may enter the LoS blockage zone in a unit time \cite{gapeyenko2018flexible}.}
	\label{fig:temporal_intensity}
	\vspace{-2mm}
\end{figure} 

\subsubsection{State Changes due to Blockage}


Consider now temporal dynamics of the blockage process. For illustrative purposes, we consider only one dynamic blockage model introduced in Section \ref{sect:system}, where UE is assumed to be static in a field of human blockers with density $\lambda_B$ moving according to RDM \cite{nain2005properties}. Particularly, we are interested in the mean duration of blocked and non-blocked intervals. Similar derivations can be performed for other dynamic models.

We start by determining the intensity of blockers entering the LoS blockage zone of UE located at the distance $x$ from BS, $\epsilon(x)$. To this aim, we draw an area around the LoS blockage zone and divide it into $i$, $i= 1,2,...7$ sub-zones as shown in Fig. \ref{fig:temporal_intensity}. Now, the generic expression binding the density of blockers in the environment with the temporal intensity of blockers entering the LoS blockage zone is 
\begin{align}\label{eqn:temporalInt_1}
	\hspace{-1mm}\epsilon(x)=\sum_{i=1}^{7}\iint\limits_{M_i}g_{i}(x, y) Pr\{E\}Pr\{T>1\}\lambda_BM_idxdy,\hspace{-1mm}
\end{align}
where $M_i$ is the area of zone $i$, $g_{i}(x,y)$ is the pdf of blockers locations in sub-zone $i$ provided by $g_{i}(x, y)=1/M_i$, $E$ is the event that blockers move towards the LoS blockage zone, and $Pr\{T>1\}=\exp(-1/\tau)$ is the probability that a blocker moves longer than a unit time without changing the movement direction. The unknown event, $E$, can be obtained by considering that there is a range of angles leading to the blocker hitting the LoS blockage zone $\eta_{i}(x,y)$. That is, $Pr\{E\}=\eta_{i}(x,y)/2\pi$. With these observations in hand, we may rewrite (\ref{eqn:temporalInt_1}) as follows
\begin{align}\label{eqn:temporalInt_2}
	\epsilon(x)&=\frac{\lambda_B e^{-1/\tau}}{2\pi}\sum_{i=1}^{7}\iint\limits_{M_i} \eta_{i}(x,y) \,dx\,dy,
\end{align}
where ranges of movements, $\eta_{i}(x,y)$, can be estimated as
\begin{align}
	&\eta_{i}(x,y) = ([x_D - x]/vt),\,i=1,3,5,7,\nonumber\\
	&\eta_{i}(x,y) = 2\cos^{-1}([x_E - x]/vt),\,i=2,6,\nonumber\\
	&\eta_{4}(x,y) = 2\tan^{-1}([x-x_E]/[y-y_E]),
\end{align}
where $x_{(\cdot)},y_{(\cdot)}$ are the coordinates in Fig.~\ref{fig:temporal_intensity}. Finally, the mean intensity of blockers crossing LoS blockage zone is
\begin{align}
	\epsilon=\int_{0}^{r_E}\epsilon(x)2x/r_E^2dx,
\end{align}
where $r_E$ is the BS service radius.


\begin{figure*}[!!t]
	\vspace{-1mm}
	\begin{align}\label{eqn:busyPeriod}
		\hspace{-3mm}F_{\eta}(x)=1\hspace{-1mm}-\hspace{-1mm}\Bigg[[1-F_{T_B}(x)]\left[1-\hspace{-2mm}\int\limits_{0}^{x}(1-F_{\eta}(x-z))\exp(-\lambda_{B,T} F_{T_B}(z))\lambda_{B,T}{}dz\right]
		+\hspace{-2mm} \int\limits_{0}^{x}(1-F_{\eta}(x-z))|de^{-\lambda_{B,T} F_{T_B}(z)}|\Bigg].
	\end{align}
	\hrulefill
	\normalsize
	\vspace{-3mm}
\end{figure*}


Alternatively to the described method, one may utilize the result of \cite{groenevelt} revealing that the inter-meeting time between a point moving according to RDM in a certain convex region $A$ and a static convex region $A_1\subset{}A$ is approximately exponentially distributed. The parameter of exponential distribution depends on the areas of $A$, $A_1$ and the speed of a moving point. This implies that the process of meetings in Poisson in nature. By utilizing the superposition property of the Poisson process \cite{kingman1993poisson} one may determine the temporal intensity of blockers crossing the LoS blockage zone.



To proceed further, we recall that the blockage process at UE is known to have an alternative renewal structure \cite{gapeyenko2017temporal}. In the same study the authors also demonstrated that under Poisson assumption of blockers entering the LoS blockage area the non-blocked interval follows exponential distribution parameter coinciding with the intensity of the Poisson process. In its turn, the blocked period may be formed by multiple blockers passing through the LoS blockage zone. By utilizing the analogy with the $M/GI/\infty$ queuing system, the duration of the blocked interval is shown to coincide with the busy period distribution. This distribution can be calculated numerically by using the results of \cite{daley1998idle} provided in (\ref{eqn:busyPeriod}). The only unknown required is the distribution of the time it takes for a blocker to cross the LoS blockage zone. This can be found by assuming a random entrance point to the LoS blockage zone as shown in \cite{gapeyenko2017temporal}. To simplify derivations, one may observe that the length of the LoS blockage zone is often significantly larger than its width. Thus, one may assume that blockers cross the LoS blockage zone following a direction perpendicular to its short side. In this case, the service time in the equivalent queuing model is deterministic and equals $2r_B/v$, where $r_B$ and $v$ are the radius and speed of the blocker.


The intensity of state changes caused by blockage, $\alpha$, can be estimated by utilizing the mean blocked and non-blocked periods.  Note that to avoid the complexity of estimating CDF in (\ref{eqn:busyPeriod}) one may resort to M/M/$\infty$ approximation for which the mean busy period is available in the closed form \cite{cohen2012single}.


\section{Service Models}\label{sect:perf}

In this section, we introduce the performance evaluation models suitable for performance assessment of traffic service performance at mmWave/THz BSs. We start by describing the traffic types and corresponding service models and metrics of interest. Then, we introduce the overall structure of the framework defining the type of models, requirements, and interfaces between them. We treat in detail the models suitable for baseline, multiconnectivity, resource reservation functionalities as well as priority-based service.
\subsection{Type of Service Models and Performance Metrics}\label{sect:metrics}

\subsubsection{Traffic Types and Corresponding Service Models}


The choice of the service model to be utilized heavily depends on the type of traffic. Recall, that performance of previous generations of cellular systems, e.g., LTE, have been evaluated silently assuming inherently adaptive traffic patterns adaptive to the network state (via application layer rate adaptation mechanisms) and thus mainly utilized the elements of stochastic geometry \cite{haenggi2012stochastic,haenggi2009stochastic}. The rationale is that the service time of these sessions is not affected by the current rate provided by the network. As a consequence, the additional assumption is that all the system resources are utilized and thus the interference is maximized for a given frequency reuse ratio. In such systems performance metrics of interest are related to principal connectivity measures of UEs, i.e., outage and coverage probabilities, spectral efficiency, Shannon channel capacity, and fairness of resource allocations, etc., as a function of UE and BS densities. Furthermore, these models are not suitable to evaluate performance in underloaded conditions at all.


Evaluating the performance of non-adaptive applications requires much more comprehensive models than those utilized for analysis of adaptive applications utilizing the tools of stochastic geometry. This applied to elastic applications whose service time is dictated by the rate they receive at the air interface (e.g. large file transfers) and applications with high, guaranteed, and constant bitrates over mmWave and THz air interfaces. The latter type of applications such as VR/AR, 8/16K, remote video gaming, telepresence \cite{hu2021vision,vannithamby2017towards,agiwal2016next} are expected to be inherent for rate-rich mmWave and THz air interfaces. These applications are inherently prone to outage events caused by blockage and micromobility and its performance analysis requires not only taking into account radio part specifics but the traffic service dynamics at BSs by joining the tools of stochastic geometry and queuing theory.


Despite in this study we will mainly concentrate on performance evaluation models for non-elastic/adaptive applications, we also briefly cover those utilized for adaptive and elastic ones. Table \ref{tab:serviceCookbook} provides the so-called service models cookbook targeted to different types of applications, where PS stands for processor sharing service disciple. For non-elastic/adaptive applications, it also provides references to the forthcoming sections, where appropriate models will be considered.


\subsubsection{Metrics of Interest}

In our work we concentrate on system-level performance metrics of interest. Contrarily to physical and link-level studies concentrating on optimizing a certain functionality of interest, e.g., \cite{zeng2016millimeter,chen2021hybrid}, these type of models characterize application layer performance as a function of lower layer mechanisms, channel specifics, and traffic type. The latter defines not only service models but performance measures of interest that are also provided in Table \ref{tab:serviceCookbook}.

System-level performance evaluation frameworks for mmWave and THz systems need to address both user- and system-centric KPIs. For user-centric KPI we consider: (i) the new session drop probability and (ii) the ongoing session drop probability. The former is defined as the probability that a new session arriving session to BS is lost due to the lack of resources needed to serve it. The ongoing session drop probability is interpreted as the probability that a session already accepted for service is lost during the ongoing service. These metrics describe the so-called session continuity of applications characterizing how reliable the provided service is and can be used to benchmark the advanced service mechanisms at BSs \cite{kovalchukov2018improved,begishev2021joint,begishev2019quantifying,begishev2021multi}.  Finally, the system-centric KPIs in these systems are mainly related to the efficiency of resource utilization at BSs.

\subsection{Methodology at the Glance}


We will consider a performance evaluation framework capable of quantifying the user- and system-centric KPI defined in Section \ref{sect:system}. The overall framework is divided into two complementary parts: (i) a queuing part specified in this section and (ii) a radio abstraction (parametrization) part introduced in Section \ref{sect:param}. The latter captures the specifics of propagation properties of mmWave/THz bands as well as additional phenomena such as micromobility and blockage. The queuing part characterizes the resource allocation dynamics at the BS and accepts the pmf of the amount of requested resources, $\{p_j\}_{j\geq0}$, and the temporal intensity of the UE stage changes between outage and non-outage conditions, $\alpha$, as the input. The former parameter is responsible for abstracting the random UE locations with respect to BSs in the deployment of interest as well as antenna and propagation specifics. The latter parameter is responsible for capturing outage prone nature of considered RATs caused by blockage and micromobility dynamics. Thus, these two parameters characterize the type of deployment providing the interface between two parts of the framework. Supplementing these two with the conventional session arrival and service characteristics that are not directly related to the radio part, the intensity of session arrivals, session services times, and required rate of sessions, one fully specifies the framework.

\begin{figure}[t!]
	\centering
	\includegraphics[width=1.0\columnwidth]{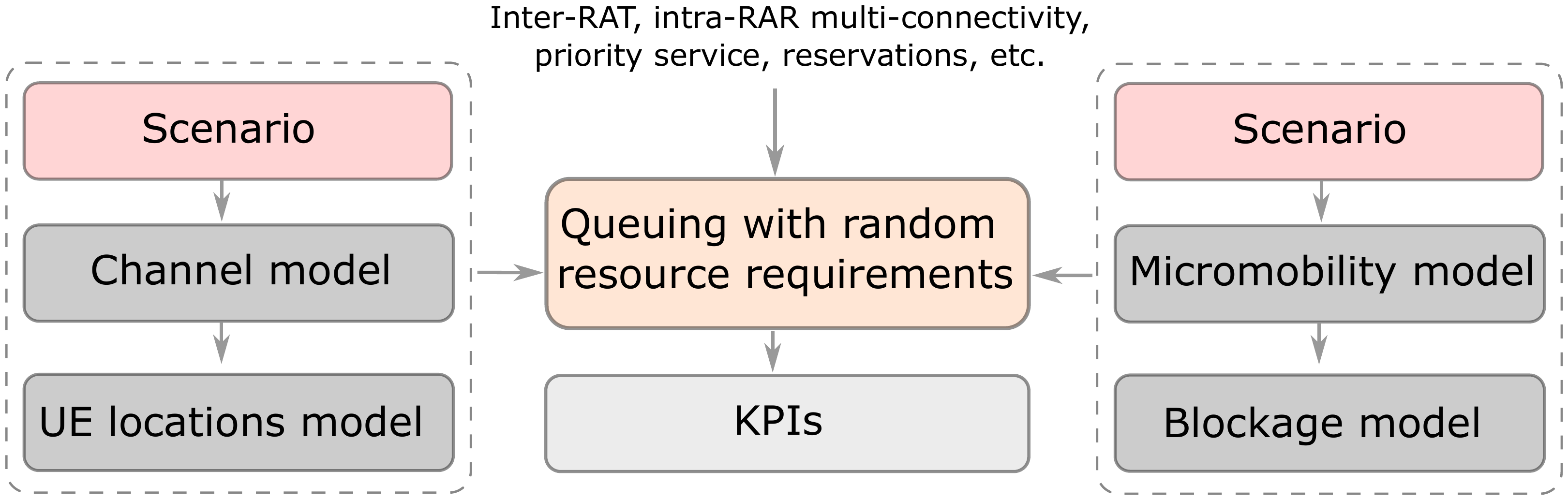}
	\caption{Illustration of the general structure of the framework for assessing KPIs of the non-elastic/adaptive traffic in mmWave/THz RATs: the core part is the queuing model representing the process of session service at mmWave/THz BSs; the rest of the model are combined to provided abstraction of the radio part specifics.}
	\label{fig:framework}
	\vspace{-2mm}
\end{figure}

The overall structure of the proposed framework is illlustrated in Fig. \ref{fig:framework}. By design, it allows for reuse of the core parts for investigating alternative mmWave/THz deployments by selecting appropriate models specified in the previous section. Here, the radio abstraction part estimates the intermediate ``interface'' parameters as a function of the considered deployment scenario, selected models as well as the system and environmental characteristics. As a result, the framework delivers KPIs of interest for selected advanced resource allocation policies at the BS, i.e., the resource reservation fraction, the number of simultaneously supported links, priorities, etc.


The queuing framework is characterized by a hierarchical structure. The baseline considered below first introduces the model with a single traffic type, no priorities and no session continuity mechanisms. Then, we proceed gradually extending the baseline to more sophisticated scenarios in incremental order of complexity, first, to resource reservation, then to multiconnectivity, further -- to priorities. Note that additional models can be built on top of these, e.g., by uniting multiconnectivity and priorities and defining additional rules to utilizing the former, one could formalize a model of a network segment with multiconnectivity capabilities servicing two or more traffic types. 

The framework is based on resource queuing systems (RQS). The main difference between RQS and conventional queuing systems is that sessions require not only a server but also some random volume of a finite resource. This difference makes it possible to take into account the heterogeneity of the session resource requirements at mmWave/THz BSs arising from the random locations of UEs and, as a consequence, the random spectral efficiency of the wireless channel associated with the data sessions.


Note that this part does not pretend to be a complete review of the RQS theory. We present only the main results that are most often used in the analysis of mmWave/THz networks. We invite the interested reader to refer to the review of the RQS in two parts ~\cite{rsmo_review1, rsmo_review2}, which describes in detail the state-of-the-art of the RQS theory.

\subsection{Baseline Resource Queuing Systems}\label{sec4-1}

\begin{figure}[t!]
	\centering
	\includegraphics[width=0.65\columnwidth]{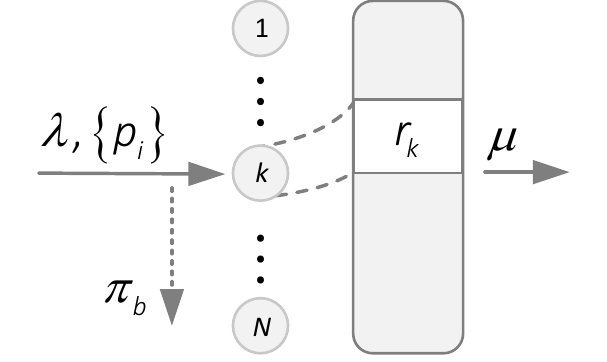}
	\caption{Illustration of the baseline resource queuing system: the principal difference with Erlang's loss model is presence of finite pool of resources that are allocated for arriving sessions.}
	\label{fig:smo}
	\vspace{-2mm}
\end{figure}

First, we consider a baseline resource model, see Fig. \ref{fig:smo}, that is, a multiserver queuing system with $N\leq \infty$ servers and a finite volume of resources, $R$. Assume that the arriving flow is Poisson with parameter $\lambda$, and service times of sessions are independent of each other, independent of the arrival process, and are exponentially distributed with parameter $\mu$. Each session requires one server and a random volume of the resources. Here, the amount of servers may represent the maximum number of active connections, if there is such a limit, while resources abstract time-frequency resources at the air interface and can be expressed in terms of resource blocks (RB). In what follows, we will assume that the distribution of resource requirements by a session is discrete and is described by the pmf $\{p_j\}_{j\geq0}$, where $p_j$ is the probability that the session requires $j$ resource units. An arriving session is lost if the amount of resource required for it exceeds the amount of unoccupied resources. At the end of the service, the session leaves the system and the total amount of occupied resources is decreased by the amount of resources allocated to the session. The schematic illustration of the considered system is shown in Fig. \ref{fig:smo}. Note that this model captures only the randomness of session resource requirements caused by UE locations, propagation model, utilized antenna arrays and session rate.

The system behavior is described by a random process $X_1(t)=\{\xi(t), \gamma(t)\}$, where $\xi(t)$ is the number of sessions in the system at the time $t$, and $\gamma(t)=(\gamma_1(t),\dots, \gamma_{\xi(t)}(t))$  is the vector that represents the number of resources allocated to each session. Let us introduce the notation for the stationary distribution of the process $X_1(t)$
\begin{align} \label{eq1}
&Q_k(r_1,\dots, r_k)=\nonumber \\ 
&=\lim_{t \to \infty} P \left\{ \xi(t)=k, \gamma_1(t)=r_1,\dots, \gamma_k(t)=r_k \right\}.
\end{align}

In \cite{naumov2016} it was shown that stationary probabilities (\ref{eq1}) can be obtained as follows
\begin{align} \label{eq2}
&Q_k(r_1,\dots, r_k)=Q_0\frac{\rho^k}{k!}\prod_{i=1}^{k}p_{r_i}, \, 1 \leq k\leq N, \sum_{i=1}^k r_i \leq R, \nonumber \\ 
&Q_0=\left( 1+\sum_{k=1}^N \sum_{r_1+...+r_k\leq R} \frac{\rho^k}{k!}\prod_{i=1}^{k}p_{r_i}, \right)^{-1},
\end{align}
where $\rho=\lambda/\mu$ is the offered load.

Note that the state space of the process $X_1(t)$ grows very quickly with the increase in $N$ and $R$, which leads to significant difficulties in calculating the stationary probabilities and the performance measures of the system, despite availability of analytical expressions in \eqref{eq2}. Moreover, the process $X_1(t)$ includes a lot of redundant information, which is unnecessary in most cases. By applying the state aggregation technique \cite{bobbio1986aggregation,ciardo1999etaqa}, to analyze the performance indicators of the system, it is sufficient to track only the number of sessions in the system and the total amount of the resource occupied by all the sessions. The stationary probabilities for the aggregated states are provided by
\begin{align} \label{eq3}
P_k(r)=\lim_{t \to \infty} P \left\{ \xi(t)=k, \sum_{i=1}^k \gamma_i(t)=r \right\}.
\end{align}

By summing up the probabilities in \eqref{eq2}, we obtain
\begin{align} \label{eq:4}
&P_k(r)=P_0\frac{\rho^k}{k!}p_r^{(k)}, \, 1 \leq k\leq N, r \leq R, \nonumber \\
&P_0=\left( 1+\sum_{k=1}^N \sum_{r=0}^R \frac{\rho^k}{k!}p_r^{(k)} \right)^{-1},
\end{align}
where $\{p_r^{(k)}\}_{r\geq 0}$ is a $k$-fold convolution of distribution $\{p_j\}_{j\geq0}$. Note that $p_r^{(k)}$ can be interpreted as the probability that $k$ sessions totally occupy $r$ resources. In practice, discrete convolutions are calculated using the recurrence relation
\begin{align} \label{eq:5}
p_r^{(k)}=\sum_{i=0}^r p_i p_{r-i}^{(k-1)}, \quad k \geq 2,
\end{align}
where $p_r^{(1)}=p_r$, $r \geq 0$.

Using the probabilities of the aggregated states $P_k(r)$, we can obtain expressions for the main characteristics of the model, namely, the new session drop probability $\pi_b$ and the average number of occupied resources $\bar{R}$ as follows
\begin{align} \label{eq:pi}
&\pi_b=1-P_0\sum_{k=0}^{N-1} \frac{\rho^k}{k!} \sum_{r=0}^R p_r^{(k+1)},\\
&\bar{R}=P_0\sum_{k=1}^{N} \frac{\rho^k}{k!} \sum_{r=1}^R rp_r^{(k)}.
\end{align}

\subsubsection{Simplified System} \label{sec4-1-1}

Consider a simplified process $X_2(t)=\{ \xi(t), \delta(t) \}$, where, differently from $X_1(t)$ considered above, the second component $\delta(t)$ denotes the total number of resources occupied by all the sessions. This simplification leads, on the one hand, to the decrease in the dimension of the state space, and, on the other hand, to the loss of important information about the queuing process. As a result, it is impossible to say exactly how many resources are released by a session upon its departure. In this case, one may utilize the following Bayesian approach. Assume that the system is in the state $(k,r)$ at the moment just before the session departure from the system. The probability that a session will release $j$ resources can be determined by the Bayes formula as $p_j p_{r-j}^{(k-1)}/p_r^{(k)}$. Then, the system of equilibrium equations for stationary probabilities $q_k(r)$ of the process $X_2(t)$ can be written in the following form
\begin{align} \label{eq:6}
&\lambda q_0 \sum_{r=0}^R p_r = \mu \sum_{r=0}^R q_1(r), \nonumber \\
&q_k(r) \left( \lambda \sum_{r=0}^{R-r} p_r + k\mu \right) = \lambda \sum_{j=0}^r q_{k-1}(j) p_{r-j} +(k+1) \times \nonumber\\
&\times\mu\sum_{j=0}^{R-r} q_{k+1}(r+j) \frac{p_j p_r^{(k)}}{p_{r+j}^{(k+1)}}, 1\leq k \leq N-1, 0 \leq r \leq R, \nonumber \\
&N\mu q_N(r) = \lambda \sum_{j=0}^r q_{N-1}(j) p_{r-j}, 0 \leq r \leq R.
\end{align}


As shown in ~\cite{naumov2016}, by substituting $P_k(r)$ from \eqref{eq3} into the system \eqref{eq:6}, one can prove that \eqref{eq:4} is its solution. Thus, the stationary probabilities of the simplified process $X_2(t)$ coincide with the probabilities of the aggregated states $P_k(r)$ of the initial process $X_1(t)$. As we will see in the following sections, the established fact is of critical importance for the analysis of more complex systems involving additional rules and service mechanisms at BSs. Generally, in more complex systems, it is not possible to obtain analytical expressions for stationary probabilities, which means that to analyze those systems one has to resort to a numerical solution of the system of equilibrium equations. However, in most practical cases it is not feasible due to the large state space of the process. Therefore, the proposed simplification makes it possible to numerically calculate performance measures of the considered class of systems in a reasonable time.


\subsubsection{Multiple Arriving Flows} \label{sec4-1-2}

Consider now a resource queuing system with multiple arriving flows of sessions. Particularly, differently from the systems considered above, we assume $L$ arriving mutually independent Poisson flows of sessions with intensities $\lambda_l$, $l=1,2,\dots,L$. Sessions of the $l$-th type are served with rates $\mu_l$, and their distribution of resource requirements is given by $\{p_{l,r}\}_{r \geq 0}$. The stationary probabilities $Q_{k_1,\dots,k_L}(r_1,\dots,r_L)$ that there are $k_l$ sessions of type $l$ occupying totally $r_l$ resources, $l=1,2,...,L$, are then provided by
\begin{align} \label{eq:7}
&Q_{k_1,\dots,k_L}(r_1,\dots,r_L)=\nonumber \\
&=	Q_0\prod_{i=1}^{L}\frac{\rho_1^{k_i}}{k_i!}\prod_{i=1}^{L}p_{r_i}^{(k_i)}, \sum_{l=1}^L k_l \leq N,  \, \sum_{l=1}^L r_l \leq R, \\
&	Q_0=\hspace{-1mm}\left( 1+\hspace{-6mm}\sum_{1 \leq k_1+\dots+k_L \leq N} \sum_{0 \leq r_1+\dots+r_L \leq R} \hspace{-7mm}Q_{k_1,\dots,k_L}(r_1,\dots,r_L) \right)^{-1}. \nonumber
\end{align}
where $\rho_l=\frac{\lambda_l}{\mu_l}$, $l=1,2,...,L$. 

Note that \eqref{eq:7} are also not overly useful for numerical calculations. However, in \cite{sopin_aggregation}, it was shown that a resource queuing system with $L$ flows of sessions is equivalent to the resource queuing system with one aggregated flow having a weighted average distribution of resource requirements. In other words, the stationary probabilities $Q_k(r)$ that there are $k$ sessions of all types in the system that totally occupies $r$ resources, are determined by
\begin{align} \label{eq:8}
&	Q_{k}(r)=Q_0\frac{\rho^{k}}{k!} \bar{p}_{r}^{(k)}, 1 \leq k \leq N,  r \leq R \nonumber\\
&	Q_0=\left( 1+\sum_{k=1}^N \sum_{r=0}^R Q_{k}(r) \right)^{-1},
\end{align}
where $\rho=\rho_1+\dots+\rho_L$, and $\bar{p}_r$ are given by
\begin{align} \label{eq:9}
	\bar{p}_r= \sum_{l=1}^L \frac{\rho_l}{\rho} p_{l,r}.
\end{align}

The mean amount of occupied resources can be found similarly to \eqref{eq:pi}, and the session drop probability of a $l$-type session takes the form of
\begin{align} \label{eq:10}
\pi_{b,l}=1- \sum_{k=0}^{N-1} \sum_{r=0}^R Q_k(r) p_{l,R-r}.
\end{align}

\subsubsection{Numerical Algorithm} \label{sec4-1-3}

The direct calculation of performance measures associated with resource queuing systems is complicated. The main reason is the need for evaluating multiple convolutions of the resource requirements distribution as discussed above. However, the authors in \cite{sopin_algorithm}, developed a recursive algorithm for calculating the normalization constant $Q_0$, which allows for efficient numerical analysis.

The algorithm proceeds as follows. First, denote 
\begin{align}
G(n,r)=\sum_{k=0}^n\frac{\rho^k}{k!} \sum_{j=0}^r p_j^{(k)}.
\end{align}

Utilizing this notation, the normalization constant in \eqref{eq:8}, interpreted as the probability that the system is empty, can be written as $Q_0=G(N,R)^{-1}$. The function $G(n,r)$ can be calculated using the following recursion
\begin{align} \label{eq:11}
&	G(n,r)=G(n-1,r)+\nonumber \\
&	+\frac{\rho}{n} \sum_{j=0}^r p_j \left( G(n-1,r-j)-G(n-2,r-j) \right),
\end{align}
with initial conditions
\begin{align} \label{eq:12}
G(0,r)=1,\quad G(1,r)=1+\sum_{j=0}^r p_j, \quad r \geq 0.
\end{align}

Using $G(n,r)$, any performance measure of the system can be calculated. For example, the session drop probability $\pi_b$ and the average amount of occupied resources $\bar{R}$ take the form
\begin{align} \label{eq:13}
&\pi_b=1- \frac{1}{G(N,R)} \sum_{j=0}^R p_j  G(N-1,R-j),\nonumber\\
&\bar{R}=R- \frac{1}{G(N,R)} \sum_{j=1}^R G(N,R-j).
\end{align}

\begin{figure*}[!!t]
	\begin{align}\label{eq:20}
	&\mathbf{D_n}(I(n,i),(n,j))= 
	\begin{cases} 
	-\left( \lambda \displaystyle\sum_{j=0}^{R-i} p_j +n\mu + n\alpha\left(1-\sum_{m=0}^i \theta_m(n,i) p_m \right)\right), \quad i=j, \\ 
	n\alpha \displaystyle\sum_{m=0}^i \theta_m(n,i) p_{j-i+m}, \quad i<j, \\
	n\alpha \displaystyle\sum_{m=i-j}^i \theta_m(n,i) p_{j-i+m}, \quad i>j,
	\end{cases}(n,i), (n,j) \in \Psi_n, \quad 1 \leq n \leq N-1,  \nonumber\\
	&\mathbf{D_N}(I(N,i),(N,j))= 
	\begin{cases} 
	-\left( N\mu + N\alpha(1-\sum_{m=0}^i \theta_m(N,i) p_m \right), \quad i=j, \\ 
	N\alpha \displaystyle\sum_{m=0}^i \theta_m(N,i) p_{j-i+m}, \quad i<j, \\
	N\alpha \displaystyle\sum_{m=i-j}^i \theta_m(N,i) p_{j-i+m}, \quad i>j,
	\end{cases}(N,i), (N,j) \in \Psi_N.
	\end{align}
	\hrulefill
	\normalsize
	\vspace{-2mm}
\end{figure*}

\subsection{Resource Queuing System with Service Interruptions} \label{sec4-2}


Consider now the extension of the system to the case of external flow of events that may potentially change the characteristics of the service process of sessions, currently served at the BS. In mmWave and THz systems these events can be utilized to capture impairments caused by the blockage process of LoS path, leading to either change in the resource requirements to maintain the target session rate \cite{begishev2019quantifying,begishev2021joint} or to outage events completely interrupting the session service process \cite{begishev2021multi}. In addition, these events can also model outage events caused by micromobility \cite{petrov2020capacity,petrov2018effect}. Due to a wide scope of the application area of these systems, in what follows, the external events are referred to as ``signals''.


Unlike the model considered in Section \ref{sec4-1}, we assume that each session is associated with a Poisson flow of signals with intensity $\alpha$. For certainty, in further exposition we consider LoS blockage process associated with UEs, where the blockage does not lead to outage conditions, i.e., in case of blockage the connection may still be maintained but more resources are required to provide the target session rate. In this case, upon signal arrival, the resources allocated for a session are all released and the session tries to occupy a new volume of resources according to the same or different probability distributions. In what follows, for certainty, we assume that the distribution remains intact. Note that this could be a pmf of session resource requirements obtained for blocked and non-blocked states and then weighted with blockage probability as discussed in Section \ref{sect:param}.

For analysis of this system, one may utilize the simplified method considered in Section \ref{sec4-1-1} that implies tracking only the total amount of the occupied resources. Accordingly, the behavior of the system is described by the stochastic process $X(t)=\{ \xi(t), \delta(t) \}$, where the first component denotes the number of sessions in the system at time $t$, and the second represents the total amount of resources occupied by all the sessions. The state space of the system is given by 
\begin{align} \label{eq:15}
\Psi=\bigcup_{k=0}^N \Psi_k, \, \Psi_k=\left\{ (k,r): 0\leq r \leq R, p_r^{(k)} > 0 \right\},
\end{align}
where the states in the subsets $\Psi_k$ are ordered according to the ascending of the number of resources. Let $I(k, r)$ denote the sequential number of the state $(k, r)$ in the set $\Psi_k$.

\subsubsection{Stationary Distribution} \label{sec4-2-1}

First, we introduce an supplementary variable facilitating our exposition in what follows. Let $\theta_i(k,r)$ be the probability that a session occupies $i$ resources, provided that $k$ sessions totally occupy $r$ resources. Then, according to Bayes' law we have
\begin{align} \label{eq:16}
	\theta_i(k,r)=\frac{p_i p_{r-i}^{(k-1)}}{p_r^{(k)}}.
\end{align}

To construct the infinitesimal generator of the process $X(t)$, consider possible transitions between the states of the system in more detail. Let the system be in the state $(k,r)$ at some time $t$. With probability $p_j$, an arriving session occupies $j$ resources, $j \leq R-r$, and the system goes to the state $(k+1,r+j)$. On the departure of a session, it releases $i$ resources with probability $\theta_i(k,r)$. In this case, the system state changes to the state $(k-1,r-i)$. Upon signal arrival, a session releases $i$ resources with probability $\theta_i(k,r)$ and requests $j$ resources with probability $p_j$. If $j \leq i$, then the session continues its service the system. If $j<i$, the system state changes to $(k,r-i+j)$ with probability $\theta_i(k,r) p_j$ while in the case of $j=i$ the state of the system does not change. Finally, if the new volume of resources requested by a session exceeds its previous volume ($j>i$), then the session remains in the system only when $j \leq R-r+i$. Otherwise, the session is lost (dropped) and the system goes to the state $(k-1,r-i)$.

The infinitesimal generator of the process $X(t)$ is a block tridiagonal matrix with diagonal blocks $\mathbf{D_0}, \mathbf{D_1}$, \dots, $\mathbf{D_N}$, superdiagonal blocks $\mathbf{\Lambda_1}$, \dots, $\mathbf{\Lambda_N}$ and subdiagonal blocks $\mathbf{M_0}, \dots, \mathbf{M_{N-1}}$, provided by
\begin{align} \label{eq:17}
&\mathbf{D_0}=-\lambda \sum_{j=0}^R p_j, \nonumber\\ 
&\mathbf{\Lambda_1}=(\lambda p_0, \dots, \lambda p_R),\nonumber\\
&\mathbf{M_0}=(\mu, \dots, \mu)^T.
\end{align}

\begin{figure*}[t!]
	\vspace{-0mm}
	\begin{align}\label{eq:sur3}
	&\text{(i)}\qquad{}\lambda Q_0 \sum_{j=0}^{R_0} p_j=\mu\sum_{j:(1,j)\in \Psi_1} Q_1(j) + \alpha \sum_{j:(1,j)\in \Psi_1}Q_1(j)\left( 1-\sum_{s=0}^{R} p_s\right),\nonumber\\ \nonumber\\
	&\text{(ii)}\qquad{}\left( \theta (R_0 - j)\lambda \sum_{j=0}^{R_0-r}p_j + k\mu +k \alpha \right)Q_k(r) = \theta (R_0 - j)\lambda\hspace{-10mm}\sum_{j \geq 0:(k-1,r-j)\in \Psi_{k-1}}\hspace{-10mm}Q_{k-1}(r-j)p_j+(k+1)\mu \times \nonumber\\
	&\qquad{}\qquad{}\qquad{}\times\hspace{-10mm}\sum_{j \geq 0:(k+1,r+j)\in \Psi_{k+1}}\hspace{-10mm}Q_{k+1}(r+j) \theta_j(k+1,j+r) +(k+1)\alpha \left( 1-\sum_{s=0}^{R-r} p_s \right) \hspace{-2mm}\sum_{j \geq 0:(k+1,r+j)\in \Psi_{k+1}}\hspace{-12mm}Q_{k+1}(r+j) \theta_j(k+1,j+r) + \\
	&\qquad{}\qquad{}+ k\alpha\hspace{-5mm}\sum_{j \geq 0:(k,j)\in \Psi_{k}}\hspace{-5mm}Q_{k}(j) \sum_{i=0}^{min(j,r)} \theta_{j-i}(k+1,j+r) p_{r-i},\quad 1 \leq n \leq N-1, \; 0 \leq r \leq R,\nonumber\\ \nonumber\\
	&\text{(iii)}\qquad{} N(\mu + \alpha)Q_N(r) =\theta (R_0 - j)\lambda \hspace{-11mm}\sum_{j \geq 0:(N-1,j)\in \Psi_{N-1}}\hspace{-11mm} Q_{N-1}(r-j)p_j+N \alpha \hspace{-8mm}\sum_{j \geq 0:(N,j)\in \Psi_{N}}\hspace{-7mm} Q_N (j)	\sum_{i=0}^{min(j,r)} \hspace{-3mm}\theta_{j-i}(N,j) p_{r-i}, \quad 1 \leq r \leq R.\nonumber
	\end{align}
	\hrulefill
	\normalsize
	\vspace{-2mm}
\end{figure*}

Omitting zero elements from the block $\mathbf{\Lambda_1}$, the number of columns in $\mathbf{\Lambda_1}$, as well as the number of rows in the block $\mathbf{M_0}$, are equal to the number of states in the subset $\Psi_1$. The blocks $\mathbf{D_n}(I(n,i),(n,j))$ and $\mathbf{D_N}(I(N,i),(N,j))$ are provided in (\ref{eq:20}). The blocks $\mathbf{\Lambda_n}(I(n-1,i),(n,j))$ are provided by
\begin{align} \label{eq:22}
	\mathbf{\Lambda_n}(I(n-1,i),(n,j))= 
	\begin{cases} 
		\lambda p_{j-i}, & i \leq j \leq R, \\ 
		0, & j<i,
	\end{cases}
\end{align}
where $(n-1,i) \in \Psi_{n-1}$, $(n,j) \in \Psi_n$,  $2 \leq n \leq N$.

Finally, the blocks $\mathbf{M_n}(I(n+1,i),(n,j))$ are specified as
\begin{align} \label{eq:23}
	\mathbf{M_n}&(I(n+1,i),(n,j))= \nonumber \\
	&=\begin{cases} 
		(n+1)\mu \theta_{i-j}(n+1,i), & j \leq i \leq R, \\ 
		0, & j>i,
	\end{cases}
\end{align}
where $(n+1,i) \in \Psi_{n+1}$, $(n,j) \in \Psi_n$, $1 \leq n \leq N-1$.

The stationary probabilities $Q_k(r)$ of the process $X(t)$ are the unique solution of the system of equilibrium equations with the normalization condition, that can be written as
\begin{align} \label{eq:24}
	\mathbf{Q}\mathbf{A}=\mathbf{0}, \, \mathbf{Q} \mathbf{1}=1,
\end{align}
where $\mathbf{Q}$ is the row vector of stationary probabilities, $\mathbf{1}$ is a column vector of ones of appropriate size. The system \eqref{eq:24} can be solved by any numerical method, including those using the special block structure of the described infinitesimal generator.

\subsubsection{Performance Metrics} \label{sec4-2-2}

In this subsection, we will proceed to the analysis of system's performance indicators including the drop probability of a session upon arrival, $\pi_b$, and the average amount of occupied resources, $\bar{R}$. These metrics can be directly calculated using stationary probabilities as
\begin{align} \label{eq:25}
&\pi_b=1-\sum_{k=0}^{N-1} \sum_{r=0}^R Q_k(r) \sum_{j=0}^{R-r} p_r,\nonumber\\
&\bar{R}=\sum_{k=0}^{N} \sum_{r=0}^R rQ_k(r).
\end{align}

In the considered system, we are also interested in the probability $\pi_t$ that a session initially accepted for service is eventually lost due to the signal arrival. In our interpretation, this implies that upon blockage, the amount of resources at BS is insufficient to maintain the target bitrate of the session. To derive this metric, we first calculate the intensity of sessions that are eventually dropped, $\nu$, as
\begin{align} \label{eq:27}
\nu=\alpha\bar{N}\sum_{k=1}^{N} \sum_{r=0}^R Q_k(r) \sum_{j=0}^r \theta_j(k,r) \left(1-\sum_{i=0}^{R-r+j} p_i \right),
\end{align}
where $\bar{N}$ is the mean number of sessions in the system
\begin{align} \label{eq:28}
	\bar{N}=\sum_{k=1}^{N} \sum_{r=0}^R kQ_k(r).
\end{align}

Then, the sought  probability $\pi_t$ can be defined as the limit of the ratio of the number of accepted sessions that have been eventually dropped to the total number of accepted sessions during the time interval of duration $T$. Accordingly, we have
\begin{align} \label{eq:29}
\pi_t=\lim_{T \to \infty} \frac{\nu T}{\lambda(1-\pi_b)T}=\frac{\nu }{\lambda(1-\pi_b)}.
\end{align}

We specifically note that the probability that a session initially accepted to the system is eventually dropped is known to drastically affect quality of user experience (QoE) of a service \cite{seitz2003itu,serral2010overview,de2012quantifying,dobrian2011understanding}. As these events may often happen in mmWave and THz communications, recently, a number of approaches for improving it have been proposed. Most of these mechanisms can be modeled by utilizing the resource queuing systems framework described above. In what follows, we consider some of these mechanisms as examples.


\subsection{Resource Reservation}\label{sec4-3}

Reserving resources for sessions that sharply increase their resource requirements due to blockage of a LoS path, is one of the mechanisms to improve session continuity in mmWave and THz systems \cite{begishev2019quantifying,moltchanov2018improving}, especially, in early rollouts of these systems, where BS will be sparsely deployed. In this case, only part of the BS resources is available for the new arriving sessions, and the rest is reserved to support the sessions already accepted for service. 

\subsubsection{Model Description} \label{sec4-3-1}

Consider a resource queuing system with $N$ servers and a finite amount of resources $R$, only part of which $R_0=(1-\gamma) R$, $0<\gamma<1$, is available for new arriving sessions. Here $ \gamma $ is the reservation coefficient and is interpreted as a fraction of reserved resources. The behavior of the system is described by a stochastic process $X(t)=\{\xi(t),\delta(t)\}$, where $\xi(t)$ is the number of sessions in the system, and $\delta(t)$ is the total the amount of the occupied resource. The state space of the system is described by \eqref{eq:15}.

\begin{figure}[t!]
	\centering
	\includegraphics[width=0.55\columnwidth]{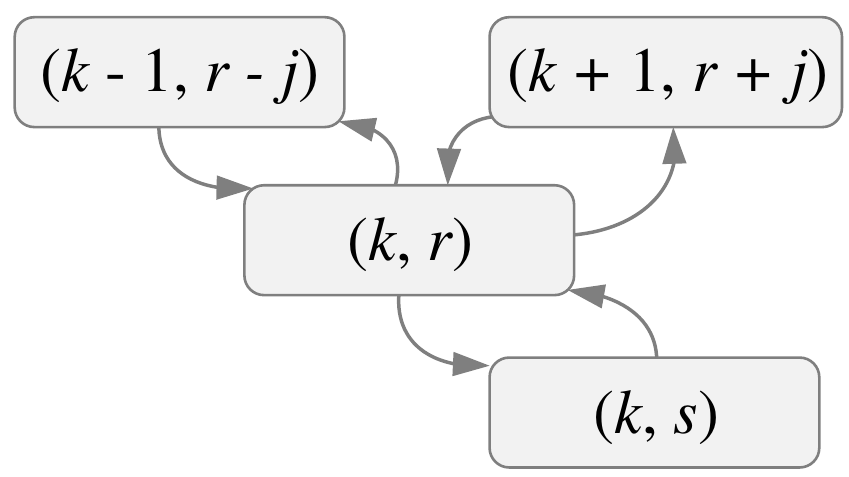}
	\caption{Feasible transitions for the central state of the model represent the session service process with resource reservation \cite{begishev2019quantifying}.}
	\label{fig:state3}
	\vspace{-2mm}
\end{figure}

Unlike the system considered in Section \ref{sec4-2}, when a session arrives to the system, it can occupy only a part of the resources, $R_0<R$. In other words, let the system be in the state $(k,r)$ and assume that an arriving session requires $j$ resources, then: (i) if $r>R_0$, any new arriving session is dropped, (ii) if $r \leq R_0$ and $j>R_0-r$, then an arriving session is dropped, and (iii) if $r \leq R_0$ and $j \leq R_0-r$, an arriving session is accepted into the system. 
Once the session is accepted for service, the whole amount of resources, $R$, becomes available for it. When a signal arrives for a session when the system is in the state $(k,r)$,  the session releases $i$ resources and tries to occupy $j$ resources. In this case, it is dropped only when $j>R-r+i$. Otherwise, the system operates similarly to the system described in Section \ref{sec4-2}. The typical subset of states and associated transitions for this system are illustrated in Fig. \ref{fig:state3}.

\subsubsection{Equilibrium Equations} \label{sec4-3-2}

By utilizing
\begin{align} \label{eq:30}
	\theta (R_0-j) = 
		\begin{cases}
			0, &\text{$j>R_0$,}\\
			1, &\text{j $\leq R_0$,}
		\end{cases}
\end{align}
the system of equilibrium equations can be written as in (\ref{eq:sur3}).

Using the system \eqref{eq:sur3}, together with the normalization condition provided in \eqref{eq:30}, one may calculate the stationary state probabilities of the system. Here, the infinitesimal generator of the process $X(t)=\{\xi(t),\delta(t)\}$ is obtained similarly to Section \ref{sec4-2} based on the system of equations \eqref{eq:sur3}. 

\subsubsection{Performance Metrics} \label{sec4-3-3}

Consider the performance metrics of the system. Since the system with resource reservation differs from the system from Section \ref{sec4-2} only in the connection admission control (CAC) functionality, all the expressions for performance metrics except for the new session drop probability remain the same. The abovementioned unknown metric, $\pi_b$, is provided by
\begin{align} \label{eq:31}
	\pi_b=1-\sum_{k=0}^{N-1} \sum_{r=0}^{R_0} Q_k(r) \sum_{j=0}^{R_0-r} p_r,
\end{align}


\subsubsection{Illustrative Example}

\begin{figure}[t!]
	\centering
	\includegraphics[width=0.8\columnwidth]{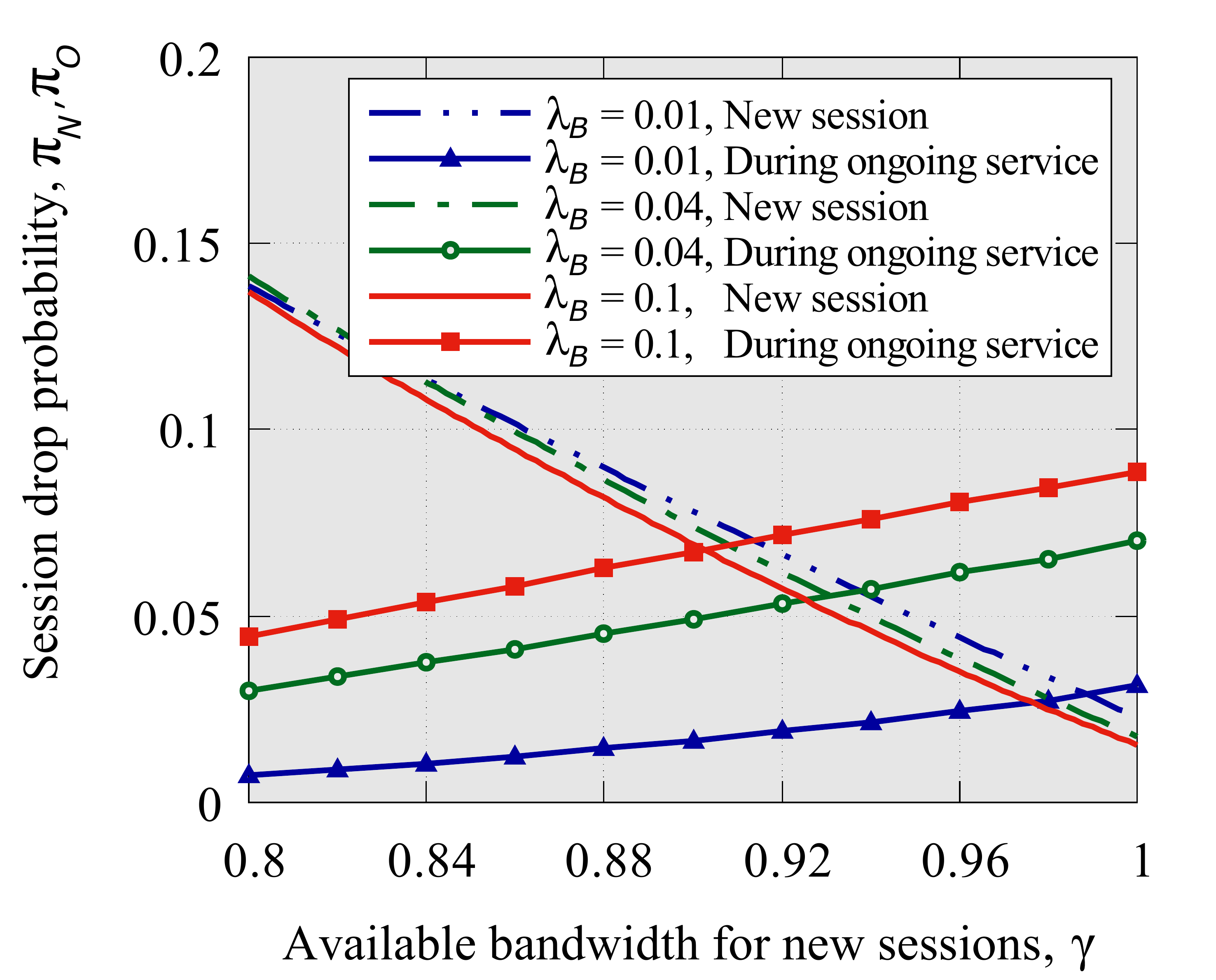}
	\caption{New and ongoing session drop probabilities: reserving resources for sessions that are already accepted for service increases the new session drop probability but increases the service reliability at mmWave/THz BSs by decreasing the ongoing session drop probability \cite{begishev2019quantifying}.}
	\label{fig:6cnew}
	\vspace{-2mm}
\end{figure}

An example of application of the abovementioned framework to the case of mmWave BS utilizing the resource reservation strategy for sessions already accepted to the system is considered in \cite{begishev2019quantifying}. Consider a single cell system with session arrivals governed by the Poisson process whose geometric locations are uniformly distributed in the BS service area. Since resource reservation may only show gains when blockage does not lead to an outage the cell service area is computed accordingly. In Fig. \ref{fig:6cnew} we show typical new and ongoing session drop probabilities as a function of the amount of resources reserved for sessions accepted for service, required session rate of $10$ Mbps and different blockers intensity in the environment, $\lambda_B$ bl/m$^2$. As one may observe, there is a clear trade-off between two considered types of session drop probabilities. Unfortunately, the new session drop probability increases at a slightly faster rate as compared to the decrease in the ongoing session drop probability. Nevertheless, by utilizing this simple strategy a network operator may control the balance between these two probabilities in early 5G mmWave systems rollouts when NR BSs are installed in hotspot areas with high traffic demands.

\subsection{Multiconnectivity and Resource Reservation} \label{sec4-4}

Another method for improving session continuity in mmWave and THz systems is standardized by 3GPP multiconnectivity operation \cite{3gpp_MC}. According to it, UE may maintain more than a single link to nearly BSs. UE may utilize all the links simultaneously via division duplex (TDD) mode or use one link for data transmission while keeping the rest as backup  option in the case the currently active link experiences outage conditions due to blockage and/or micromobility. These connectivity strategies and associated resource allocation policies fall within the class of resource queuing networks -- an extension of resource queuing systems considered in the previous sections. Below, we describe the solution method for the case, where both multiconnectivity and resource reservation are simultaneously utilized. 


\begin{figure}[b!]
	\vspace{-0mm}
	\centering
	\includegraphics[width=1.0\columnwidth]{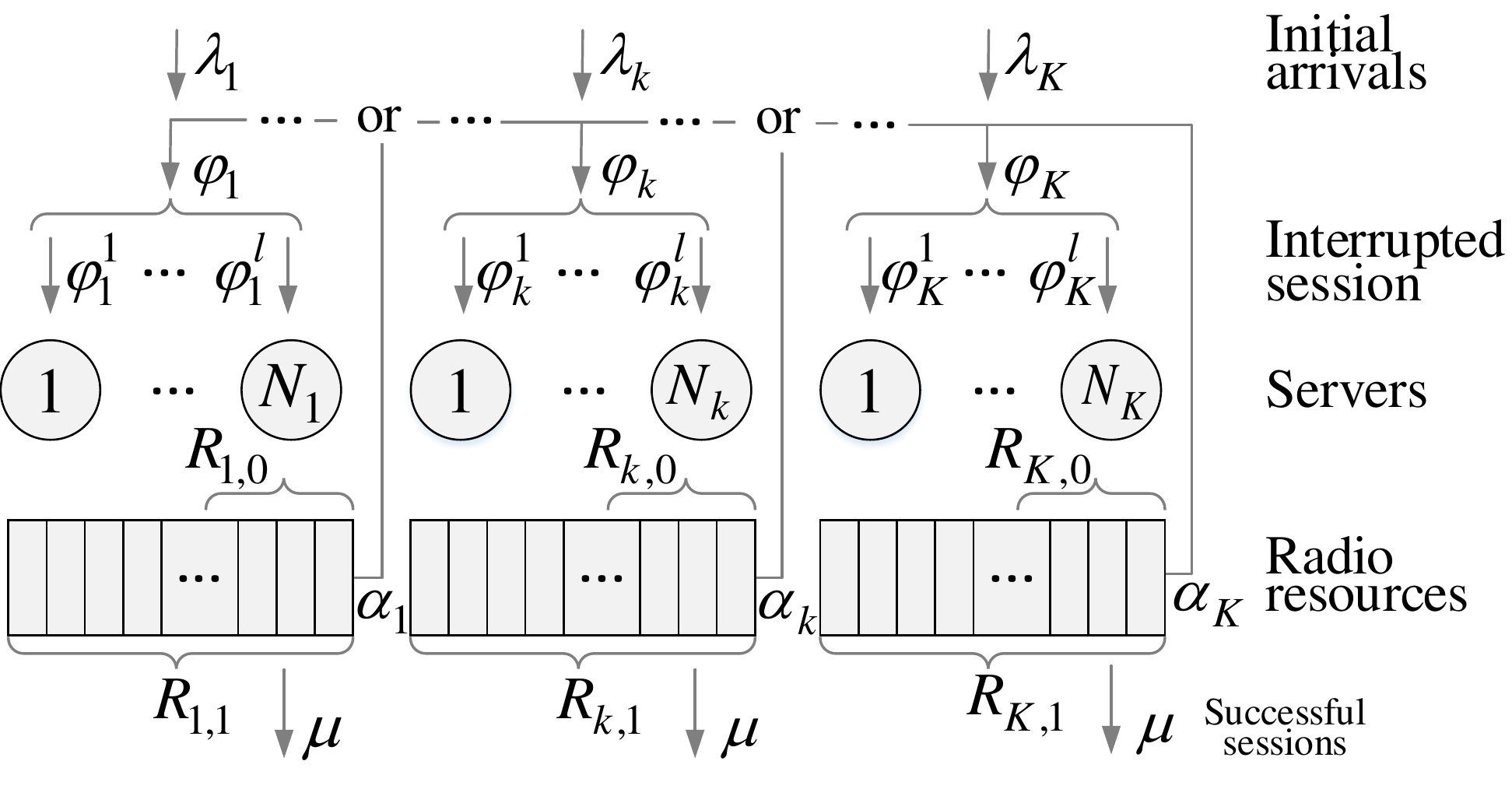}
	\caption{Illustration of the resource queuing network: upon blockage active sessions are handed over to other BS available in the area inducing dependence in the session service process \cite{begishev2021joint}.}
	\label{fig:queue}
	\vspace{-0mm}
\end{figure}

\subsubsection{Model Description} \label{sec4-4-1}

Consider a queuing network consisting of $K$ BSs. The $k$-th BS has $N_k$ servers and $R_{k,1}$ resources. The $k$-th BS receives a Poisson flow of sessions with intensity $\lambda_k$, $k=1,2,\dots,K$, and $\lambda = \sum_{k=1}^{K}\lambda_k$. Each session arriving to the $k$-th BS is characterized by a random resource requirements with pmf $\{p_{0,r}\}_{r>0}$. For new arriving sessions, only a part of the resources is available, $R_{k,0}=R_{k,1}(1-\gamma)$, $0<\gamma<1$. Similarly to the previous section, if an arriving session finds that there are not enough resources in the system for its service, then this session is dropped. Service times are assumed to be exponentially distributed with the parameter $\mu$. Further, each session which is currently in service at the $k$-th BS is associated with a Poisson flow of signals with the intensity of $\alpha_k$, $k=1,2,\dots,K$,  that represents channel state transitions between outage and connectivity states that might be caused by either blockage or micromobility or both. Upon a signal arrival, the session releases the occupied resources and generates a new value of resource requirements according to pmf $\{p_{1,r}\}_{r>0}$. If these resources are not available at the current BS, the session is routed to BS $k$ with probability $1/(K-1)$. Rerouted sessions are called ``secondary'' sessions. We also introduce the ``level'' a secondary session as the number of reroutes a session experienced. Secondary sessions form an additional Poisson arriving flow to each BS with intensities $\varphi_k$, $k=1,2,\dots,K$. All available resources at BSs are available for secondary sessions. If there is an insufficient amount of resources available, the session is dropped. Due to the memoryless property of the exponential distribution, the residual service time of a secondary session also follows an exponential distribution with the same parameter $\mu$. The illustration of the considered model is sketched in Fig. \ref{fig:queue}.

\begin{figure}[!b]
	\vspace{-0mm}
	\centering
	\includegraphics[width=1.0\columnwidth]{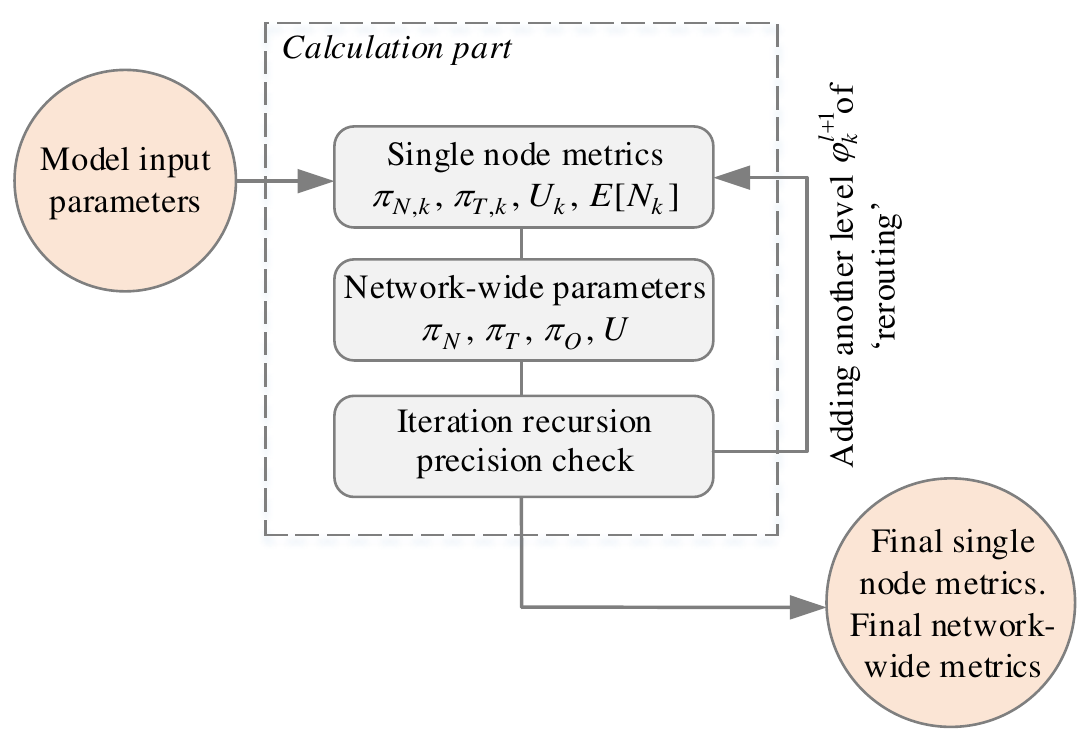}
	\caption{Iterative algorithm for calculating performance metrics: at each stage the flow variables between BSs are updated the the difference in performance metrics at the current and previous iteration is assessed to decide on the convergence \cite{begishev2021joint}.}
	\label{fig:algorithm}
\end{figure}


To analyze the described model, we rely upon the network decomposition method, which is commonly utilized in the analysis of complex queuing networks \cite{jia2009solving,kuehn1979approximate,gershwin1987efficient}. As usual, the key assumption here is that the service process of sessions at each BS does not depend on the service processes at other BSs in the network. Based on the decomposition approach, we develop a recurrent algorithm for the evaluation of the performance metrics as shown in Fig. \ref{fig:algorithm} and discussed below. First, the system parameters are initialized assuming the arrival intensity of the secondary sessions to be zero, $\varphi_k=0$, $k=1,2,\dots,K$. Then, the stationary probabilities of each BS are calculated, from which the network-wide measures are obtained. Further, the arrival intensities of the secondary sessions are recalculated by taking into account the next level of session rerouting. These new values of the arrival intensities $\varphi_k=0$, $k=1,2,\dots,K$ are used at the next iteration of the algorithm. The algorithm stops as the difference between the network characteristics at the consequent iterations becomes less than a certain specified level of accuracy.

\subsubsection{Single BS Performance} \label{sec4-4-2}

Consider BS with two Poisson flows of sessions: the flows of primary and secondary sessions with rates $\lambda_k$ and $\varphi_k$, respectively. By utilizing the memoryless property of exponential distribution we observe that the sojourn time of the session at the BS is also exponential with the parameter $\mu+\alpha_k$, $i=1,2,\dots,K$. Here, we again apply the simplified approach, where only the total amount of the occupied resources is tracked at BS. Then, by taking into account the number of primary and secondary sessions, the behavior of the system is described by a three-dimensional process $X(t)=\{\xi_1(t),\xi_2(t),\delta(t)\}$, where $\xi_1(t)$ is the number of primary sessions in the system at time $t$, $\xi_2(t)$ is the number of secondary sessions at time $t$, and $\delta(t)$ is the total amount of the resource occupied by all the sessions. The state space of the process is thus given by
\begin{align}
	&\Psi = \hspace{-4mm}\bigcup_{0\leq n_1+n_2 \leq N} \hspace{-4mm}\Psi_{n_1,n_2},  \nonumber\\
	&\Psi_{n_1,n_2} =\nonumber\\
	&\hspace{-3mm}=\left\{(n_1,n_2,r):0\leq r \leq R_1, \hspace{-3mm}\sum_{i=0}^{\min(r,R_0)}\hspace{-2mm}p_{0,i}^{(n_1)}p_{1,r-i}^{(n_2)}>0\right\},
	\end{align}
where, the superscript denotes the order to convolution.

Stationary probabilities $Q_{n_1,n_2}(r)$ that the system is in the state $(n_1,n_2,r) \in \Psi$ are defined as
\begin{align} \label{eqn:statprob}
	Q_{n_1,n_2}(r)=\nonumber\\
	&\hspace{-7mm}=\lim_{t \to \infty} P\{ \xi_1(t)=n_1, \xi_2(t)=n_2, \delta(t)=r \}.
\end{align}

Consider the amount of resources that a session releases upon departure. Denote by $\beta_{0,j}(n_1,n_2,r)$ the probability that, $j$ resources are released as a result of the primary session departure provided that the system is in the state $(n_1,n_2,r)$. For $r\leq R_{0}$, the probabilities $\beta_{0,j}(n_1,n_2,r)$ can be calculated using the Bayes law
	\begin{align} \label{eq:beta0_r0}
	\beta_{0,j}(n_1,n_2,r)=\frac{p_{0,j}\sum_{i=0}^{r-j}p_{0,i}^{(n_1-1)}p_{1,r-j-i}^{(n_2)}}{\sum_{i=0}^{r}p_{0,i}^{(n_1)}p_{1,r-i}^{(n_2)}}.
	\end{align}

Similarly, define by $\beta_{1,j}(n_1,n_2,r)$ the probability that as a result of the secondary session departure, $j$ resources are released provided that the system is in the state
$(n_1,n_2,r)$,
	\begin{align} \label{eq:beta1_r0}
	\beta_{1,j}(n_1,n_2,r)=\frac{p_{1,j}\sum_{i=0}^{r-j}p_{0,i}^{(n_1)}p_{1,r-j-i}^{(n_2-1)}}{\sum_{i=0}^{r}p_{0,i}^{(n_1)}p_{1,r-i}^{(n_2)}}.
	\end{align}

In the case of $r>R_0$, estimation of $\beta_{0,j}(n_1,n_2,r)$ and $\beta_{1,j}(n_1,n_2,r)$ is complicated as these probabilities depend on the arrival order of primary and secondary sessions, which cannot be determined from the state of the process. However, since the arrivals of the primary and secondary sessions are independent of each other and taking into account the known number of sessions of each type, any permutation of them in terms of the arrival time is equiprobable. Thus, the probability that the latest primary session takes the $k$-th place is
\begin{align}
	\binom{k-1}{n_1 -1} \Big/ \binom{n_1 + n_2}{n_1},
\end{align}
where $\binom{k-1}{n_1 -1}$ is the number of ways to place $n_1 - 1$ the primary session in the first $k-1$ places (since the last one will take the $k$-th place), and $\binom{n_1 + n_2}{n_1}$ is the overall number of ways to place $n_1$ primary sessions. The probability that $n_1$ primary sessions and $k-n_1$ secondary sessions occupy $i \leq R_0$ is calculated using the convolution
\begin{align}
\sum\limits_{s=0}^{i}p_{0,s}^{(n_1)}p_{1,i-s}^{(k-n_1)},
\end{align}

Further, the probability that $n_1$ primary and $n_2$ secondary sessions occupy $r$ resources, provided that the latest primary session takes the $k$-th place, is estimated as
\begin{align}
	\sum\limits_{i=0}^{\min(r,R_0)} p_{1,r-i}^{(n_2+n_1-k)} \sum\limits_{s=0}^{i}p_{0,s}^{(n_1)}p_{1,i-s}^{(k-n_1)}.
\end{align}

Finally, the probability that $n_1$ primary and $n_2$ secondary sessions together occupy $r$ resources has the following form
    \begin{align}
    \sum\limits_{k=n_1}^{n_1+n_2}\frac{\binom{k-1}{n_1 -1}}{\binom{n_1 + n_2}{n_1}}\sum\limits_{i=0}^{min(r,R_0)} p_{1,r-i}^{(n_2+n_1-k)} \sum\limits_{s=0}^{i}p_{0,s}^{(n_1)}p_{1,i-s}^{(k-n_1)}.
    \end{align}
    

Estimation of  $\beta_{0,j}(n_1,n_2,r)$ and $\beta_{1,j}(n_1,n_2,r)$ in the case $r>R_0$ is not an easy computational problem. To reduce the computational complexity, one may utilize approximation obtained by neglecting the order of session arrivals, i.e.,
\begin{align}\label{eq:beta_0}
&	\beta_{0,j}(n_1,n_2,r)=\frac{p_{0,j}\sum\limits_{i=0}^{\min(r-j,R_0-j)}p_{0,i}^{(n_1-1)}p_{1,r-j-i}^{(n_2)}}{\sum\limits_{i=0}^{\min(r,R_0)}p_{0,i}^{(n_1)}p_{1,r-i}^{(n_2)}},\nonumber\\
&	\beta_{1,j}(n_1,n_2,r)=\frac{p_{1,j}\sum\limits_{i=0}^{\min(r-j,R_0)}p_{0,i}^{(n_1)}p_{1,r-j-i}^{(n_2-1)}}{\sum\limits_{i=0}^{\min(r,R_0)}p_{0,i}^{(n_1)}p_{1,r-i}^{(n_2)}}.
\end{align}

\begin{figure}[t!]
	\centering
	\includegraphics[width=0.9\columnwidth]{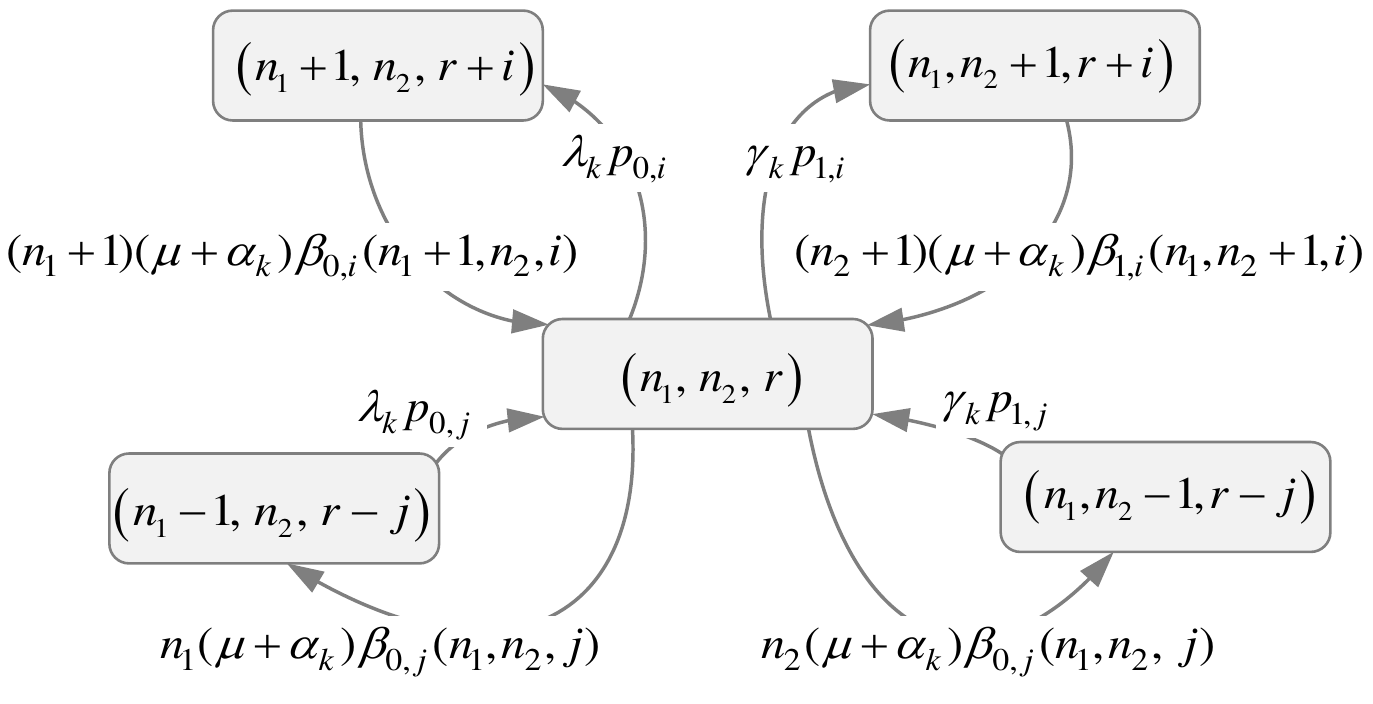}
	\caption{Typical set of states and associated transition intensities for model representing session service process with multiconnectivity and resource reservation \cite{begishev2021joint}.}
	\label{fig:state4}
	\vspace{-2mm}
\end{figure}


By utilizing $\beta_{0,j}(n_1,n_2,r)$ and $\beta_{1,j}(n_1,n_2,r)$ the system of equilibrium equations for the process $X(t)$ takes the form of (\ref{eq:equilibrium}). The typical subset of states and associated transition intensities are illustrated in Fig. \ref{fig:state4}.

\begin{figure*}[!!t]
	\vspace{-0mm}
	\begin{align}\label{eq:equilibrium}
		&\text{(i)}\qquad{}Q_0\left[ \lambda_k \sum_{j=0}^{R_0}p_{0,j}+\varphi_k\sum_{j=0}^{R_1}p_{1,j}\right]=(\mu+\alpha_k)\left[\sum_{j:(1,0,j)\in \Psi_{1,0}}\hspace{-3mm}Q_{1,0}(j)+\sum_{j:(0,1,j)\in \Psi_{0,1}}Q_{0,1}(j)\right],\nonumber\\ \nonumber\\
		&\text{(ii)}\qquad{}Q_{n_1,n_2}(r)\left[ \lambda_k\sum_{j=0}^{R_0-r}p_{0,j}+\varphi_k\sum_{j=0}^{R_1-r}p_{1,j}+(n_1+n_2)(\mu+\alpha_k)\right]= \lambda_k\hspace{-7mm}\sum_{j:(n_1-1,n_2,r-j)\in \Psi_{n_1-1,n_2}}\hspace{-12mm}p_{0,j}Q_{n_1-1,n_2}(r-j)+ \nonumber\\
		&\hspace{11mm}+\varphi_k\hspace{-12mm}\sum_{j:(n_1,n_2-1,r-j)\in \Psi_{n_1,n_2-1}}\hspace{-12mm}p_{1,j}Q_{n_1,n_2-1}(r-j)+(n_1+1)(\mu+\alpha_k)\hspace{-12mm}\sum_{j:(n_1+1,n_2,r+j)\in \Psi_{n_1+1,n_2}}\hspace{-12mm}Q_{n_1+1,n_2}(r+j)\beta_{0,j}(n_1+1,n_2,r+j)+\nonumber\\
		&\hspace{11mm}+(n_2+1)(\mu+\alpha_k)\hspace{-12mm}\sum_{j:(n_1,n_2+1,r+j)\in \Psi_{n_1,n_2+1}}\hspace{-12mm}Q_{n_1,n_2+1}(r+j)\beta_{1,j}(n_1,n_2+1,r+j), \quad n_1+n_2<N, \; r\leq R_0,  \nonumber \\ \nonumber\\
		&\text{(iii)}\qquad{}Q_{n_1,n_2}(r)\left[\varphi_k\sum_{j=0}^{R_1-r}p_{1,j}+(n_1+n_2)(\mu+\alpha_k)\right]=  \varphi_k\hspace{-7mm}\sum_{j:(n_1,n_2-1,r-j)\in \Psi_{n_1,n_2-1}(r-j)}\hspace{-14mm}p_{1,j}Q_{n_1,n_2-1}(r-j)+  \\
		&\hspace{12mm}+(n_1+1)(\mu+\alpha_k)\hspace{-12mm}\sum_{j:(n_1+1,n_2,r+j)\in \Psi_{n_1+1,n_2}}\hspace{-12mm}Q_{n_1+1,n_2}(r+j)\beta_{0,j}(n_1+1,n_2,r+j) + \nonumber\\
		&\hspace{12mm}+(n_2+1)(\mu+\alpha_k)\hspace{-12mm}\sum_{j:(n_1,n_2+1,r+j)\in \Psi_{n_1,n_2+1}}\hspace{-12mm}Q_{n_1,n_2+1}(r+j)\beta_{1,j}(n_1,n_2+1,r+j), \quad n_1+n_2<N, \; r>R_0,\nonumber \\ \nonumber\\
		&\text{(iv)}\qquad{}(n_1+n_2)(\mu+\alpha_k)Q_{n_1,n_2}(r)=\nonumber \\
		&\hspace{12mm}=\lambda_k\hspace{-12mm}\sum_{j:(n_1-1,n_2,r-j)\in \Psi_{n_1-1,n_2}}\hspace{-12mm}p_{0,j}Q_{n_1-1,n_2}(r-j)+\varphi_k\hspace{-6mm}\sum_{j:(n_1,n_2-1,r-j)}\hspace{-6mm}p_{1,j}Q_{n_1,n_2-1}(r-j), \quad n_1+n_2=N, \; r\leq R_0.\nonumber \\ \nonumber\\
		&\text{(v)}\qquad{}(n_1+n_2)(\mu+\alpha_k)Q_{n_1,n_2}(r)=\varphi_k\hspace{-12mm}\sum_{j:(n_1,n_2-1,r-j)\in \Psi_{n_1,n_2-1}}\hspace{-12mm}p_{1,j}Q_{n_1,n_2-1}(r-j),\quad n_1+n_2=N, \; r>R_0. \nonumber
	\end{align}
	\hrulefill
	\normalsize
	\vspace{-3mm}
\end{figure*}

\subsubsection{Performance Metrics} \label{sec4-4-3}

The system of equilibrium equations defined in (\ref{eq:equilibrium}) needs to be solved numerically. Due to the fact that the number of equations in the system can reach $N(N+1)R_1/2$, we recommend using special libraries for sparse matrices and associated iterative methods, i.e., Gauss-Seidel method \cite{he2014fundamentals,latouche1999introduction}. 

Once the stationary state distribution is obtained, performance metrics immediately follow. The new session drop probability $\pi_{b,k}$ at BS $k$ and the probability $\pi_{s,k}$ that rerouting leads to the drop of a session take the form
\begin{align}\label{eq:pib}
	&\pi_{b,k}=1-\hspace{-9mm}\sum_{0\leq n_1+n_2\leq N-1}\sum_{r\leq R_{0,k}:(n_1,n_2,r)\in \Psi_{n_1,n_2}}\hspace{-12mm}Q_{n_1,n_2}(r)\sum_{j=0}^{R_{0,k}-r}p_{0,j},\nonumber \\
	&\pi_{s,k}=1-\hspace{-9mm}\sum_{0\leq n_1+n_2\leq N-1}\sum_{r:(n_1,n_2,r)\in \Psi_{n_1,n_2}}\hspace{-10mm}Q_{n_1,n_2}(r)\sum_{j=0}^{R_{1,k}-r}p_{1,j}.
\end{align}
	
The intensities of secondary sessions are given by
\begin{align}\label{eq:phi}
	&\varphi_k=\sum_{v=1}^{\infty}\varphi_{k}^{v}, \nonumber\\
	&\varphi_k^1=\sum_{i=1}^{K}\lambda_{i} (1-\pi_{b,i})\frac{\alpha_i}{\mu+\alpha_i}\varphi_{i,k}^{0},\nonumber\\
	&\varphi_{k}^{v}=\sum_{i=1}^{K}\varphi_{i}^{v-1}(1-\pi_{s,i})\frac{\alpha_i}{\mu+\alpha_i}\varphi_{i,k}^{v-1}, v>1,
\end{align}
where $v$ denotes the level of the second session.


The network-wide counterparts of these metrics are
\begin{align}\label{eq:pb}
	\pi_b=\sum_{k=1}^{K}\frac{\lambda_k}{\lambda}\pi_{b,k},\, \pi_s=\sum_{k=1}^{K}\frac{\varphi_k}{\varphi}\pi_{s,k},
\end{align}
and they need to be utilized as shown in Fig. \ref{fig:algorithm}.

Finally, the probability that the initially accepted for service session is eventually dropped is given by 
\begin{align} \label{eq:piO}
	\pi_O=\lim_{t\rightarrow\infty} \frac{\varphi\pi_s t}{\lambda(1-\pi_b) t} = \frac{\varphi\pi_s}{\lambda(1-\pi_b)},
\end{align}
where the numerator is the mean number of accepted sessions that have been dropped during time $t$, and the denominator is the overall number of accepted sessions during time $t$.

\subsubsection{Illustrative Example}

\begin{figure}[t!]
	\centering
	\includegraphics[width=0.8\columnwidth]{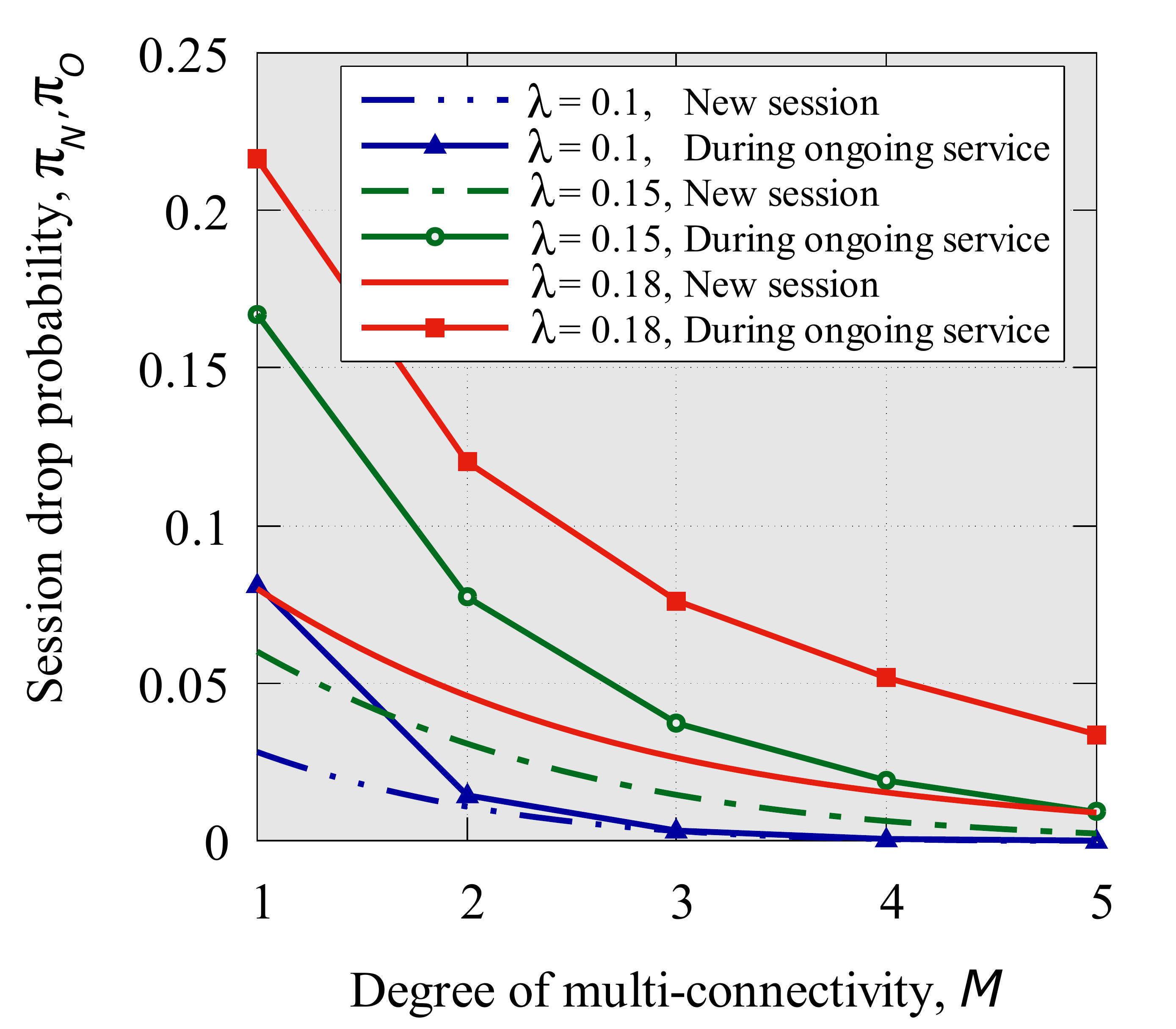}
	\caption{New and ongoing session drop probabilities: increasing the number of mmWave/THz BS accessible for UE drastically imrpoves service reliability including both new and ongoing session drop probability \cite{begishev2021joint}.}
	\label{fig:fig9anew}
	\vspace{-2mm}
\end{figure}

The new and ongoing session drop probabilities in presence multiconnectivity obtained using the framework discussed above are shown in Fig. \ref{fig:fig9anew} for the session rate of $10$ Mbps. Following \cite{begishev2021joint} we consider a circular deployment with a number of BSs evenly spaced over the circumference. As one may observe, increasing the degree of multiconnectivity leads to the dramatic reduction in both new and ongoing session drop probability. This is in contrast to resource reservation that improves ongoing session drop probability at the expense of new session loss probability inducing a trade-off between these two metrics. However, this gain comes at the expense of increased implementation complexity, signaling load and the decreased energy efficiency of UEs. On top of this, the efficient use of this functionality requires dense mmWave BS deployments that are only feasible at later stages of 5G rollouts. Using the described framework one may also capture the joint effects of resource reservation and multiconnectivity. For in-depth analysis of this scenario we refer to \cite{begishev2021joint}.

\subsection{Multiple Flows with Priorities}\label{sect:priorities}

Finally, we consider how to incorporate session priorities in the considered framework. To this aim, consider a resource queuing system with two types of sessions, (e.g., URLLC and eMBB), pre-emptive priority service discipline, $N$ servers and resource volume $R$. Here, we refer to prioritized sessions as first type sessions. Other sessions are referred to as second type sessions. Both types of sessions arrive according to mutually independent Poisson processes with parameters $\lambda_1$ and $\lambda_2$, respectively, and their service times are exponentially distributed with intensities $\mu_1$ and $\mu_2$. Besides, the resource requirements distributions are $\{p_{1,r}\}$ and $\{p_{2,r}\}$, $r \geq 0$.

The behavior of the considered system can be described by four-dimensional Markov process $\left( \xi_1(t), \delta_1(t), \xi_2(t), \delta_2(t) \right)$, where $\xi_i(t)$ is the number of $i$-th type sessions in the system and $\delta_i(t)$ is the total number of resources occupied by $i$-th type sessions at time $t$. However, the straightforward approach to obtain steady-state distribution of the number of sessions via numerical solution of the system of equilibrium equations is not feasible. The reason is the state space explosion with the number of states reaching $(N+1)(N+2)(R+1)(R+2)/4$. To decrease computational complexity of the solution, one may resort to the approximation by deducing performance metrics via marginal distributions of both types of sessions in the system. To demonstrate the basic principles, denote by $Q_i(n_i,r_i)$ the marginal distribution that $n_i$, $i=1,2$, sessions of $i$th type totally occupy $r_i$ resources. Observe that under the pre-emptive priority service discipline, the service process of the first type of sessions is not affected by the service process of the second type of sessions. Thus, the marginal distribution $Q_1(n_1,r_1)$ can be obtained by solving \eqref{eq:4} for the system with only first type of sessions.

To determine $Q_2(n_2,r_2)$ knowing $Q_1(n_1,r_1)$ one may utilize the following idea. Observe that when there are $n_2$ second type sessions that totally occupy $r_2$ resources, the number of priority sessions and resources occupied by them is given by the marginal distribution $Q_1(n_1,r_1)$ given that there are no more than $N-n_2$ of them, and they totally occupy no more than $R-r_2$ resources. Applying this idea, one can estimate the following auxiliary probabilities that depend on the number of second type sessions in the system and volume of resources occupied by them: (i) the probability that an arriving session of the second type is dropped and (ii) the probability that an arriving session of the first type interrupts the service process of $k$ sessions of the second type. With these probabilities in hand, one may further derive the system of equilibrium equations for the marginal distribution $Q_2(n_2,r_2)$, which is then solved numerically. The performance metrics of interest include drop probabilities of both types of sessions as well as the probability that a session of the second type will be interrupted during the service will immediately follow.


\subsubsection{Illustrative Example}

The described framework has been applied in \cite{sopin2020resource} to evaluate priority-based service between URLLC and eMBB sessions served as mmWave BS. We consider a single cell system with session arrivals governed by the Poisson process whose geometric locations are uniformly distributed in the cell service area. We illustrate high-priority and low-priority session drop probabilities as well as low-priority session interruption probability in Fig. \ref{fig:interup} as a function of high-priority session arrival intensity $\lambda_1$. As expected the high-priority traffic, e.g., URLLC, is well-isolated from the low-priority one. However, there is a complex interplay between low-priority session drop probability and session interruption probability, where the latter first decreases and then, starting from $\lambda_1\approx{}17$ sessions/s, increases. To reduce the latter undesirable effect of dropping sessions already accepted to service one may apply resource reservation for low-priority traffic. This type of system can also be analyzed within the proposed framework by combining the priority model with the resource reservation model considered above. Note that priority service, and thus, the proposed framework can also be applied to enable slicing at the air interface \cite{sallent2017radio,koucheryavy2021quantifying}.

\subsection{Multiconnectivity with Reservation for Adaptive Sessions}\label{sect:adaptive}

So far we have considered scenarios with intra-RAT multiconnectivity, where this functionality is utilized exclusively for mmWave or for THz RATs. A viable option is to utilize intra-RAT multiconnectivity, where UE may support active links with LTE or sub-6 GHz NR BSs along and with mmWave/THz BS simultaneously. Being inherently unaffected by blockage microwave BS in this scenario may provide an option for temporal offloading of sessions from mmWave/THz BS in case of blockage or micromobility. To address this case, consider a scenario with one BS operating in microwave band and several mmWave/THz BSs in its coverage area \cite{moltchanov2021performance}. Sessions from mmWave BSs are rerouted to the microwave BS in case of an outage. Due to inherent rate mismatch between microwave and mmWave/THz band we consider inherently adaptive traffic (also known as a full buffer in 3GPP specifications) that utilizes application layer adaptation to changing rate at the air interface. Since in this deployment the microwave BS is a natural bottleneck, we concentrate on performance assessment of sessions bottlenecked at this BS. To protect original traffic from the rerouted one, explicit reservations for both traffic types are utilized.

\subsubsection{Model Description} 

The microwave BS is modeled as a resource queuing system with $R$ units of resources (PRBs). A part of the resources $R_1$ is reserved for sessions that initially arrive at the microwave BS, and another part $R_2$, $R_1+R_2<R$, is reserved for sessions that are rerouted from mmWave BS in case of outage. In what follows, we refer to the former and latter as first and second type of sessions, respectively. Assume that both types of sessions arrive according to Poisson processes with intensities $\lambda_1$ and $\lambda_2$, respectively. Service times are exponentially distributed with intensities $\mu_1$ and $\mu_2$. Since sessions are adaptive in nature, the resources are divided equally between them, accounting for the reserved resource volumes and minimal requirements $x_1$ and $x_2$.

The behavior of the considered system can be described by a two-dimensional Markov process $X(t)=\{\xi_1(t),\xi_2(t)\}$, where $\xi_i$ is the number of $i$-type sessions in the system. The first type sessions may occupy no more than $R-R_2$ resources, while sessions of the second type -- $R-R_1$ resources. The rest of the resources is shared by both types of sessions. Then, the maximum number of the first type sessions is $K_1=\lfloor (R-R_2)/x_1 \rfloor$ and the maximum number of the second type sessions is $K_2=\lfloor (R-R_1)/x_2 \rfloor$. The set of states $\Psi$ is
\begin{equation}
\Psi \hspace{-1mm}=\hspace{-1mm} \left\{ (k_1,k_2)\hspace{-1mm}: k_1x_1+k_2x_2 \leq R, k_1 \leq K_1, k_2 \leq K_2 \right\}.
\end{equation}

\subsubsection{Equilibrium Equations}

The stationary probabilities $Q(k_1,k_2)$ can be found as the solution of the following set of equilibrium equations
\begin{align} \label{eq:elset}
		&Q(k_1,k_2)\left[ \lambda_1 I(x_1(k_1+1)+x_2k_2 \leq R, k_1<K_1)+ \right. \nonumber \\
		&+\lambda_2 I(x_1k_1+x_2(k_2+1) \leq R, k_2<K_2)+ \nonumber \\
		&\left.+k_1\mu_1 +k_2\mu_2 \right]=\lambda_1 I(k_1>0)Q(k_1-1,k_2)+ \nonumber \\
		&+\lambda_2 I(k_2>0)Q(k_1,k_2-1)+ \nonumber \\
		&+(k_1+1)\mu_1 I(x_1(k_1+1)+x_2k_2 \leq R, k_1<K_1)+ \nonumber \\
		&+(k_2+1)\mu_2 I(x_1k_1+x_2(k_2+1) \leq R, k_2<K_2),
\end{align}
where $I(\cdot)$ is the indicator function.

Since the Markov chain is reversible, it is easy to see that the solution of \eqref{eq:elset} has the following product form
\begin{align}
		&Q(k_1,k_2)=Q_0\frac{\rho_1^{k_1}}{k_1!}\frac{\rho_2^{k_2}}{k_2!}, \nonumber \\
		&Q_0=\left( \sum_{(k_1,k_2)\in \Psi} \frac{\rho_1^{k_1}}{k_1!}\frac{\rho_2^{k_2}}{k_2!} \right)^{-1},
\end{align}
where $\rho_i=\lambda_i/\mu_i$, $i=1,2$, is the load of $i$-type sessions.

\subsubsection{Performance Metrics}

Once the stationary-state probabilities are obtained, the session drop probabilities can be calculated. Let $\Psi_{b,i} \subset \Psi$, $i=1,2$, be the subset of states, where arriving $i$-type session is dropped, i.e.,
\begin{align}
\Psi_{b,i}&=\left\{ (k_1,k_2) \in \Psi: k_i=K_i \right\}  \cup \nonumber \\
& \cup  \left\{ (k_1,k_2) \in \Psi : k_1x_1+k_2x_2>R-x_i \right\}.
\end{align}
	
Then, the new session drop probability $\pi_b$ expressed by 
	\begin{align}
		&\pi_b=\sum_{(k_1,k_2) \in \Psi_{b,1}} Q(k_1,k_2).
\end{align}

Observe that the drop of the sessions initially arriving to microwave BS affects the ongoing session drop probability of sessions temporarily offloaded from mmWave/THz BSs. Also note that sessions originally arriving to the mmWave/THz BSs may suffer multiple outage events and thus temporal reroutes to microwave BS until their service completion. Thus, we first characterize the section type sessions drop intensity $\nu$ as
\begin{align}
\nu=\lambda_2 \sum_{(k_1,k_2) \in \Psi_{b,2}} Q(k_1,k_2).
\end{align}
	
Then, the ongoing session drop probability of sessions that originally arrive to mmWave/THz BS is given by the ratio of the drop intensity and arrival intensity, i.e.,
\begin{align}
\pi_t=\nu/\lambda_{\bullet},
\end{align}
where $\lambda_{\bullet}$ is the total arrival intensity at all mmWave BSs.

Finally, having defined $r_i(k_1,k_2)$ as the number of resources occupied by each $i$-type session provided that there are $k_1$ 1-type and $k_2$ 2-type sessions in the system, one can evaluate the average bitrate achieved for both type of sessions for a given fairness criterion using standard stochastic geometry technique \cite{moltchanov2021performance}, see also Section \ref{sect:approaches}.

\subsection{Performance Models for Elastic Traffic}\label{sec:servcoefs}

The least investigated type of traffic in mmWave/THz cellular systems is an elastic one, where application level rate adaptation is utilized similarly to adaptive traffic but the session itself requires a certain amount to bytes to be transferred. In other words, the service time for this traffic type depends on the rate provided to sessions during the service. Consider now a scenario with a mmWave/THz BS serving elastic sessions, whose service times depend on the volume of occupied resources. Assume that all the resources of the BS are equally divided between all sessions. Note that depending on the selected MCSs, the resulting bitrate may differ among sessions. Such types of service models are significantly more complex as compared to those utilized for adaptive traffic and similarly to non-elastic/adaptive traffic requires the use of queuing models with PS service discipline \cite{yashkov2007processor}.

\subsubsection{Model Description} 

Consider single a single server queue with PS discipline and unit service rate. Sessions arrive according to the Poisson process with intensity $\lambda$, and the session volumes (e.g., file size) have exponential distribution with parameter $\mu$. Each session is associated with a random service rate coefficient $v$ with values from the set $V=\{v_1,v_2,\dots,v_L\}$. Without loss of generality, we assume that values in set $V$ are arranged in the ascending order, i.e. $v_1<v_2<\dots<v_L$. Service rate coefficients of different sessions are assumed to be independent identically distributed (iid) random variables with probability distribution $\{p_{v_i}\}, i=1,2,\dots,L$. These coefficients abstract the impact of MCSs by inducing different service rates to the currently served sessions. They are determined on arrival time instant of a session by utilizing a propagation model and remain unchanged during the service time.

The system operates as follows. If there are $n$ sessions with service rate coefficients $u_1,u_2,\dots,u_n$, each session receives $\frac{1}{n}$ of the processing rate, and the service rate of the $i$th session is $\frac{u_i}{n}$. Note that the total service intensity in this state is $\frac{1}{n} \sum_{i=1}^n u_i$. The number of simultaneously served sessions is no more than $N$, which is determined as the maximum number of sessions such that $\frac{v_1}{N}$ is still greater or equal to the minimum required bitrate. Based on these facts, the behavior of the system can be described by the number of sessions and the vector of their service rate coefficients. However, to simplify the description, one can use a state aggregation approach similar to the one applied for the resource queuing systems. According to it, the system's behavior is described by two-dimensional Markov process $X(t)=\{\xi(t), \delta(t)\}$, where $\xi(t)$ is the number of sessions and $\delta(t)$ is the sum of their service rate coefficients. The set of states is then
\begin{equation}
\Psi=\left\{ (n,u): 0 \leq n \leq N, p_u^{(n)}>0 \right\},
\end{equation} 
where $p_u^{(n)}$ is the probability that the sum of service rate coefficients of $n$ sessions is equal to $u$. As in previous sections, these probabilities can be evaluated by utilizing $n$-fold convolution of the distribution $\{p_{v_i}\}, i=1,2,\dots,L$.

Similarly to the resource queuing systems, we cannot exactly determine the exact amount of resources freed upon the departure of a session. Contrarily, we can utilize the Bayesian approach to evaluate probabilities $\theta_i(n,u)$ that the service rate coefficient of a departing session is $i$, given that before the departure, the sum of $n$ serving rate coefficients was $u$
\begin{equation}
\theta_i(n,u)=\frac{p_i p_{u-i}^{(n-1)}}{p_u^{(n)}}, i \leq u, (n,u) \in \Psi.
\end{equation} 

\subsubsection{Equilibrium Equations}

The stationary probabilities $q_n(u)$ that there are $n$ sessions in the system, and the sum of their service rate coefficients is $u$, can be found as the solution of the following balance equations
\begin{align}\label{eq:surservrate}
		&\lambda q_0=\mu \sum_{l=1}^L v_l q_{v_l}(v_l); \nonumber \\
		&\left( \lambda+\frac{\mu u}{n} \right) q_n(u)=\lambda \sum_{l=1}^L p_{v_l}q_{n-1}(u-v_l) + \nonumber \\
		&+\frac{\mu}{n+1} \sum_{l=1}^L (u+v_l)q_{n+1}(u+v_l) \theta_{v_l}(n+1,u+v_l), \\
		& n=1,2,..,N-1, (n,u) \in \Psi; \nonumber \\
		& \frac{\mu u}{n}q_N(u)=\lambda \sum_{l=1}^L p_{v_l} q_{N-1}(u-v_l), (N,u) \in \Psi. \nonumber
\end{align}

The system \eqref{eq:surservrate} together with the normalization condition can be solved by any appropriate numerical method. Note that by introducing the proper order in the set of states, the generator matrix can be written in block-tridiagonal form suitable for numerical algorithms \cite{meyer2000matrix}.

\subsubsection{Performance Metrics}

Once the stationary state distribution is obtained performance metrics can be calculated. For example, an arriving session is dropped if it finds that there are $n$ sessions in the system. Therefore, the session drop probability $\pi$ is provided by
	\begin{equation}
		\pi=\sum_{(N,u) \in \Psi} q_N(u).
\end{equation}  

Other important metrics are characteristics of the session serving time. The moments of the session serving time can calculated by employing the phase method \cite{asmussen}. For more details please refer to \cite{sopin2021coefs}. We note that the proposed approach can be further extended similarly to the resource queuing systems to include the outage events caused by blockage and/or micromobility as well as to the support of resource reservation and multiconnectivity.

\section{Conclusions and Future Challenges}\label{sect:concl}

MmWave and THz communications systems are expected to build the foundation of future cellular access providing the extreme amount of resources at the air interface. These systems are primarily targeted to support principally new applications such as AR/VR, holographic telepresence, characterized by high bitrate non-elastic/adaptive traffic patterns. However, intrinsic characteristics of mmWave and THz frequency bands such as extreme path losses, blockage, and micromobility makes provisioning of QoS guarantees to these applications an extremely complex task inherently requiring complex mechanisms to maintain session continuity.

In this paper, we provided a tutorial on mathematical analysis of mmWave and THz deployments with non-elastic/adaptive traffic patterns. Particularly, we extended the conventional methods of stochastic geometry that are inherently limited to systems supporting adaptive (full buffer) traffic only, to the case of composite models capturing both radio specifics of mmWave and THz bands as well as traffic service dynamics at BSs. These two parts of the composite framework are interrelated to each other via a well-defined interface parameters. The two main blocks in the framework are: (i) radio part abstraction and (ii) service model. The former is specified in terms of radio part sub-models in Section \ref{sect:system} and is then abstracted by utilizing the predefined set of parameters suitable for the service model as shown in Section \ref{sect:param}. To this aim, we also provided a comprehensive review of analytically tractable models of various components utilized for building the modeling scenarios including deployment, propagation, antenna, blockage, micromobility, beamsearching , traffic and service models under different system and environmental conditions. The service models utilized at the last stage of the framework are provided in Section \ref{sect:perf}. We note that depending on the investigators’ needs some of the radio part models can be dropped, e.g., if the antenna arrays are not extremely directional one may drop micromobility model out of consideration or when the density of human crowd is not significant the blockage model can be omitted. Additionally, one is also free to select the required models based on the needs, e.g., the required accuracy and/or the resulting framework complexity. That is, we provide our readers with a set of building blocks for specifying their own models suitable for a given scenario.

The modular structure of the framework allows for a potential reuse of its core parts for studying various mmWave and THz deployments (including combinations of them) characterized by different scenario geometry, blocker types and their mobility, antenna arrays, micromobility patterns of applications, network associations, reliability and rate improvement mechanisms. Finally, we considered several applications including systems with resource reservation, multiconnectivity, and priorities between arriving sessions. 

\subsection{Future Challenges In Performance Modeling}


The emergence of bandwidth-rich access interfaces in 5G/6G systems utilizing mmWave/THz bands brings principally new challenges to system designers. As compared to previous generation of cellular systems the access becomes truly unreliable with blockage and micromobility leading to frequent and these challenges cannot be solved by utilizing physical- or link-layer functionalities. To ensure reliable service, future 5G/6G RAT system will include networking functionalities, where UEs will be serviced collectively by a set of BS by utilizing multiconnectivity functionality Along these lines, is the future integrated access and backhaul (IAB) architecture recently proposed by 3GPP \cite{3gpp_iab}. By utilizing state-of-the-art mechanisms such as multiconnectivity, multi-beam directional antennas as well as being inherent based on multi-hop topology these system will form the basis for future reliable access in 5G/6G networks.


As indicated in radio part ``cookbook'' provided in Table \ref{tab:radioCookbook}, the radio models including propagation, antenna array, blockage, micromobility, and beamalignment models have been well developed so far. These components can be joined together to abstract the radio part as demonstrated in Section \ref{sect:param} to provide the input to appropriate service models. Contrarily, the service part cookbook provided in Table \ref{tab:serviceCookbook} is half-empty implying that critical service models having essential networking feature are still missing. This is specifically the case of non-elastic/adaptive and elastic applications with minimum rate guarantees. Thus, in general, we may conclude that the mathematical theory of  future 5G/6G cellular system lacks appropriate service models. Below, we provide few examples of research directions related to the field of advanced queuing theory that would allow to build new service models.


One of the natural generalizations of resource queuing systems is the addition of the waiting buffer, as some type of user sessions may tolerate waiting. However, unlike conventional queues, the main feature of resource queuing systems -- heterogeneity of session requirements -- cause many new previously unexplored problems. As an example, consider the case when, at some moment of time, there are $r$ occupied resources and $R-r$ are available for new sessions. If a new session arrives and requires $r_1>R-r$ resources, it cannot be served immediately and, thus, joins  queue. However, if another session arrives and requires $r_2 \leq R-r$ resources, the system should decide, whether to accept the last arrived session or to force it to wait until the first session is accepted for service. Another problem is associated with moments of session departures. If a "heavy" session departs from the system, then more than one sessions from the queue can be accepted for service. The number of sessions that can replace the departing "heavy" session strongly depends on the session service strategy.


Analysis of resource queuing systems with waiting and external events representing outages caused by blockage and/or micromobility arrivals that change the resource requirements of sessions also has its own specifics. For example, the LoS blocked/non-blocked states can change even for session that currently wait for service, so that the blockage and micromobility will affect not only those sessions that are currently being served. Besides, the waiting buffer provides several opportunities for processing of the interrupted sessions. For example, upon the change of LoS state, a session increased its resource requirements so that unoccupied resources of the system are not enough to meet the new requirements. The session can be dropped, as it is done in resource queuing systems without buffer, or placed to the waiting buffer, or even placed to the special prioritized buffer. The theory of resource queuing systems should be further developed to capture these additional specifics.

The research on the queuing systems with random service rate coefficients suitable for modeling elastic traffic with minimum rate requirements described in Section \ref{sec:servcoefs} has just been initiated. Similarly to the models for non-elastic/adaptive applications, those models can be expanded to include evens signaling outages caused by mmWave/THz specific blockage and micromobility that would change the rate coefficients dynamically during the service process of sessions. Besides, the waiting buffer can be easily incorporated, since the serving rate coefficients do not affect the acceptance procedure.

Finally, yet another direction is supplementing the resource queuing systems with random service rate coefficients, where a part of resources is reserved for non-elastic/adaptive sessions, and the rest -- for elastic or adaptive sessions.

\balance
\bibliographystyle{IEEEtran}

\end{document}